\documentclass[aps,prd,onecolumn,eqsecnum, nofootinbib]{revtex4-2}
\usepackage{bm}
\usepackage{amssymb}
\usepackage{amsmath}
\usepackage{amsfonts}
\usepackage{graphicx}
\usepackage{hyperref}
\DeclareSymbolFont{EulerScript}{U}{eus}{m}{n}
\SetSymbolFont{EulerScript}{bold}{U}{eus}{b}{n}
\DeclareSymbolFontAlphabet\scrpt{EulerScript}
\newcommand{\gothg}{\mathfrak{g}} 
\newcommand{\elltide}{{\mbox{\scriptsize $\ell$-tide}}} 
\newcommand{\ellmass}{{\mbox{\scriptsize $\ell$-mass}}} 
\newcommand{\elltot}{{\mbox{\scriptsize $\ell$-tot}}} 
\newcommand{\stf}[1]{{\langle #1 \rangle}}
\newcommand{\pn}{{\mbox{\sc pn}}} 
\newcommand{\W}{{\scrpt W}} 
\newcommand{\m}{{\sf m}} 
\allowdisplaybreaks
\begin{document}
\title{Tidally induced multipole moments of a charged material body}  
\author{Victoria Leaker}
\email{vleak060@uottawa.ca}
\author{Tristan Pitre}
\email{tpitre@uoguelph.ca}
\author{Eric Poisson}
\email{epoisson@uoguelph.ca}
\affiliation{Department of Physics, University of Guelph, Guelph, Ontario, N1G 2W1, Canada} 
\date{December 3, 2024}
\begin{abstract}
We define and calculate the mass multipole moments of a material body of mass $M$ and electric charge $Q$ tidally deformed by a particle of mass $m \ll M$ and charge $q \ll Q$ placed at a distance $r_0$ from the body. Given $Q/M$ and $r_0$, we choose $q/m$ so that the gravitational attraction between body and particle is balanced by the electrostatic repulsion; the system can then be maintained in a static state. The multipole moments are defined in a setting in which the body's self-gravity is allowed to be strong, but the mutual gravity between  body and companion is required to be weak. In this setting, the body is described in full general relativity, in terms of a perturbed metric and electromagnetic potential characterized by tidal constants, and the mutual gravity is described within the post-Newtonian approximation to general relativity, in terms of objects with a multipole structure. Matching the different descriptions of the same field delivers a relation between the tidal constants and the multipole moments. In our implementation of this program, the calculation is performed in full Einstein-Maxwell theory (as a linearized perturbation of the unperturbed field), without appeal to a post-Newtonian approximation. After the fact we take $M/r_0$ to be small and carry out an expansion of the metric in powers of $M/r$ to obtain the multipole moments and associated Love numbers. The calculations are performed for a body made up of a perfect fluid with a uniform ratio of charge to mass densities, governed by a polytropic equation of state. We show that the Love numbers of a charged body in a situation of balanced gravitational and electrostatic forces are negative. The statement remains true even when $Q/M$ is very small, and we conclude that the tidal deformability of a charged body is radically different from that of an uncharged object, for which the Love numbers are positive.  
\end{abstract} 
\maketitle
\tableofcontents

\section{Introduction and summary} 
\label{sec:intro} 

In this paper we define and calculate the mass multipole moments of a material body of mass $M$ and electric charge $Q$ tidally deformed by a particle of mass $m \ll M$ and charge $q \ll Q$ placed at a distance $r_0$ from the body. For given $Q/M$ and $r_0$ we choose $q/m$ so that the gravitational attraction between body and particle is balanced by the electrostatic repulsion; the system can then be maintained in a static state. Before we present our results and explain why we chose to spend time and effort on this largely academic problem --- after all, few stars are electrically charged and few binary systems are static --- we first describe the context behind this work.

\subsection{Context: Tidal dynamics in binary inspirals}

The tidal dynamics of compact bodies in the late stages of a binary inspiral can disclose important information regarding each body's internal structure, which manifests itself as a subtle modulation of the gravitational-wave signal \cite{flanagan-hinderer:08}. While the vast majority of events measured by the LIGO, Virgo, and KAGRA instruments implicate black holes, which have no discernable internal structure, a few detections feature a neutron star, whose internal structure is determined by the poorly understood properties of nuclear matter at extreme densities \cite{ozel-freire:16, oertel-etal:17, baym-etal:18}. In such cases, a measurement of the tidal deformability of a neutron star can produce insightful constraints on the equation of state of nuclear matter beyond the saturation density. In particular, a measurement carried out for GW170817 \cite{GW170817:17, GW170817:18, narikawa-uchikata-tanaka:21} produced an upper bound for the tidal deformability which favors a relatively soft equation of state that gives rise to a relatively small neutron star \cite{landry-essick-chatziioannou:20}. These exciting developments are reviewed in Ref.~\cite{chatziioannou:20}, and prospects for future measurements are summarized in Ref.~\cite{pacilio-maselli-fasano-pani:22}. Future detections, in particular, will access the regime of dynamical tides \cite{hinderer-etal:16, steinhoff-etal:16}, in which an approach to resonance can significantly enhance the tidal response of a neutron star and deliver yet more insights into the nature of nuclear matter at extreme densities \cite{andersson-ho:18, schmidt-hinderer:19, poisson:20a, steinhoff-etal:21, andersson-pnigouras:21, passamonti-andersson-pnigouras:21, passamonti-andersson-pnigouras:22, williams-pratten-schmidt:22, pratten-schmidt-williams:22, ho-andersson:23, mandal-etal:23, mandal-etal:24, pitre-poisson:24, hegade-ripley-yunes:24a, hegade-ripley-yunes:24b, katagiri-yagi-cardoso:24}.  

Black holes have no internal structure to elucidate, but the statement that they possess a vanishing tidal deformability \cite{damour-nagar:09, binnington-poisson:09} continues to fascinate \cite{porto:16, cardoso-etal:17, chirenti-posada-guedes:20, hui-etal:21a, chia:21, charalambous-dubovsky-ivanov:21a, letiec-casals:21, letiec-casals-franzin:21, iteanu-etal:24, kehagias-riotto:24}. A hidden ladder symmetry governing the perturbations of black holes was recently unearthed and proposed as an explanation for this phenomenon \cite{charalambous-dubovsky-ivanov:21b, hui-etal:21b, rai-santoni:24, combaluzier-szteinsznaider-etal:24}.

\subsection{Context: Love numbers, tidal constants, Wilson coefficients}

The tidal deformability of compact objects (neutron stars and black holes) is usually described in the literature in terms of {\it Love numbers}, or frequency-dependent Love functions. These are dimensionless and scalefree measures of the deformability, and are meant to be parameters that appear as observables in models of gravitational-wave signals. A difficulty that arises in a survey of this literature is that many distinct things have been named Love numbers (or functions). Another difficulty concerns the definition of the parameters that appear as tidal observables in gravitational-wave models; what is their precise link with the various notions of Love numbers that have been introduced? These issues underline the importance of providing a precise operational definition to Love numbers, a point that was emphasized in Refs.~\cite{gralla:18, poisson:21a}. 

While a precise definition of Love numbers can be elusive, the more primitive notion of {\it metric tidal constant} is easier to define. The tidal constants are parameters that occur in the exterior metric of a tidally deformed body, which characterize the body's response to an applied tidal field. The metric is presented in a given coordinate system, the perturbation is calculated in a given gauge, and given those choices, a tidal constant is an overall multiplicative factor in front of a decaying solution to the perturbation equations\footnote{The definition applies to both linear and nonlinear perturbations of a compact body. In the nonlinear case, the perturbation equations at each new order in perturbation theory involve the same differential operators as in the linearized case, together with source terms constructed from the solution at the preceding order. The general solution to these equations is the sum of a particular solution to the sourced equations and a general solution to the sourcefree (homogeneous) equations, which are identical to the linearized equations. A tidal constant is an overall multiplicative factor in front of the decaying piece of this solution.}; for a multipole of order $\ell$, this solution comes with a leading term of order $r^{-(\ell+1)}$ in an expansion in inverse powers of $r$. Because the tidal constant comes in front of an entire solution to the perturbation equations, its definition is unambiguous; the complete gauge fixing of the metric confers it with a gauge-invariant meaning. The numerical value of each tidal constant is determined by connecting the exterior metric with the body's interior metric at the stellar surface; this provides the link between the tidal constants and the body's equation of state. This is the calculation that is performed when the tidal deformability of a neutron star is related to the properties of its internal structure (see, for example, Ref.~\cite{postnikov-etal:10}). In the case of a black hole, the metric tidal constants vanish \cite{damour-nagar:09, binnington-poisson:09}.  

The key question that comes next is how to link the metric tidal constants, which carry information about the internal structure of a compact body, to gravitational-wave observables. This question is challenging because the metric of a tidally deformed body is limited by construction to a small region of spacetime that surrounds the body and is situated deep within the near zone (with distances from the body that are small compared with a typical wavelength of the gravitational radiation), while gravitational waves are measured far into the wave zone (with distances that are large compared with the wavelength). The information about the internal structure must therefore be propagated from the near zone to the wave zone.
 
The most developed method to describe the dynamics of tidally deformed bodies and calculate the emitted gravitational waves combines the mature tools of post-Newtonian theory \cite{blanchet:24} with powerful techniques of effective field theory; it treats each compact body as an effective point particle \cite{goldberger-rothstein:06a, goldberger-rothstein:06b, damour-nagar:10, porto:16, bini-damour-geralico:20, levi:20, henry-faye-blanchet:20a, henry-faye-blanchet:20b, henry-faye-blanchet:20c}. In this description, each body moves on a world line and possesses an action in which additional terms are inserted within the familiar $S = -M \int\, d\tau$ to account for the tidal deformation. (Here $M$ is the body's mass, and $\tau$ is proper time on the world line.) These are constructed from the spacetime's Riemann tensor (and derivatives), and {\it Wilson coefficients} are introduced to account for the body's internal degrees of freedom, which are ``integrated out'' in the effective theory. Because these additional terms are highly singular on the world line, the theory must be regularized and the Wilson coefficients subjected to a renormalization flow. Variation of the effective action with respect to the world line gives rise to equations of motion for the body, and variation with respect to the metric produces an energy-momentum tensor that acts as a source in the Einstein field equations. The solution to these equations, once evaluated in the wave zone, describes gravitational waves emitted by a binary system of tidally deformed bodies; the waves are parametrized by the Wilson coefficients. These methods provide a global description of the spacetime metric, from deep inside the near zone to far away into the wave zone, but they do not provide a direct link with the body's internal structure. For this, a mapping between the metric tidal constants and the Wilson coefficients must be established. While it appears to be well understood that the Wilson coefficients associated with a black hole must vanish (as the tidal constants do) \cite{kol-smolkin:12, hui-etal:21a, charalambous-dubovsky-ivanov:21a}, the mapping has not yet been obtained for material bodies.

\subsection{Context: Tidally induced multipole moments of compact objects}
\label{subsec:multipole}

An approach to provide operational meaning to the metric tidal constants, and eventually bring them into contact with gravitational-wave observables, was initiated in Ref.~\cite{poisson:21a}. It treats each compact body as an extended object with strong internal gravity, but the mutual gravity between bodies is required to be weak. In such a setting, each body is described in full general relativity in terms of a perturbed metric characterized by tidal constants, but the mutual gravity is described within the post-Newtonian approximation to general relativity, in terms of objects with a multipole structure. Matching the different descriptions of the same gravitational field delivers a relation between the tidal constants and the multipole moments. The tidally induced multipole moments of each body, a property of the post-Newtonian field, are then related to the tidal field created by the companion body, and a meaningful notion of {\it Love numbers} is obtained through this relationship. The Love numbers are defined by the multipole moments, they are related to the tidal constants by the matching procedure, and we have a direct link between the body's internal structure and the tidally induced multipole moments. Note that this method provides us with a sound definition of multipole moments for individual bodies in a dynamical spacetime\footnote{In the exact formulation of general relativity, multipole moments can be defined only for an entire spacetime when the spacetime is stationary and asymptotically flat \cite{geroch:70, hansen:74}.}; the post-Newtonian approximation supplies an essential ingredient in this construction. 

The approach described in the preceding paragraph returns precise definitions for the mass multipole moments ${\cal Q}^{(\ell)}$ of a tidally deformed body, and for the tidal multipole moments ${\cal E}^{(\ell)}$ that provide a description of the tidal environment in which the body is immersed; here $\ell$ is the multipolar order. When the tidal deformation is sufficiently small that it can be described as a linearized perturbation of a spherical body, and when the tidal field varies so slowly that it can be idealized as static (nonlinear and dynamical situations were also considered in Ref.~\cite{poisson:21a}), the {\it Love number} $k_\ell$ is defined precisely through the relation 
\begin{equation}
{\cal Q}^{(\ell)} = -\frac{2 (\ell-2)!}{(2\ell-1)!!}\, k_\ell\, R^{2\ell+1}\, {\cal E}^{(\ell)}, 
\label{Q_vs_E_intro} 
\end{equation}
in which $R$ is the body's unperturbed radius. In addition, the approach permits a determination of $k_\ell$ in terms of the tidal constants that appear in the exterior metric of the deformed body. Thus, while the multipole moments are creatures of the post-Newtonian spacetime (which describes the weak mutual gravity between bodies), the Love numbers --- through their link with the tidal constants --- are a creature of general relativity in its strong-field aspects.

The matching of metrics that gives rise to the relationship between ${\cal Q}^{(\ell)}$ and the tidal constants was carried out through the first post-Newtonian order in Ref.~\cite{poisson:21a}. To obtain the multipole moments of a tidally deformed black hole, however, the matching would have to be pushed at least through the fifth post-Newtonian order, an objective that is currently out of reach. The reason is that for $\ell = 2$ --- the lowest multipolar order --- the factor of $R^{2\ell+1}$ in Eq.~(\ref{Q_vs_E_intro}) becomes $R^5$, which is equal to $(2M)^5$ for a Schwarzschild black hole; the five powers of $M$ are necessarily associated with a quantity of the fifth post-Newtonian order. To overcome this rather formidable obstacle, it was envisioned in Ref.~\cite{poisson:21b} to endow the black hole with a small electric charge, to let it be part of a binary system with a charged particle, and to contrive the system into a static state through a careful balance between gravitational attraction and electrostatic repulsion. In this setup, the tidal deformation of the black hole is created by the charged particle, and it can be calculated exactly in full general relativity (as a perturbation of the Reissner-Nordstr\"om black hole). A post-Newtonian expansion of the resulting metric, carried out to all orders, then delivers the mass multipole moments of the black hole, with the outcome that they all vanish. It then follows from  Eq.~(\ref{Q_vs_E_intro}) that $k_\ell = 0$ for a charged black hole, and we arrive at an operational meaning behind the statement that Love numbers vanish for a black hole. It was conjectured in Ref.~\cite{poisson:21b} that the black-hole charge was an inconsequential device in this calculation; the charge could be as small as one wishes (though not identically zero, by virtue of the requirement of balanced forces), and it would not change the intrinsic properties of the black hole by much. The implication was that if the tidally induced multipole moments vanish for a very small charge, they should continue to vanish for a strictly zero charge.

\subsection{This work}
\label{subsec:intro_work}

It appeared to us important to test the conjecture by extending the calculation of Ref.~\cite{poisson:21b} to a material body instead of a black hole. The conjecture would be verified if the tidally induced multipole moments of a body with a very small charge could be shown to be very close to those of a body with vanishing charge. We carry out these computations in this paper, and ascertain whether the conjecture is correct. 

We consider a material body of mass $M$, charge $Q$, and radius $R$ tidally deformed by a particle of mass $m \ll M$ and charge $q \ll Q$ situated at a distance $r_0$ from the body. We select the parameters in such a way that the gravitational attraction between body and particle is precisely balanced by the electrostatic repulsion, so that the system can be maintained in a static state. For concreteness we take the body to be made up of a perfect fluid with a uniform ratio $\beta := \rho_e/\rho_m$ of charge to mass densities, and we choose its equation of state to be of the polytropic form $p \propto \rho_m^{1+1/n}$, where $n$ is a constant polytropic index. (Here, $\rho_m$ is the density of baryonic mass within the fluid, $\rho_e$ is the density of charge, and $p$ is the fluid pressure.) The perturbation to the metric and electromagnetic potential is calculated to first order in $m$ and $q$, both inside and outside the body; the calculation is performed exactly, without appeal to a post-Newtonian approximation. After the fact we take $M/r_0$ to be small and carry out an expansion of the metric in powers of $M/r$; the expansion delivers the moments ${\cal Q}^{(\ell)}$ and ${\cal E}^{(\ell)}$, and Eq.~(\ref{Q_vs_E_intro}) defines the Love numbers $k_\ell$ of a tidally deformed, charged star.

\begin{figure}
\includegraphics[width=0.49\linewidth]{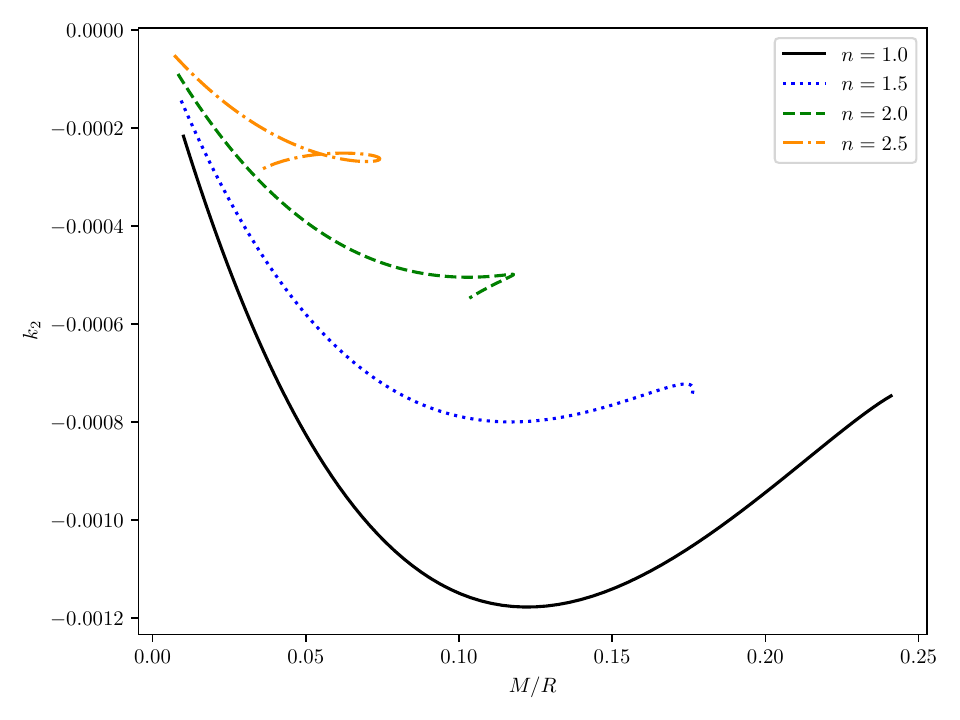}
\includegraphics[width=0.49\linewidth]{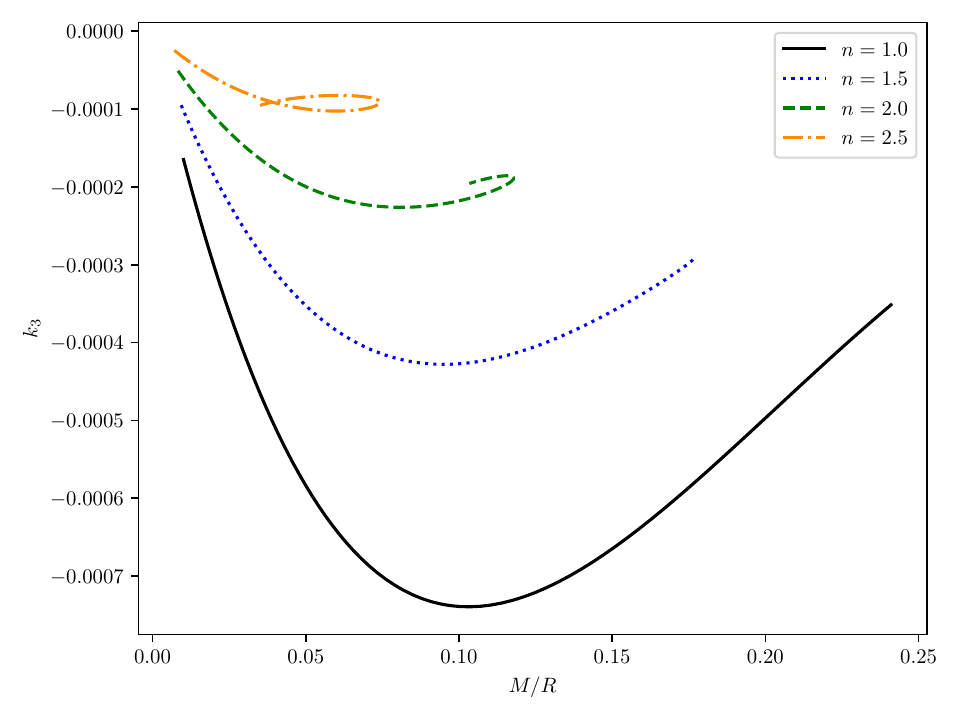}
\caption{Love numbers $k_2$ (left) and $k_3$ (right) plotted as functions of the stellar compactness $M/R$. These are computed for a body with a ratio $\beta = 0.2$ of charge to mass densities, for selected values of the polytropic index $n$. The Love numbers are negative, and they approach zero as $M/R \to 0$. For larger values of $n$ we see that the Love numbers are multi-valued functions of the compactness. This has to do with the fact $M/R$ first increases and then decreases along the sequence of equilibrium configurations; the Love numbers are single-valued when presented as functions of the central density $\rho_m(r=0)$.}
\label{fig:fig1} 
\end{figure}

\begin{figure}
\includegraphics[width=0.49\linewidth]{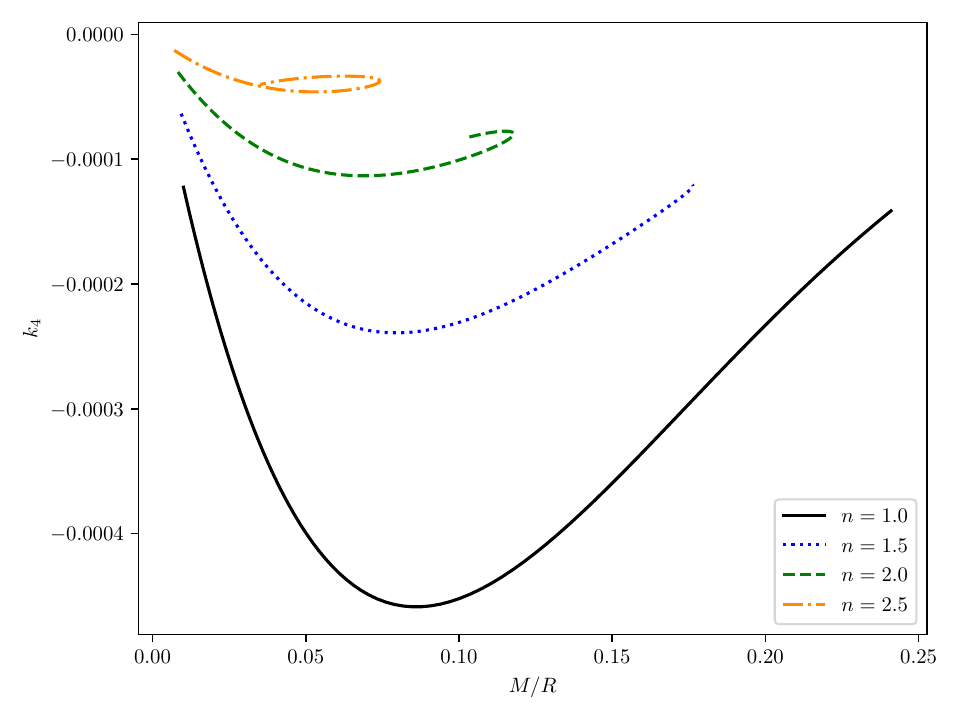}
\includegraphics[width=0.49\linewidth]{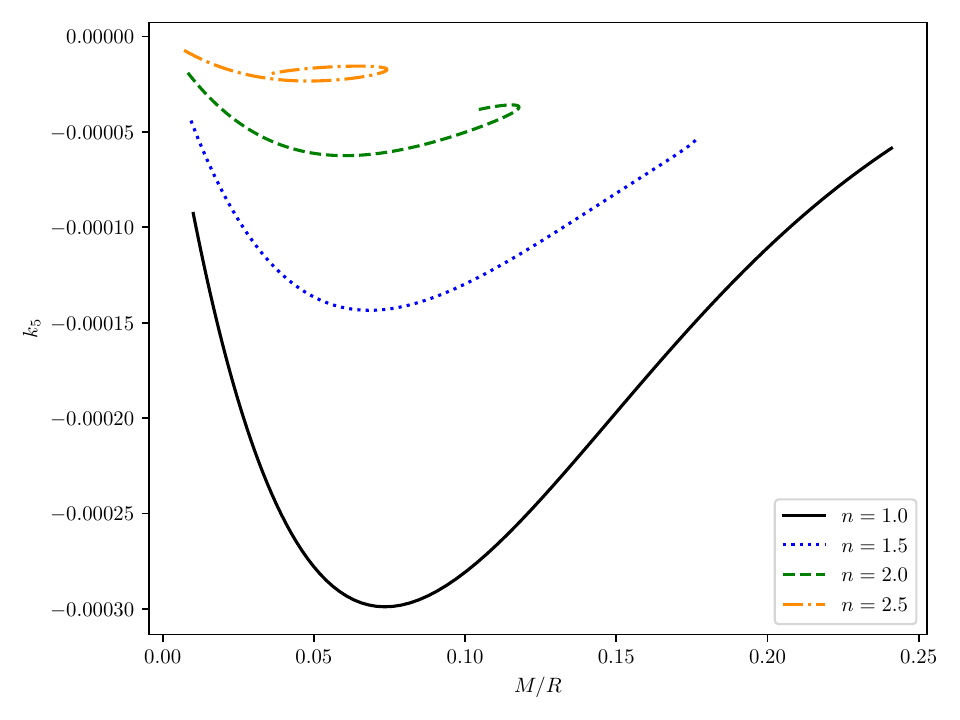}
\caption{Love numbers $k_4$ (left) and $k_5$ (right) plotted as functions of the stellar compactness $M/R$. These are computed for $\beta = 0.2$ and selected values of the polytropic index $n$.}
\label{fig:fig2} 
\end{figure}

We display a sample of our main results in Figs.~\ref{fig:fig1} and \ref{fig:fig2}, which present plots of $k_\ell$ as a function of the stellar compactness $M/R$, for $\ell = \{2, 3, 4, 5\}$. The Love numbers are computed for polytropic models with selected values of the index $n$, and for a ratio $\beta = 0.2$ of charge to mass densities. A quick glance at the figures discloses that {\it the conjecture is decidedly not verified: the Love numbers of a charged star in a situation of balanced gravitational and electrostatic forces are negative}, while those of an uncharged material body are positive. The Love numbers were also computed for smaller values of $\beta$, and the results presented in Sec.~\ref{subsec:love_results} indicate that the formal limit $\beta \to 0$ appears to exist (in spite of the fact that it can only be reached by sending $q/m$ to infinity); the Love numbers stay negative and the conclusion is not altered. Our numerical results also indicate that when $M/R$ is small, the Love numbers behave as $k_\ell = -(\mbox{constant}) M/R$ and therefore go to zero in the limit $M/R \to 0$; this is unlike the Love numbers of an uncharged polytrope, which approach a nonzero constant in the limit of zero compactness. (Here, the constant depends on $\ell$ and the choice of polytropic index.) We find, therefore, that the tidally induced multipole moments of an electrically charged body are nothing like those of an uncharged body.

The conclusion is not too surprising in retrospect. Consider a Newtonian version of this problem. We have a system that consists of a body of mass $M$ and charge $Q$ with a particle of mass $m$ and charge $q$. In units in which $G = 1$ and $4\pi \epsilon_0 = 1$, the gravitational and electrostatic forces are balanced when $mM = qQ$. Consider now a fluid element within the body, of mass $\delta M$ and charge $\delta Q$, with $\delta Q = \beta\, \delta M = (Q/M)\, \delta M$. The net force on this fluid element exerted by the particle is proportional to $m\, \delta M - q\, \delta Q = (m - qQ/M)\, \delta M = 0$. The situation of balanced forces therefore creates a vanishing deformation of the body, and we get ${\cal Q}^{(\ell)} = 0$, so that $k_\ell = 0$. This is very unlike the situation for an uncharged body, which is subjected to the particle's gravitational force only, and which undergoes a nonvanishing deformation. 

The relativistic version of this calculation, detailed in this paper, does not produce a vanishing result for $k_\ell$, but one that is negative and vanishes in the Newtonian limit, $M/R \to 0$. There are two key differences. First, in the relativistic setting the forces acting on each fluid element are no longer precisely balanced; this can be attributed to the fact that the distribution of electrostatic field energy contributes to gravity, and that the distribution inside and outside the body are different. Second, in general relativity the mass multipole moments account also for the deformation in the distribution of field energy. These effects are explored in Appendix \ref{sec:Newtmodel}, in which we formulate an augmented Newtonian model in which the electrostatic field energy is allowed to act as a source of gravity. The calculations presented there are far simpler than the relativistic computations described in the bulk of the paper, and the model is very successful at capturing the essence of the phenomenon: it also returns Love numbers that are negative and proportional to $M/R$ when the compactness is small.

\subsection{Structure of the paper}

We begin in Sec.~\ref{sec:newton} with a review of the Newtonian definitions for the tidal multipole moments ${\cal E}^{(\ell)}$, the mass multipole moments ${\cal Q}^{(\ell)}$, and the Love numbers $k_\ell$. In Sec.~\ref{sec:pN} we promote these definitions to the post-Newtonian setting described in Sec.~\ref{subsec:multipole}, and explain how the multipole moments can be defined in a situation that involves any number of strongly gravitating bodies in relative motion under a weak mutual gravitational interaction. These definitions specify what will be computed in the remaining sections of the paper. The calculational strategy is summarized in Sec.~\ref{sec:strategy}; we map out the entire task and outline the steps that are carried out in the technical sections of the paper. These then follow. In Secs.~\ref{sec:exterior} and \ref{sec:interior} we describe the unperturbed spacetime of a charged body, in Secs.~\ref{sec:tidal_exterior} and \ref{sec:tidal_interior} we calculate the perturbation created by the particle, in Secs.~\ref{sec:harmonic_exterior} and \ref{sec:harmonic_interior} we construct harmonic coordinates to describe the perturbed spacetime (an essential element of the post-Newtonian setting), and finally, in Sec.~\ref{sec:love} we collect our results and compute the Love numbers displayed in Figs.~\ref{fig:fig1} and \ref{fig:fig2}.

Appendix \ref{sec:Newtmodel} details the augmented Newtonian model mentioned previously, and Appendix \ref{sec:Cshell} provides some insights into a constant $C$ that occurs in the transformation to harmonic coordinates. While the first appendix is a valuable read that gets quickly to the essence of our results, the second appendix is technical and will be of interest to the strongly devoted.   

\section{Newtonian considerations}
\label{sec:newton}

We begin our discussion of the tidally induced moments of a charged material body by reviewing the situation in Newtonian gravity. This will serve to introduce our notation and conventions, and to trace the way to a relativistic generalization in the following section. Many elements of this review are imported from the textbook by Poisson and Will \cite{poisson-will:14}, hereafter referred to as {\it Gravity}. Throughout the paper we set $G=1$ and $c=1$, where $G$ is the gravitational constant and $c$ the speed of light. 

\subsection{Tidal potential}

We consider a body of mass $M$ and radius $R$ whose center of mass is placed at the origin of the coordinate system. Remote objects exert tidal forces on this body. Assuming that each object is placed at a large distance from the body, we expand the external potential $U^{\rm ext}$ in powers of $r/d \ll 1$, in which $r := |\bm{x}|$ is the distance from the body's center of mass, and $d$ is the typical distance to an external object. The $\ell$-th order term in the Taylor expansion of the external potential is
\begin{equation}
U_\elltide = -\frac{1}{(\ell-1)\ell}\, r^\ell\, {\cal E}_L \Omega^L,
\label{tidal_potential1} 
\end{equation} 
where
\begin{equation}
{\cal E}_L := -\frac{1}{(\ell-2)!} \partial_L U^{\rm ext} \biggr|_{\bm{x}=\bm{0}}
\label{EL_def} 
\end{equation}
is a tidal multipole moment, and $\bm{\Omega} := \bm{x}/r$ is the radial unit vector. The multi-index $L$ contains a number $\ell$ of individual indices, so that $\partial_L U^{\rm ext} := \partial_{a_1} \partial_{a_2} \cdots \partial_{a_\ell} U^{\rm ext}$, and we use the notation $\Omega^L := \Omega^{a_1} \Omega^{a_2} \cdots \Omega^{a_\ell}$. The Cartesian tensor ${\cal E}_L$ is symmetric and tracefree in all its indices. The normalization of the tidal moments in Eq.~(\ref{EL_def}) follows the conventions of Binnington and Poisson \cite{binnington-poisson:09}, which were inherited from Zhang \cite{zhang:86}. It differs from the normalization adopted in {\it Gravity} by the factor of $(\ell-2)!$.

When the tidal environment is axially symmetric, the tidal moments are necessarily proportional to $e_\stf{L}$, which denotes the tracefree projection of $e_L := e_{a_1} e_{a_2} \cdots e_{a_\ell}$  --- see Eqs.~(1.153) and (1.154) of {\it Gravity}; $\bm{e}$ is a unit vector in the direction of the symmetry axis. We write
\begin{equation}
{\cal E}_L = \frac{(2\ell-1)!!}{\ell!}\, {\cal E}^{(\ell)}\, e_\stf{L},
\label{EL_vs_Eell1}
\end{equation}
and use this relation as a definition for the reduced multipole moments ${\cal E}^{(\ell)}$.

The identity [{\it Gravity}, Eq.~(1.160a)] $e_\stf{L} \Omega^L = [\ell!/(2\ell-1)!!] P_\ell(\cos\theta)$,  where $\cos\theta := \bm{e} \cdot \bm{\Omega}$ and $P_\ell$ is a Legendre polynomial, implies that
\begin{equation}
{\cal E}_L \Omega^L = {\cal E}^{(\ell)} P_\ell(\cos\theta).
\label{EL_vs_Eell2}
\end{equation}
Equation (\ref{tidal_potential1}) can therefore be rewritten as
\begin{equation}
U_\elltide = -\frac{1}{(\ell-1)\ell}\, {\cal E}^{(\ell)}\, r^\ell P_\ell(\cos\theta). 
\label{tidal_potential2} 
\end{equation} 
We recognize $r^\ell P_\ell(\cos\theta)$ as an elementary solution to Laplace's equation. 

\subsection{Response potential}

The tidal forces exerted by the remote objects deform the body from its original spherical state. The deformation is described by the mass multipole moments
\begin{equation}
{\cal Q}^L := \int \rho\, x^\stf{L}\, dV,
\label{QL_def}
\end{equation}
where $\rho$ is the body's mass density and $x^\stf{L}$ is the tracefree projection of $x^L := x^{a_1} x^{a_2} \cdots x^{a_\ell}$. The corresponding term in the body's potential is [{\it Gravity}, Eqs.~(1.149), (1.156), and (1.157)] 
\begin{equation}
U_\ellmass = \frac{(-1)^\ell}{\ell!}\, {\cal Q}^L \partial_L \frac{1}{r}
= \frac{(2\ell-1)!!}{\ell!}\, r^{-(\ell+1)} {\cal Q}_L \Omega^L.
\label{resp_potential1}
\end{equation}

An axisymmetric tidal environment will produce an axisymmetric deformation, and the mass multipole moments will necessarily be proportional to $e_\stf{L}$. In parallel with Eq.~(\ref{EL_vs_Eell1}) we write 
\begin{equation}
{\cal Q}_L = \frac{(2\ell-1)!!}{\ell!}\, {\cal Q}^{(\ell)}\, e_\stf{L},
\label{QL_vs_Qell1}
\end{equation}
so that
\begin{equation}
{\cal Q}_L \Omega^L = {\cal Q}^{(\ell)} P_\ell(\cos\theta); 
\label{QL_vs_Qell2}
\end{equation}
we use Eq.~(\ref{QL_vs_Qell1}) as a definition for the reduced multipole moments ${\cal Q}^{(\ell)}$. Substituting within Eq.~(\ref{resp_potential1}), we find that the response potential becomes
\begin{equation}
U_\ellmass = \frac{(2\ell-1)!!}{\ell!}\, {\cal Q}^{(\ell)}\, r^{-(\ell+1)} P_\ell(\cos\theta). 
\label{resp_potential2}
\end{equation}
We recognize $r^{-(\ell+1)} P_\ell(\cos\theta)$ as another elementary solution to Laplace's equation. 

\subsection{Love numbers} 

To first order in the body's deformation, the mass multipole moments ${\cal Q}_L$ are proportional to the tidal forces exerted by the remote bodies, which are characterized by the tidal multipole moments ${\cal E}_L$. The Love number $k_\ell$ is the (dimensionless) proportionality constant. It is conventional to express the relationship as
\begin{equation}
{\cal Q}_L = -\frac{2 (\ell-2)!}{(2\ell-1)!!}\, k_\ell\, R^{2\ell+1}\, {\cal E}_L,
\label{Q_vs_E1}
\end{equation}
or as
\begin{equation}
{\cal Q}^{(\ell)} = -\frac{2 (\ell-2)!}{(2\ell-1)!!}\, k_\ell\, R^{2\ell+1}\, {\cal E}^{(\ell)}, 
\label{Q_vs_E2}
\end{equation}
where $R$ is the body's radius. Equation (\ref{Q_vs_E1}) is the same statement as Eq.~(2.267) of {\it Gravity}, after taking into account the different normalizations of the tidal multipole moments.

Combining Eqs.~(\ref{tidal_potential2}), (\ref{resp_potential2}), and (\ref{Q_vs_E2}), we have that the $\ell$-pole contribution to the total body potential is
\begin{equation}
U_\elltot = -\frac{1}{(\ell-1)\ell} \Bigl( r^\ell + 2 k_\ell R^{2\ell+1} r^{-(\ell+1)} \Bigr)
{\cal E}^{(\ell)} P_\ell(\cos\theta).
\label{tot_potential}
\end{equation}
This is the same statement as Eq.~(2.269) of {\it Gravity}. 

\subsection{Computation of tidal moments}

As a simple and relevant example of a tidal environment, we examine one produced by a single object of mass $m$ situated at a distance $r_0$ from the reference body. The object's position vector is $\bm{x}_0 = r_0\, \bm{e}$, where $r_0 := |\bm{x}_0|$, and $\bm{s} := \bm{x} - \bm{x}_0 = s\, \bm{n}$ is the separation between the object and a field point at $\bm{x}$; in the second equality the vector $\bm{s}$ was factorized into a length $s := |\bm{s}|$ and a unit vector $\bm{n}$. 

For this situation the external potential is $U^{\rm ext} = m/s$. Making use of Eq.~(1.156) of {\it Gravity}, we have that
\begin{equation}
\partial_L s^{-1} = (-1)^\ell (2\ell-1)!!\, s^{-(\ell+1)} n_\stf{L},
\end{equation}
and evaluating this at $\bm{x} = \bm{0}$ returns
\begin{equation}
\partial_L s^{-1} \Bigr|_{\bm{x}=\bm{0}} = (2\ell-1)!!\, r_0^{-(\ell+1)} e_\stf{L};
\end{equation}
the factor of $(-1)^\ell$ is eliminated thanks to the fact that $\bm{n} = -\bm{e}$ when $\bm{x}=\bm{0}$.

Returning to Eq.~(\ref{EL_def}), we have that the tidal multipole moments are given by
\begin{equation}
{\cal E}_L = -\frac{(2\ell-1)!!}{(\ell-2)!} \frac{m}{r_0^{\ell+1}}\, e_\stf{L}.
\label{EL_Newt1}
\end{equation}
And according to Eq.~(\ref{EL_vs_Eell1}), the reduced moments are
\begin{equation}
{\cal E}^{(\ell)} = -(\ell-1) \ell\, \frac{m}{r_0^{\ell+1}}.
\label{EL_Newt2}
\end{equation}

\section{Post-Newtonian considerations}
\label{sec:pN}

The post-Newtonian generalization of the setup described in Sec.~\ref{sec:newton} was presented in great detail in Refs.~\cite{poisson:21a, poisson:21b}. Here we summarize the main points and define post-Newtonian versions of the tidal multipole moments ${\cal E}^{(\ell)}$ and mass multipole moments ${\cal Q}^{(\ell)}$.

\subsection{Spacetime zones} 

\begin{figure}
\includegraphics[width=0.5\linewidth]{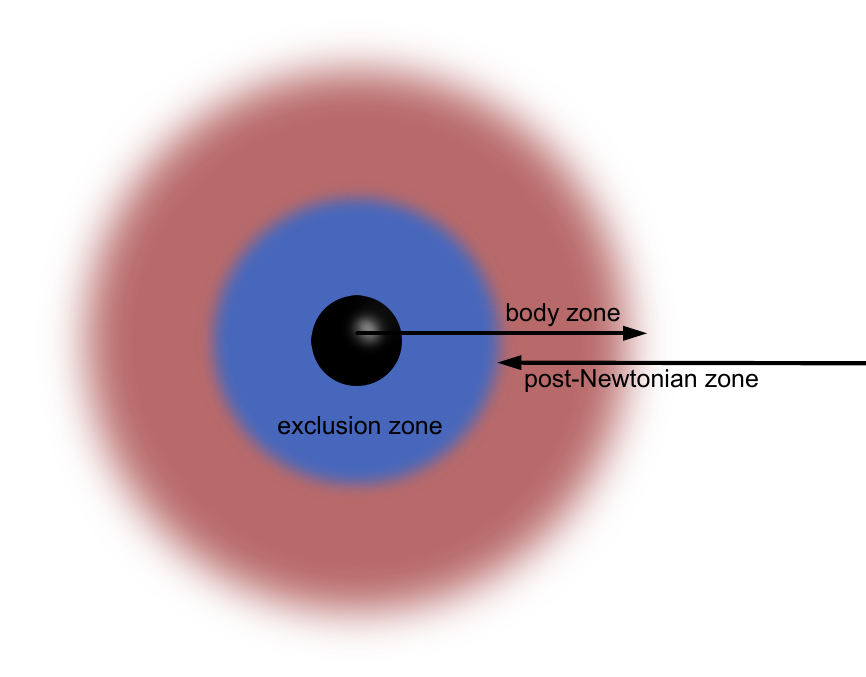}
\caption{Spacetime partitioned into zones. The body is shown at the center, in black. The post-Newtonian zone, where gravity is weak, is shown in white. The exclusion zone, where gravity is strong, is shown in blue. The body zone, where gravity is described by a tidal deformation of the body's metric, is shown in red. The body zone is contained deep within the near zone, in which the time derivative of the metric is much smaller than the spatial derivatives. The post-Newtonian zone extends to the wave zone, in which all derivatives are comparable and the radiative nature of the field asserts itself.} 
\label{fig:fig3} 
\end{figure}

We take the self-gravity of the material body to be strong, so that an adequate description of its gravitational field requires general relativity in its exact formulation. Sufficiently far from the body, however, gravity is weaker, and it can be adequately described by a post-Newtonian expansion carried out to a specified order. As shown in Fig.~\ref{fig:fig3}, we choose to work in the overlap between the {\it body zone} and the {\it post-Newtonian zone}, a region of space outside the {\it exclusion zone}. In the entirety of the body zone the gravitational field is described by a tidal deformation of a strongly gravitating body; the metric is calculated in full general relativity. In the post-Newtonian zone gravity is weak, and the metric is calculated through a post-Newtonian expansion. The exclusion zone is a region around the body in which gravity is too strong to be adequately captured by a post-Newtonian approximation. In the overlap between the body and post-Newtonian zones, the region of interest in this discussion, we have a tidally deformed body whose gravity is adequately described by a post-Newtonian approximation.

We are interested in the post-Newtonian metric of a tidally deformed body, defined in the overlap between the body and post-Newtonian zones. This region is limited by an inner boundary at $\bar{r} = \bar{r}_{\rm in} \gg M$, beyond which gravity is too strong for a post-Newtonian approximation, and by an outer boundary at $\bar{r} = \bar{r}_{\rm max} \ll d$, beyond which the multipole expansion of the tidal field breaks down. Here $\bar{r}$ is the harmonic radial coordinate to be introduced below, $M$ is the mass of the reference body, and $d$ is the typical distance to external objects.

\subsection{Field equations}

The fundamental variables in the post-Newtonian approximation to general relativity are the components of the ``gothic inverse metric'' $\gothg^{\alpha\beta} := \sqrt{-g} g^{\alpha\beta}$, where $g := \mbox{det}[g_{\alpha\beta}]$. In the near zone\footnote{In the near zone, time derivatives of $\gothg^{\alpha\beta}$ are much smaller than spatial derivatives; they are placed on the right-hand side of the field equations and make up part of the source term. In the wave zone, time derivatives are comparable to spatial derivatives; they are placed on the left-hand side of the equations, which become a set of wave equations for the inverse gothic metric.} these satisfy a Poisson equation of the form (refer, for example, to Chapter 6 of {\it Gravity} \cite{poisson-will:14}) 
\begin{equation}
\nabla^2 \gothg^{\alpha\beta} = S^{\alpha\beta},
\label{LLeqn} 
\end{equation}
where $\nabla^2$ is the usual Laplacian operator of three-dimensional flat space, and $S^{\alpha\beta}$ is a source term constructed from the matter's energy-momentum tensor, as well as $\gothg^{\alpha\beta}$ and its derivatives. The equation comes with the harmonic coordinate conditions $\partial_\beta \gothg^{\alpha\beta} = 0$, and the quasi-Lorentzian coordinates $(t, x, y, z)$ are therefore harmonic coordinates. The radial coordinate introduced above is defined in the usual way by $\bar{r}^2 := x^2 + y^2 + z^2$, with the understanding that the reference body is situated at the spatial origin of the coordinate system.

In practice Eq.~(\ref{LLeqn}) is integrated by formally expanding $\gothg^{\alpha\beta}$ in powers of $G$ and $c^{-2}$ and performing iterations ($G$ is the gravitational constant, and $c$ is the speed of light), wherein $S^{\alpha\beta}$ is computed at the end of a given iteration to be used as a source for the next iteration. The iterations are continued until $\gothg^{\alpha\beta}$ has reached the required degree of accuracy. 

In our implementation of the post-Newtonian algorithm, Eq~(\ref{LLeqn}) is integrated in the region of interest only, the overlap between body and post-Newtonian zones. The solution, however, must be informed by the conditions outside this region. It must incorporate information about the body inside the exclusion region, and information about the external objects outside the body zone. At each iteration this information is fed through solutions to the inhomogeneous version of Eq.~(\ref{LLeqn}), Laplace's equation $\nabla^2 \gothg^{\alpha\beta} = 0$.

\subsection{Post-Newtonian multipole moments} 

To be concrete, let us focus the discussion on $W := -\gothg^{tt}$, let us assume that the gravitational field is axisymmetric around the $z$-axis, and let us imagine that after each iteration, $W$ is decomposed in Legendre polynomials $P_\ell(\cos\theta)$, with $\cos\theta := z/\bar{r}$. Solutions to $\nabla^2 W = 0$ come in two guises. First we have a growing solution proportional to $\bar{r}^\ell P_\ell(\cos\theta)$; this describes a tidal field, and therefore incorporates information about the external objects. Second we have a decaying solution proportional to $\bar{r}^{-(\ell+1)} P_\ell(\cos\theta)$; this describes the multipole structure of a body situated at $\bm{x} = \bm{0}$, and therefore incorporates information about the reference body. The tidal moments ${\cal E}^{(\ell)}$ and the mass moments ${\cal Q}^{(\ell)}$ can be defined in terms of these elementary solutions to Laplace's equation.

At the leading, Newtonian order (Box 7.5 of {\it Gravity}) we have that $W = 1 + 4 U + O(1\pn)$, where $U$ is the Newtonian potential and the notation $O(1\pn)$ describes neglected terms of the first post-Newtonian order. A contribution to $W$ coming from an $\ell$-pole tidal moment is therefore given by
\begin{equation}
W_\elltide = -\frac{4}{(\ell-1)\ell}\, {\cal E}^{(\ell)}[0\pn]\, \bar{r}^\ell P_\ell(\cos\theta)
+ O(1\pn), 
\label{W_tidal_N} 
\end{equation} 
where ${\cal E}^{(\ell)}[0\pn]$ is the Newtonian contribution to the $\ell$-pole tidal moment; this equation can be compared with Eq.~(\ref{tidal_potential2}). Similarly, a contribution to $W$ coming from an $\ell$-pole mass moment is given by
\begin{equation}
W_\ellmass = \frac{4(2\ell-1)!!}{\ell!}\, {\cal Q}^{(\ell)}[0\pn]\, \bar{r}^{-(\ell+1)} P_\ell(\cos\theta)
+ O(1\pn), 
\label{W_mass_N}
\end{equation}
where ${\cal Q}^{(\ell)}[0\pn]$ is the Newtonian contribution to the $\ell$-pole mass moment; this equation can be compared with Eq.~(\ref{resp_potential2}).

Each subsequent iteration of the field equations brings updated information about the tidal field and the body's multipole structure. Because the functional forms of $W_\elltide$ and $W_\ellmass$ are necessarily restricted to be solutions to Laplace's equation, they must continue to be proportional to
$\bar{r}^\ell P_\ell(\cos\theta)$ and $\bar{r}^{-(\ell+1)} P_\ell(\cos\theta)$, respectively. The updated information, therefore, is necessarily in the form of corrections to the tidal and mass moments. At the end of the iteration procedure we obtain
\begin{subequations}
\label{W_tidal_mass}
\begin{align}
W_\elltide &= -\frac{4}{(\ell-1)\ell}\, {\cal E}^{(\ell)}\, \bar{r}^\ell P_\ell(\cos\theta), \\
W_\ellmass &= \frac{4(2\ell-1)!!}{\ell!}\, {\cal Q}^{(\ell)}\, \bar{r}^{-(\ell+1)} P_\ell(\cos\theta),
\label{Wellmass1} 
\end{align}
\end{subequations}
where the multipole moments are now accurate through the final post-Newtonian order. If we imagine an iterative process continued indefinitely, and if we assume that it converges, we have tidal and mass moments defined at {\it all post-Newtonian orders}.

The Newtonian relations between tensorial and reduced moments can be exported to the post-Newtonian context. We have
\begin{equation}
{\cal E}_L = \frac{(2\ell-1)!!}{\ell!}\, {\cal E}^{(\ell)}\, e_\stf{L}, \qquad
{\cal Q}_L = \frac{(2\ell-1)!!}{\ell!}\, {\cal Q}^{(\ell)}\, e_\stf{L}, 
\label{L_vs_ell}
\end{equation}
just as in Eqs.~(\ref{EL_vs_Eell1}) and (\ref{QL_vs_Qell1}), respectively. 

\subsection{Nonlinear and derivative expansions} 

In a situation in which the body is known to be spherical in isolation, the mass moments ${\cal Q}^{(\ell)}$ are created entirely by the body's tidal deformation. We can therefore expect that these will be related to the tidal moments ${\cal E}^{(\ell)}$. In general, the relationship can be horribly complicated. The post-Newtonian setting, however, simplifies the relation, because of the assumptions that the mutual gravity between body and companion is weak and the orbital motion is slow. These properties imply that ${\cal Q}^{(\ell)}$ can be expressed as a simultaneous expansion in (i) powers of the tidal moments (which reflects the weak mutual gravity) and (ii) time derivatives of the tidal moments (which reflects the small orbital velocity). Such an expansion was constructed explicitly for ${\cal Q}_{ab}$ at the first post-Newtonian order in Ref.~\cite{poisson:21a}; it included a static term proportional to ${\cal E}_{ab}$, a dynamical term proportional to $\ddot{{\cal E}}_{ab}$ (with the overdots indicating two differentiations with respect to time), and a nonlinear term proportional to ${\cal E}_{c\langle a} {\cal E}^c_{\ b\rangle}$ (with the angular brackets denoting a tracefree projection).  

\subsection{Love numbers} 

To leading order in the nonlinear and derivative expansion, the mass moments will simply be proportional to the tidal moments, and the Newtonian relation of Eq.~(\ref{Q_vs_E2}) can be ported to the post-Newtonian setting. We have
\begin{equation}
{\cal Q}^{(\ell)} = -\frac{2 (\ell-2)!}{(2\ell-1)!!}\, k_\ell\, R^{2\ell+1}\, {\cal E}^{(\ell)}
+ \mbox{dynamical, nonlinear terms}.
\label{Q_vs_E}
\end{equation}
In the new interpretation of this relation, the multipole moments are calculated to a specified post-Newtonian order, but the combination $k_\ell\, R^{2\ell+1}$ is computed in full general relativity, by connecting the interior and exterior metrics of the tidally deformed body. If we again imagine an iteration procedure carried out indefinitely, we have multipole moments computed to all post-Newtonian orders linked by a fully relativistic Love number.

Inserting Eq.~(\ref{Q_vs_E}) within Eq.~(\ref{Wellmass1}), we obtain 
\begin{equation}
W_\ellmass = -\frac{4}{(\ell-1)\ell}\, 2 k_\ell R^{2\ell+1} {\cal E}^{(\ell)}\,
\bar{r}^{-(\ell+1)} P_\ell(\cos\theta), 
\label{Wellmass2} 
\end{equation}
a new expression for $W_\ellmass$, valid to leading order in the simultaneous nonlinearity and time-derivative expansion introduced previously. This expression supplies the relativistic Love number $k_\ell$ with an official definition. 

\section{Strategy: the path ahead}
\label{sec:strategy} 

\subsection{The task} 

We wish to calculate the tidal deformation of a body of mass $M$, charge $Q$, and radius $R$ perturbed by a particle of mass $m$ and charge $q$ situated at a distance $r_0$ from the body. We wish to describe this deformation in terms of the mass multipole moments ${\cal Q}^{(\ell)}$ defined officially by Eq.~(\ref{Wellmass1}). The binary system is maintained in a static state by ensuring that the gravitational and electrostatic forces acting on the particle are balanced, and the perturbation is computed to first order in $m$ and $q$. In this situation, the dynamical terms in Eq.~(\ref{Q_vs_E}) vanish identically, and the nonlinear terms are neglected. The proportionality relation between the mass and tidal moments becomes exact, and the deformation is characterized entirely by the Love numbers $k_\ell$. Our end goal is to calculate these Love numbers.

The definition of the mass moments ${\cal Q}^{(\ell)}$, and therefore of the Love numbers $k_\ell$, rests in an essential way on the post-Newtonian construction detailed in Sec.~\ref{sec:pN}. To calculate these things we must compute $W := \sqrt{-g}\, g^{tt}$ in harmonic coordinates, decompose it in Legendre polynomials $P_\ell(\cos\theta)$, and identify the terms proportional to $\bar{r}^\ell$ and $\bar{r}^{-(\ell+1)}$ in a post-Newtonian expansion. In our actual implementation of this program, we compute the perturbation created by the charged particle by integrating the Einstein-Maxwell equations in full general relativity, without approximation beyond the linearization with respect to $m$ and $q$. The post-Newtonian context is incorporated after the fact, by taking $r_0$ to be large compared with $M$, and by formally expanding $W$ in powers of $M/\bar{r}$. This allows us, starting from a fully relativistic expression for the perturbed metric, to identify the relevant terms in $W$ so as to define precisely the tidal and mass multipole moments, and obtain the associated Love numbers. 

\subsection{Unperturbed spacetime}

The computation is very long, and it requires many steps. We begin in Sec.~\ref{sec:exterior} with a description of the unperturbed spacetime outside the charged body, in the absence of a perturbing particle. The metric there is given by the familiar Reissner-Nordstr\"om solution to the Einstein-Maxwell equations. We then insert a test particle of mass $m$ and charge $q$ at a radius $r_0$ in this spacetime, and calculate the force $F$ required of an external agent to keep the particle in place at this position. In Sec.~\ref{sec:interior} we construct a solution to the Einstein-Maxwell equations that describes the body's unperturbed interior. For concreteness we take the matter to consist of a perfect fluid with a uniform ratio of charge to mass densities, governed by a polytropic equation of state. The equilibrium configurations, in which the fluid pressure makes up for unbalanced gravitational and electrostatic forces within the fluid, are constructed numerically. Such models of charged bodies were previously presented in Ref.~\cite{acena-etal:24}.

\subsection{Perturbation}

The particle is then allowed to perturb the metric and electromagnetic potential. To ensure that the complete energy-momentum tensor is conserved in the background spacetime --- an integrability condition for the perturbation equations --- we demand that $F=0$. The force required of the external agent vanishes, and we have balanced gravitational and electrostatic forces acting on the particle.

In Sec.~\ref{sec:tidal_exterior} we calculate the perturbation of the exterior fields. We first perform a decoupling of the equations that govern the perturbations in the metric and vector potential, and then obtain exact closed-form solutions to these equations. The perturbation is characterized by two dimensionless constants, $p_\ell$ and $q_\ell$, which are attached to decaying solutions to the decoupled equations; these tidal constants form building blocks to construct the Love numbers $k_\ell$. In Sec.~\ref{sec:tidal_interior} we turn to the interior problem. We formulate perturbation equations for the metric, vector potential, and matter variables, and integrate these numerically for our polytropic stellar models. Connecting the interior and exterior solutions at the stellar surface produces numerical values for the tidal constants $p_\ell$ and $q_\ell$.

\subsection{Harmonic coordinates}

The perturbed metric constructed in Secs.~\ref{sec:exterior}, \ref{sec:interior}, \ref{sec:tidal_exterior}, and \ref{sec:tidal_interior} is presented in the familiar $(t,r,\theta,\phi)$ coordinates, in which $r$ is an areal radius. It is also presented in the Regge-Wheeler gauge \cite{regge-wheeler:57} (see also Ref.~\cite{martel-poisson:05} for a more recent presentation), in which some components of the metric perturbation are set to zero. The definition of the Love numbers $k_\ell$, however, relies on a metric presented in harmonic coordinates $X^\Omega = (X^0, X^1, X^2, X^3)$, with each coordinate (viewed as a scalar field) satisfying the wave equation
\begin{equation}
g^{\alpha\beta} \nabla_\alpha \nabla_\beta X^\Omega = 0
\label{harm_cond}
\end{equation}
in the perturbed spacetime; here $\nabla_\alpha$ is the covariant-derivative operator compatible with the perturbed metric.

In Sec.~\ref{sec:harmonic_exterior} we carry out the coordinate transformation for the exterior part of the spacetime. We do this in two stages. First, we obtain harmonic coordinates for the unperturbed, Reissner-Nordstr\"om spacetime. If the metric were describing a charged black hole, the harmonic radial coordinate $\bar{r}$, defined by $\bar{r}^2 := (X^1)^2 + (X^2)^2 + (X^3)^2$, would be related to the areal radius $r$ simply by $\bar{r} = r - M$. The relationship is more complicated when the metric describes a material body; there are additional terms with a logarithmic dependence on $r$, and these come with a dimensionless parameter $C$ that depends on the internal structure of the charged fluid. In the second stage we carry out a gauge transformation from the initial Regge-Wheeler gauge to a final harmonic gauge, defined so that the coordinates $X^\Omega$ stay harmonic in the perturbed spacetime. The end result is a perturbed metric presented in harmonic coordinates. Remarkably, all calculations can be performed analytically, and the final metric is written in terms of standard functions. In addition to the tidal constants $p_\ell$ and $q_\ell$, this metric is characterized by two dimensionless constants, $r_\ell$ and $s_\ell$, which originate in the gauge transformation. The metric also depends on $C$, a property of the unperturbed spacetime in harmonic coordinates.

In Sec.~\ref{sec:harmonic_interior} we transform the interior metric to harmonic coordinates, and again we do this in two stages. In the first we construct the interior harmonic coordinates for the unperturbed stellar structure. This is done numerically for a given choice of equation of state, and connecting the interior and exterior coordinates at the stellar surface returns a numerical value for the parameter $C$. In the second stage we perform a transformation from the Regge-Wheeler gauge of Sec.~\ref{sec:tidal_interior} to the harmonic gauge, again defined so that the coordinates $X^\Omega$ stay harmonic in the perturbed spacetime. Joining the interior and exterior metrics at the stellar surface returns numerical values for the constants $r_\ell$ and $s_\ell$. At this stage the perturbed spacetime and electromagnetic field is completely determined and presented in harmonic coordinates.

\subsection{Love numbers}

With these results in hand, we conclude in Sec.~\ref{sec:love} with the calculation of the Love numbers $k_\ell$. We obtain $W := \sqrt{-g}\, g^{tt}$ in harmonic coordinates, and examine its expression in the interval $R < r < r_0$, which represents the region between the body and the perturbing particle. Taking $r_0 \gg M$, we carry out a post-Newtonian expansion in powers of $M/\bar{r}$. The term proportional to $\bar{r}^\ell P_\ell(\cos\theta)$ provides us with the tidal multipole moment ${\cal E}^{(\ell)}$, and the term proportional to $\bar{r}^{-(\ell+1)} P_\ell(\cos\theta)$ delivers the mass moments ${\cal Q}^{(\ell)}$. The Love number $k_\ell$ is then obtained via Eq.~(\ref{Q_vs_E}). We find that it is given by 
\begin{equation}
k_\ell = p_\ell + q_\ell + r_\ell + t_\ell (L/R)^{2\ell+1},
\label{love_vs_constants}
\end{equation}
where $p_\ell$ and $q_\ell$ are the previously encountered tidal constants, $r_\ell$ is one of the two constants associated with the transformation from Regge-Wheeler gauge to harmonic gauge, and $t_\ell$ is yet another constant that has to do with the conversion from areal radius $r$ to harmonic radius $\bar{r}$; we also have that $L := (M^2-Q^2)^{1/2}$. All these ingredients are required in the computation of $k_\ell$, and we present a sample of our numerical results at the end of Sec.~\ref{sec:love}. 

A central lesson of this long calculation is that a distinction must be made between the Love numbers (which are precisely defined in the post-Newtonian setting detailed in Sec.~\ref{sec:pN}) and the tidal constants (which appear as coefficients in front of decaying solutions to the perturbation equations). These are not the same, and the correct relationship is expressed by Eq.~(\ref{love_vs_constants}).   

\section{Unperturbed spacetime: exterior}
\label{sec:exterior}

In this section we review the Reissner-Nordstr\"om (RN) solution to the Einstein-Maxwell equations, which describes the exterior spacetime of any static, spherically symmetric, and electrically charged body. We also examine a test particle held at a fixed spatial position in this spacetime, and calculate the force required of an external agent to keep the particle at this position.  

\subsection{Metric and vector potential} 

The metric of the RN spacetime is given by
\begin{equation}
ds^2 = -f\, dt^2 + f^{-1}\, dr^2 + r^2(d\theta^2 + \sin^2\theta\, d\phi^2), 
\label{metric_RN} 
\end{equation}
where
\begin{equation}
f := 1 - \frac{2M}{r} + \frac{Q^2}{r^2},
\label{f_RN}
\end{equation}
with $M$ denoting the body's gravitational mass and $Q$ its total charge. The electromagnetic vector potential is given by 
\begin{equation}
A_\alpha = -\frac{Q}{r}\, \partial_\alpha t, 
\label{vecpot_RN} 
\end{equation}
and the electromagnetic field tensor is
\begin{equation}
F_{\alpha\beta} = \nabla_\alpha A_\beta - \nabla_\beta A_\alpha.
\label{F_tensor} 
\end{equation} 
For a static observer, it describes a radial electric field that behaves as $Q/r^2$. The field produces an energy-momentum tensor
\begin{equation}
T^{\alpha\beta}_{\rm field} = \frac{1}{4\pi} \Bigl( F^{\alpha\mu} F^\beta_{\ \mu}
- \frac{1}{4} g^{\alpha\beta}\, F^{\mu\nu} F_{\mu\nu} \Bigr). 
\label{T_em} 
\end{equation}

The metric of Eq.~(\ref{metric_RN}) and the vector potential of Eq.~(\ref{vecpot_RN}) are solutions to Maxwell's equations
\begin{equation}
M^\alpha := \nabla_\beta F^{\alpha\beta} = 0
\label{Meqns_unperturbed} 
\end{equation}
and the Einstein field equations
\begin{equation}
E^{\alpha\beta} := G^{\alpha\beta} - 8\pi T^{\alpha\beta}_{\rm field} = 0,
\label{Eeqns_unperturbed}
\end{equation}
where $G^{\alpha\beta}$ is the Einstein tensor. Equation (\ref{Meqns_unperturbed}) comes with a zero right-hand side because the RN solution applies to the body's exterior, which is free of charge. The other set of Maxwell's equations, $\nabla_\alpha F_{\beta\gamma} + \nabla_\gamma F_{\alpha\beta} + \nabla_\beta F_{\gamma\alpha} = 0$, is automatically satisfied when the field tensor is expressed in terms of a vector potential.

\subsection{Test charge} 

We insert, at a position $r = r_0$ and $\theta = 0$ in the RN spacetime, a point particle of mass $m$ and electric charge $q$. We take $m \ll M$ and $q \ll Q$, and in this section we treat the particle as a test mass and charge. The particle's world line is described by the parametric equations $x^\alpha = Z^\alpha(\tau)$, where $\tau$ is proper time, and its velocity vector is $u^\alpha := dZ^\alpha/d\tau$; the only nonvanishing component is $u^t = f_0^{-1/2}$, where $f_0 := 1-2M/r_0 + Q^2/r_0^2$.

The particle comes with an energy-momentum tensor
\begin{equation}
T^{\alpha\beta}_{\rm part} = m \int u^\alpha u^\beta\, \delta(x, Z)\, d\tau,
\label{T-part} 
\end{equation}
in which $\delta(x,Z) := \delta(x-Z)/\sqrt{-g}$ is a scalarized, four-dimensional Dirac distribution. The only nonvanishing component is 
\begin{equation}
T^{\ \ \ \ t}_{{\rm part}\ t} = -m \frac{\sqrt{f_0}}{r_0^2}\, \delta(r-r_0) \delta(\cos\theta-1) \delta(\phi - \phi_0),
\label{T-part-tt} 
\end{equation}
where $\phi_0$ is the particle's (arbitrarily assigned) azimuthal position in the RN spacetime.

The particle also comes with a current density
\begin{equation}
j^\alpha_{\rm part} = q \int u^\alpha\, \delta(x,Z)\, d\tau, 
\label{j-part} 
\end{equation}
with
\begin{equation}
j^t_{\rm part} = \frac{q}{r_0^2}\, \delta(r-r_0) \delta(\cos\theta-1) \delta(\phi - \phi_0)
\label{j-part-t} 
\end{equation}
its only nonvanishing component.

The force required of an external agent to keep the particle in place is
\begin{equation} 
F^\alpha = m u^\beta \nabla_\beta u^\alpha - q F^\alpha_{\ \beta} u^\beta,
\end{equation} 
and its only nonvanishing component is $F^r$. Its covariant magnitude is $F := \pm (g_{\alpha\beta} F^\alpha F^\beta)^{1/2}$, with the sign chosen so that $\mbox{sign}(F) = \mbox{sign}(F^r)$. We find that 
\begin{equation}
F = \frac{1}{r_0^2} \biggl[ \frac{m(M - Q^2/r_0)}{\sqrt{f_0}} - qQ \biggr]. 
\label{force} 
\end{equation}
Below we shall choose $q/m$ so that $F$ vanishes: the gravitational attraction between body and particle shall be balanced by the electrostatic repulsion. 

\section{Unperturbed spacetime: interior}
\label{sec:interior}

In this section we examine the structure of an electrically charged body in general relativity,  and construct its interior metric and vector potential. The body is assumed to be static and spherically symmetric, and to consist of a perfect fluid with a uniform ratio of charge to mass densities, governed by a polytropic equation of state. Our developments below are virtually identical to those of Ref.~\cite{acena-etal:24}, and were carried out independently. 

\subsection{Charged relativistic fluid}

For our purposes it is sufficient to adopt the simplest model of a charged fluid in general relativity. The fluid comes with a matter current $n^\alpha$ and a charge current $j_{\rm fluid}^\alpha$, which are each assumed to be conserved: 
\begin{equation}
\nabla_\alpha n^\alpha = 0, \qquad \nabla_\alpha j_{\rm fluid}^\alpha = 0.
\label{currents1} 
\end{equation}
We take the matter and charge to move with the same velocity $u^\alpha$, and assign to the fluid a particle-mass density $\rho_m$ and a charge density $\rho_e$, so that
\begin{equation}
n^\alpha = \rho_m u^\alpha, \qquad
j_{\rm fluid}^\alpha = \rho_e u^\alpha.
\label{currents2}
\end{equation}
Compatibility with the conversation laws of Eq.~(\ref{currents1}) is ensured with
\begin{equation}
\rho_e = \beta \rho_m, \qquad \beta = \mbox{constant}.
\label{rhoe_vs_rhom} 
\end{equation}
We take the charged matter to be unpolarized.

The fluid is further described by an energy-momentum tensor, 
\begin{equation}
T^{\alpha\beta}_{\rm fluid} = \mu\, u^\alpha u^\beta + p \bigl( g^{\alpha\beta} + u^\alpha u^\beta \bigr),
\label{T_fluid} 
\end{equation}
in which $\mu = \rho_m + \epsilon$ is the energy density and $p$ is the pressure; $\epsilon$ is the density of internal (thermodynamic) energy. The force density exerted on the fluid is given by
\begin{equation}
\nabla_\beta T^{\alpha\beta}_{\rm fluid} = u^\alpha \Bigl[ u^\beta \nabla_\beta \mu
+ (\mu + p) \nabla_\beta u^\beta \Bigr] + \Bigl[ (\mu+p) a^\alpha
+ \bigl( g^{\alpha\beta} + u^\alpha u^\beta \bigr) \nabla_\beta p \Bigr],
\label{divT1} 
\end{equation}
where $a^\alpha := u^\beta \nabla_\beta u^\alpha$ is the fluid's acceleration. The first group of terms on the right of Eq.~(\ref{divT1}) is aligned with $u^\alpha$; the second group is orthogonal to the velocity. 

The fluid creates an electromagnetic field $F_{\alpha\beta}$. The field's dynamics is governed by Maxwell's equations,
\begin{equation}
M^\alpha = \nabla_\beta F^{\alpha\beta} - 4\pi j_{\rm fluid}^\alpha = 0; 
\label{maxwell} 
\end{equation}
the homogeneous equations $\nabla_\alpha F_{\beta\gamma} + \nabla_\gamma F_{\alpha\beta}
+ \nabla_\beta F_{\gamma\alpha} = 0$ imply the existence of a vector potential $A_\alpha$ such that $F_{\alpha\beta} = \nabla_\alpha A_\beta - \nabla_\beta A_\alpha$. The field comes with the energy-momentum tensor of Eq.~(\ref{T_em}), and both sets of Maxwell's equations imply
\begin{equation}
\nabla_\beta T^{\alpha\beta}_{\rm field} = -F^\alpha_{\ \beta}\, j_{\rm fluid}^\beta.  
\label{divT2}
\end{equation}
The right-hand side is recognized as the Lorentz force density; by virtue of Eq.~(\ref{currents2}) we have that it is orthogonal to $u^\alpha$.  

The fluid's dynamics is subjected to the statement of energy-momentum conservation,
\begin{equation}
\nabla_\beta \bigl( T^{\alpha\beta}_{\rm fluid} + T^{\alpha\beta}_{\rm field} \bigr) = 0.
\label{en-mom-cons} 
\end{equation}
The component in the direction of $u^\alpha$ is $u^\beta \nabla_\beta \mu + (\mu + p) \nabla_\beta u^\beta = 0$, or
\begin{equation}
u^\beta \nabla_\beta \mu - \frac{\mu+p}{\rho_m}\, u^\beta \nabla_\beta \rho_m = 0 
\label{first_law} 
\end{equation}
after making use of Eq.~(\ref{currents1}). This is recognized as the first law of thermodynamics, $d\mu = h\, d\rho_m$, with $h := (\mu+p)/\rho_m$ denoting the specific enthalpy. The first law allows the fluid to possess equations of state of the barotropic form $p = p(\rho_m)$ and $\mu = \mu(\rho_m)$, which we shall select below. It is noteworthy that the first law does not feature a term describing work done by the electromagnetic field; this has to do with the fact that the charge and matter are taken to move together. 

The component of the conservation equation orthogonal to $u^\alpha$ gives rise to
\begin{equation}
(\mu+p) a^\alpha + \bigl( g^{\alpha\beta} + u^\alpha u^\beta \bigr) \nabla_\beta p 
- \rho_e F^\alpha_{\ \beta} u^\beta = 0.
\label{euler}
\end{equation}
This is Euler's equation for a charged, relativistic fluid.

The gravitational field, represented by the metric $g_{\alpha\beta}$, is determined by the Einstein field equations,
\begin{equation}
E^{\alpha\beta} := G^{\alpha\beta} - 8\pi \bigl( T^{\alpha\beta}_{\rm fluid}
+ T^{\alpha\beta}_{\rm field} \bigr) = 0. 
\label{Ein}
\end{equation}
Equation (\ref{en-mom-cons}) is implied by the Einstein equations and the Bianchi identities. 

\subsection{Static and spherically symmetric body}

Next we specialize the equations to the description of a static and spherically symmetric body. The metric in the body's interior is written as
\begin{equation}
ds^2 = -e^{2\psi}\, dt^2 + f^{-1}\, dr^2 + r^2 \bigl( d\theta^2 + \sin^2\theta\, d\phi^2 \bigr),
\label{metric_int} 
\end{equation}
where $f := 1-2m/r$; $\psi$ and $m$ are functions of $r$ only. The interior electromagnetic field is represented by the vector potential
\begin{equation}
A_\alpha = -\Upsilon\, \partial_\alpha t,
\end{equation}
with $\Upsilon$ depending on $r$ only. The charged fluid is at rest in the spacetime, and it possesses the velocity vector
\begin{equation} 
u^\alpha = (e^{-\psi}, 0, 0, 0).
\end{equation} 
The matter variables, such as the mass density $\rho_m$, energy density $\mu$, charge density $\rho_e$, and pressure $p$, depend on $r$ only. 

Outside the body the metric and potential are given by the Reissner-Nordstr\"om solution of Sec.~\ref{sec:exterior}. The interior and exterior solutions are joined smoothly at the body's surface, where $r = R$. This implies that $m(R) = M - Q^2/(2R)$, $\psi(R) = \frac{1}{2} \ln(1-2M/R+Q^2/R^2)$, and $\Upsilon(R) = Q/R$. 

\subsection{Charge inside radius $r$; electric field}

Let $\Sigma$ be the piece of the hypersurface $t = \mbox{constant}$ that corresponds to the interval $[0, r)$ of the radial coordinate, and let $S$ be its boundary, a two-dimensional surface of constant $t$ and $r$. The charge inside radius $r$ is given by
\begin{equation}
q(r) = \int_\Sigma j^\alpha\, d\Sigma_\alpha
= -\int_\Sigma j^\alpha u_\alpha\, d\Sigma
= \int_\Sigma \rho_e\, d\Sigma,
\end{equation}
where $d\Sigma_\alpha = -u_\alpha\, d\Sigma$ is the directed volume element on $\Sigma$. With the metric of Eq.~(\ref{metric_int}) we have that $d\Sigma = f^{-1/2} r^2 \sin\theta\, dr d\theta d\phi$, and integration over the angles yields
\begin{equation}
q(r) = 4\pi \int_0^r r^2 f^{-1/2} \rho_e\, dr.
\label{q_in_r}
\end{equation}

It is convenient to introduce an electric-field variable $E(r)$ defined by
\begin{equation}
E := \frac{q}{r^2}.
\label{E_def}
\end{equation}
We use Gauss's law to relate this to the scalar potential $\Upsilon$. Maxwell's equations and Stokes's theorem imply that $q(r)$ can also be expressed as
\begin{equation}
q(r) = \frac{1}{4\pi} \int_\Sigma \nabla_\beta F^{\alpha\beta}\, d\Sigma_\alpha
= \frac{1}{8\pi} \oint_S F^{\alpha\beta}\, dS_{\alpha\beta}
= -\frac{1}{4\pi} \oint_S F^{\alpha\beta} u_\alpha r_\beta\, dS,
\end{equation}
where $dS_{\alpha\beta} = -2u_{[\alpha} r_{\beta]}\, dS$ is the directed surface element on $S$, with $r_\alpha$ denoting the unit normal vector (defined to be spacelike, orthogonal to $u^\alpha$, and pointing out of $S$). With $r_\alpha = (0,f^{-1/2},0,0)$ and $dS = r^2\sin\theta\, d\theta d\phi$, integration yields
\begin{equation}
q(r) = -r^2 e^{-\psi} f^{1/2}\, \Upsilon',
\end{equation}
in which a prime indicates differentiation with respect to $r$. Comparison with Eq.~(\ref{E_def}) reveals that the electric field is related to the potential via
\begin{equation}
E = -e^{-\psi} f^{1/2}\, \Upsilon'.
\label{E_vs_pot}
\end{equation}

\subsection{Structure equations}

The entire content of Maxwell's equations is in Gauss's law, which was invoked in the preceding subsection. It gives rise to
\begin{equation}
q' = 4\pi r^2 f^{-1/2} \rho_e,
\label{q_prime}
\end{equation}
a differential equation for $q(r)$ that follows directly from Eq.~(\ref{q_in_r}). The relationship with the electric field is provided by Eq.~(\ref{E_def}), and the field's own relationship with the scalar potential, displayed in Eq.~(\ref{E_vs_pot}), converts Eq.~(\ref{q_prime}) to a first-order differential equation for $\Upsilon'$. The formulation of Eq.~(\ref{q_prime}), however, is more convenient for the purposes of computing the body's internal structure.

The Einstein field equations produce
\begin{equation}
m' = 4\pi r^2 \mu + \frac{1}{2} r^2 E^2
\label{m_prime}
\end{equation}
and
\begin{equation}
\psi' = \frac{m + 4\pi r^3 p - \frac{1}{2} r^3 E^2}{r^2 f};
\label{psi_prime} 
\end{equation}
these equations determine the mass function $m(r)$ and gravitational potential $\psi(r)$. The presence of $E^2$ in these equations comes with a clear interpretation: $E^2/8\pi$ is a positive addition to the energy density, and a negative addition to the pressure.

The Euler equation reduces to
\begin{equation}
p' = -\frac{\mu+p}{r^2 f} \Bigl( m + 4\pi r^3 p - \tfrac{1}{2} r^3 E^2 \Bigr)
+ f^{-1/2} \rho_e E;
\label{p_prime} 
\end{equation}
this is the condition of hydrostatic equilibrium. The last term in Eq.~(\ref{p_prime}) is recognized as the force density exerted by the electric field on the fluid's charge distribution.  

The structure equations (\ref{q_prime}), (\ref{m_prime}), (\ref{psi_prime}), and (\ref{p_prime}) must be supplemented with Eq.~(\ref{rhoe_vs_rhom}) for the charge-to-mass ratio, as well as equations of state 
that relate the energy density $\mu$ and pressure $p$ to the mass density $\rho_m$. The equations also come with boundary conditions. The charge $q(r)$ and mass $m(r)$ must vanish at $r=0$, while $p$ and $\psi$ approach constant central values. At the body's boundary $r = R$, we have that $q(R) = Q$, $m(R) =  M - Q^2/(2R)$, $\psi(R) = \frac{1}{2} \ln(1-2M/R+Q^2/R^2)$, and $p(R) = 0$. 

\subsection{Polytropes}
\label{subsec:polytrope} 

At this stage we make a specific choice of equations of state, which (as we observed previously) we can take to be of the barotropic form $\mu = \mu(\rho_m)$ and $p = p(\rho_m)$. Our body shall be a polytrope with
\begin{equation}
p = K \rho_m^{1+1/n}, \qquad
\epsilon = n p, \qquad
\label{EOS}
\end{equation}
where $K$ and $n$ are constants; we recall that $\mu = \rho_m + \epsilon$. 

We let $\rho_c := \rho_m(r=0)$ be the central value of the mass density, and $p_c := p(r=0)$ be the central value of the pressure, as determined by the equation of state. We introduce the ratio
\begin{equation}
b := p_c/\rho_c
\end{equation}
as a measure of how relativistic the fluid is. In view of the relation $\rho_c = b^n/K^n$, we may use $b$ a substitute for the central density, which is typically used to parametrize a sequence of equilibrium configurations. In this description, for fixed values of $K$ and $n$, the constants $b$ and $\beta$ provide the parametrization for a two-parameter family of configurations for the charged fluid.

We introduce a Lane-Emden variable $\vartheta$ with the assignment
\begin{equation}
\rho_m = \rho_c\, \vartheta^n.
\label{theta_def}
\end{equation}
The definition implies that $\vartheta = 1$ at $r = 0$, while $\vartheta = 0$ at $r = R$. We have that
\begin{equation}
p = b \rho_c\, \vartheta^{n+1}, \qquad
\mu = \rho_c (1 + n b \vartheta) \vartheta^n.
\end{equation}
The new variable is linearly related to the specific enthalpy $h := (\mu+p)/\rho_m$. 

Next we introduce the dimensionless variables $\chi$, $\sigma$, and $\zeta$, defined by
\begin{equation}
m = \frac{4\pi}{3} (1+nb) \rho_c\, r^3\, \chi, \qquad
q = \frac{4\pi}{3} \beta \rho_c\, r^3\, \sigma, \qquad
r^2 = r_0^2\, \zeta,
\label{dimless}
\end{equation}
where
\begin{equation}
r_0^2 := \frac{3}{2\pi} \frac{(n+1)b}{(1+nb) \rho_c}
\label{r0_def}
\end{equation}
has the dimension of a length squared. From integrating Eqs.~(\ref{q_prime}) and (\ref{m_prime}) near $r=0$ we infer that $\chi(r=0) = 1$ and $\sigma(r=0) = 1$.

The structure equations become
\begin{subequations}
\label{structure}
\begin{align}
\frac{d\vartheta}{d\zeta} &= -\frac{1 + (n+1)b\vartheta}{f}
\Bigl( \chi + \frac{3b}{1+nb} \vartheta^{n+1} \Bigr)
+ \beta^2 \frac{\sigma}{(1+nb) f^{1/2}} \biggl[ 1
+ \frac{(n+1)b}{1+nb} \frac{1 + (n+1)b\vartheta}{f^{1/2}}\, \zeta\sigma \biggr], \\
\frac{d\chi}{d\zeta} &= -\frac{3}{2\zeta} \biggl[ \chi - \frac{(1+nb\vartheta)\vartheta^n}{1+nb} \biggr]
+ \beta^2 \frac{(n+1)b}{2(1+nb)^2}\, \sigma^2, \\
\frac{d\sigma}{d\zeta} &= -\frac{3}{2\zeta} \Bigl( \sigma - f^{-1/2} \vartheta^n \Bigr), \\
\frac{d\psi}{d\zeta} &= \frac{(n+1)b}{f} \Bigl( \chi + \frac{3b}{1+nb} \vartheta^{n+1} \Bigr)
- \beta^2 \frac{(n+1)^2 b^2}{(1+nb)^2} \frac{\zeta\sigma^2}{f},
\end{align}
\end{subequations}
where $f = 1 - 4(n+1)b\, \zeta\chi$. They are to be integrated from $\zeta=0$, at which $\vartheta = 1$, $\chi=1$, $\sigma=1$, and $\psi = \psi_c$ (an arbitrary constant). Integration proceeds until the stellar surface at $\zeta = \zeta_s$, at which $\vartheta = 0$, $\chi = \chi_s$, $\sigma = \sigma_s$, and $\psi = \psi_s$. A clever implementation of the structure equations actually uses $\vartheta$ instead of $\zeta$ as the independent variable. This has the great advantage of making the range of integration known in advance, since it is necessarily given by $1 \geq \vartheta \geq 0$.

Integration returns $\zeta_s$, $\chi_s$, and $\sigma_s$, and the body's global quantities are given by
\begin{equation}
R = r_0\, \zeta_s^{1/2}, \qquad
M = \frac{4\pi}{3} (1 + nb) \rho_c R^3 \biggl[ \chi_s
+ \beta^2 \frac{(n+1)b}{(1+nb)^2}\, \zeta_s \sigma_s^2 \biggr], \qquad
Q = \frac{4\pi}{3} \beta \rho_c R^3\, \sigma_s.
\end{equation}
Recalling that $\rho_c = b^n/K^n$, we have that
\begin{equation}
R = R_{\rm unit} \frac{b^{(1-n)/2}}{(1+nb)^{1/2}}\, \zeta_s^{1/2}, \qquad
M = M_{\rm unit} \frac{b^{(3-n)/2}}{(1+nb)^{1/2}}\, \zeta_s^{3/2}
\biggl[ \chi_s + \beta^2 \frac{(n+1)b}{(1+nb)^2}\, \zeta_s \sigma_s^2 \biggr],
\end{equation}
where
\begin{equation}
R_{\rm unit} :=\biggl[ \frac{3}{2\pi} (n+1) \biggr]^{1/2} K^{n/2}, \qquad
M_{\rm unit} := \frac{4\pi}{3} \biggl[ \frac{3}{2\pi} (n+1) \biggr]^{3/2} K^{n/2}
\end{equation}
are respectively radius and mass units that are independent of $b := p_c/\rho_c$. The body's compactness is
\begin{equation}
M/R = 2(n+1) b\, \zeta_s \biggl[ \chi_s + \beta^2 \frac{(n+1)b}{(1+nb)^2}\, \zeta_s \sigma_s^2 \biggr],
\label{compactness}
\end{equation}
and the global charge-to-mass ratio is
\begin{equation}
Q/M = \frac{\beta \sigma_s}{1+nb} 
\biggl[ \chi_s + \beta^2 \frac{(n+1)b}{(1+nb)^2}\, \zeta_s \sigma_s^2 \biggr]^{-1}. 
\end{equation}

\subsection{Numerical integrations}

\begin{figure}
\includegraphics[width=0.45\linewidth]{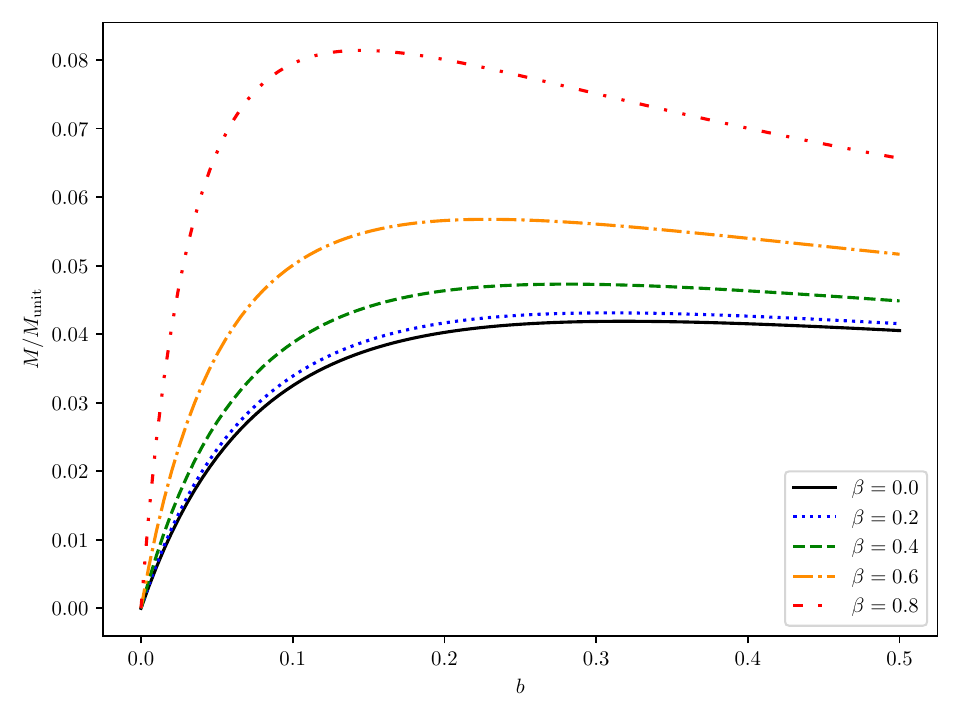}
\includegraphics[width=0.45\linewidth]{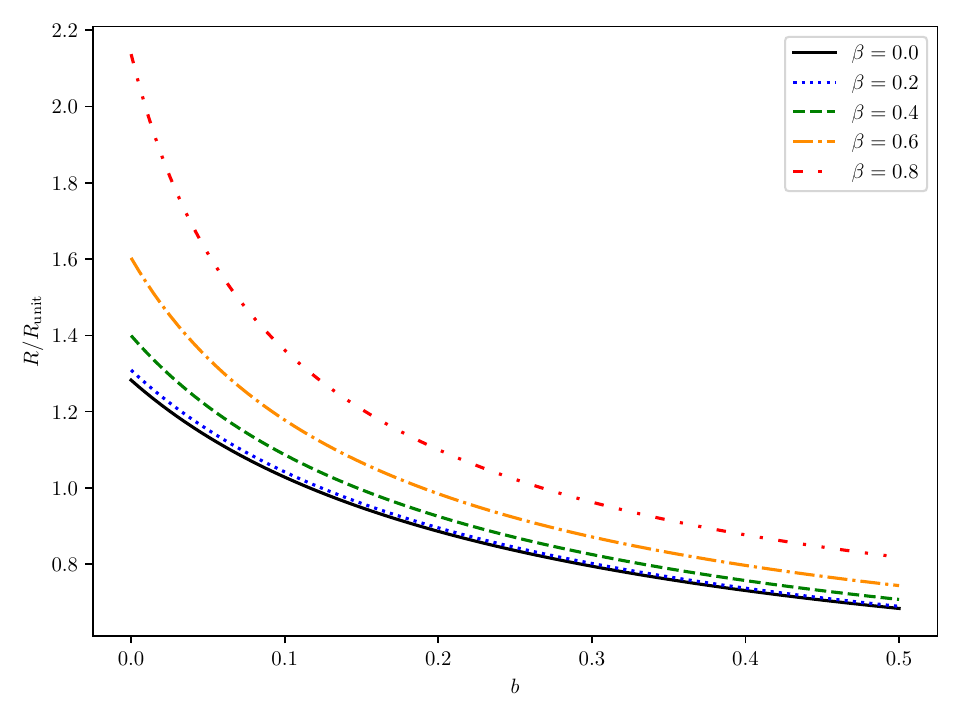}
\caption{Left: $M/M_{\rm unit}$ as a function of $b := p_c/\rho_c$ for a polytrope with $n=1$ and selected values of $\beta := \rho_e/\rho_m$. Right: $R/R_{\rm unit}$ as a function of $b$ for the same polytropes.} 
\label{fig:fig4} 
\end{figure}

\begin{figure}
\includegraphics[width=0.45\linewidth]{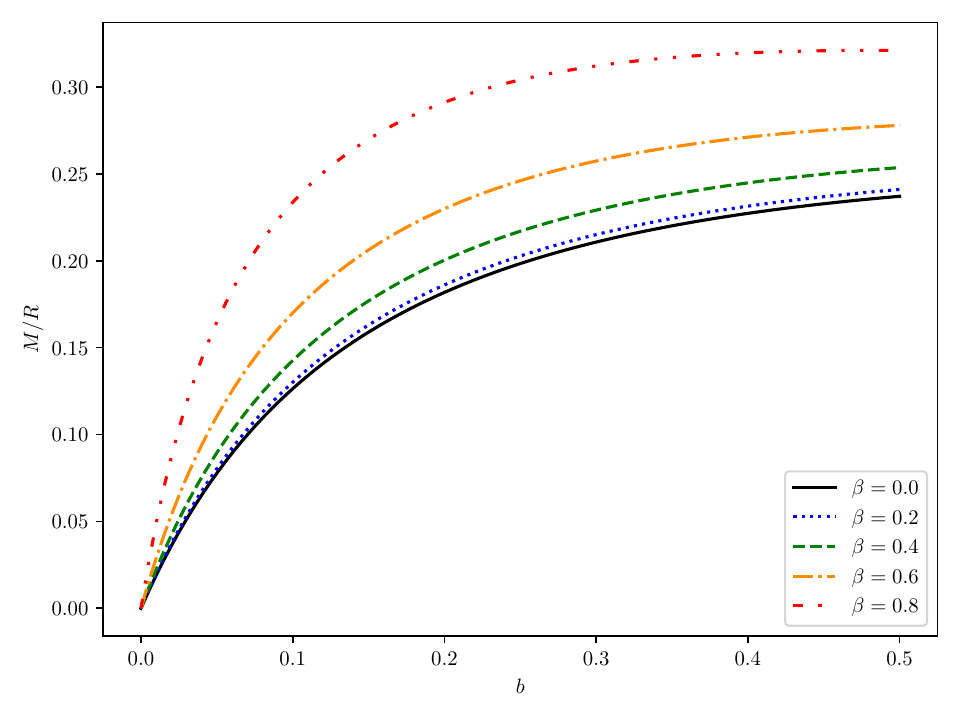}
\includegraphics[width=0.45\linewidth]{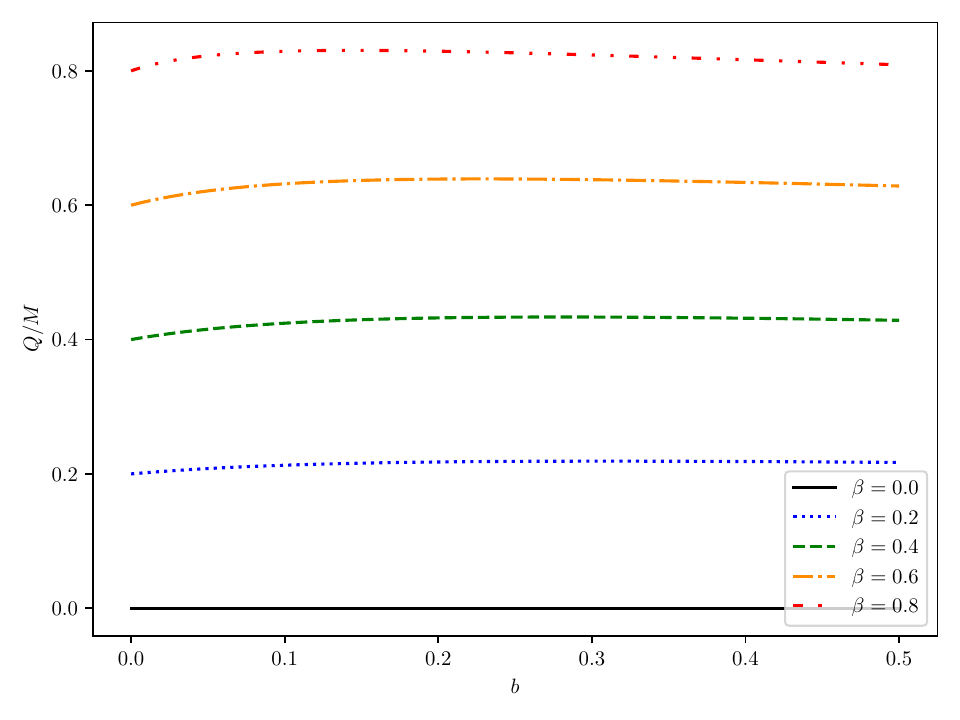}
\caption{Left: $M/R$ as a function of $b$ for a polytrope with $n=1$ and selected values of $\beta := \rho_e/\rho_m$. Right: $Q/M$ as a function of $b$ for the same polytropes.} 
\label{fig:fig5} 
\end{figure}

We carry out numerical integrations of Eqs.~(\ref{structure}), using $\vartheta$ as the independent variable. We first select $n = 1$ for the polytropic index\footnote{A polytropic equation of state with $n=1$ makes up a crude but representative approximation for the equation of state of nuclear matter at the high densities found inside a neutron star.} and sample several values of $\beta := \rho_e/\rho_m$. For each one  we compute $M/M_{\rm unit}$, $R/R_{\rm unit}$, $M/R$, and $Q/M$ as functions of $b := p_c/\rho_c$. The results are displayed in Figs.~\ref{fig:fig4} and \ref{fig:fig5}. 

The left panel of Fig.~\ref{fig:fig4} reveals that the gravitational mass $M$ first increases with $b$, but eventually reaches a maximum and then decreases; this behavior is familiar from uncharged stellar models. The value $b_{\rm max}$ at which $M$ reaches its maximum decreases with increasing values of $\beta$. In the case of uncharged stars, the maximum signals the onset of an instability to radial perturbations, and it may be that the same is true in the case of charged stars. Provided that this is correct, we see that as $\beta$ increases toward unity, the onset of instability occurs earlier and earlier on the sequence of equilibria.

\begin{figure}
\includegraphics[width=0.7\linewidth]{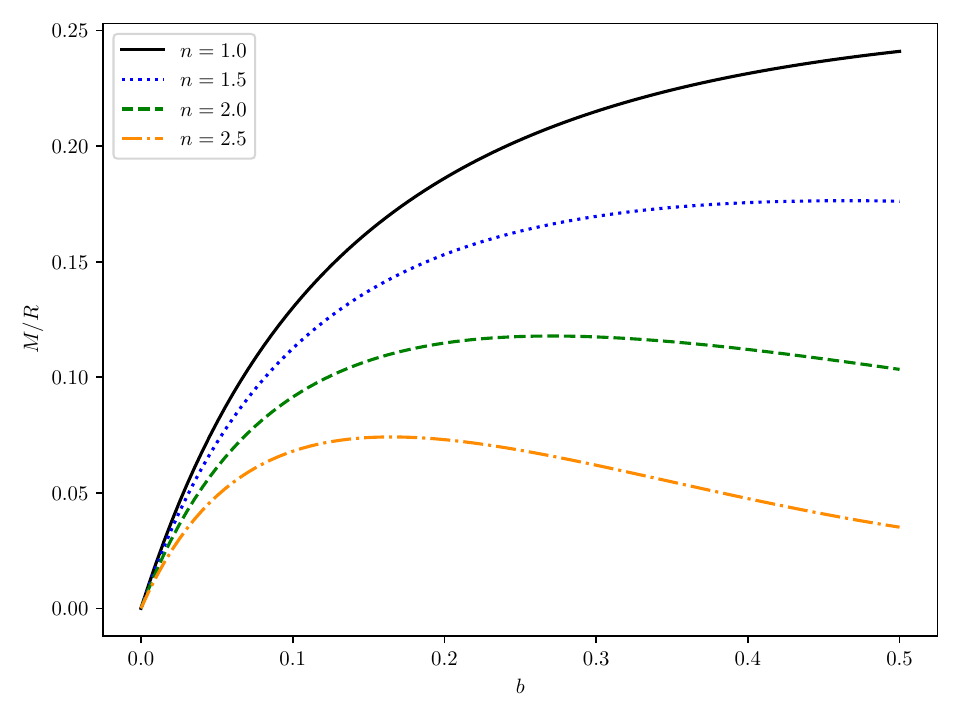}
\caption{$M/R$ as a function of $b := p_c/\rho_c$ for polytropes with different values of $n$, computed for $\beta := \rho_e/\rho_m = 0.2$.}  
\label{fig:fig6} 
\end{figure} 

In the right panel of Fig.~\ref{fig:fig4} we observe that $R$ decreases monotonically with increasing $b$. The left panel of Fig.~\ref{fig:fig5} reveals that $M/R$ is monotonically increasing in the considered interval; beyond this interval, the compactness eventually reaches a maximum and starts decreasing. In the right panel of Fig.~\ref{fig:fig5} we see that $Q/M$, the global charge-to-mass ratio, stays everywhere close to $\beta := \rho_e/\rho_m$, the local measure of charge to mass, deviating by at most 10\%. The main contribution to this deviation comes from the body's binding energy, the difference between the total gravitational mass $M$ and the material mass given by the spatial integral of the mass density $\rho_m$.

In Fig.~\ref{fig:fig6} we plot the stellar compactness $M/R$ as a function of $b := p_c/\rho_c$ for polytropic models with different values of $n$, all with $\beta = 0.2$. We see that for a given value of $b$, the compactness decreases with increasing values of $n$: a softer equation of state gives rise to a less compact star. We observe also that $M/R$ is not monotonic in $b$. This feature is responsible for the looping behavior witnessed in the Love numbers of Figs.~\ref{fig:fig1} and \ref{fig:fig2}.  

\section{Perturbation: exterior}
\label{sec:tidal_exterior}

We return to the exterior portion of the spacetime, and allow the particle of mass $m$ and charge $q$ to create a perturbation in the metric and vector potential. We take $m \ll M$, $q \ll Q$, and recall that the particle is situated at $r = r_0$ and $\theta = 0$.

\subsection{Perturbation equations}

The source of the perturbation is the particle's energy-momentum tensor of Eq.~(\ref{T-part-tt}) and the current density of Eq.~(\ref{j-part-t}). We decompose these quantities in Legendre polynomials, making use of the identity
\begin{equation}
\delta(\cos\theta-1) =\sum_{\ell=0}^\infty \frac{1}{2} (2\ell+1) P_\ell(\cos\theta).
\label{delta_identity} 
\end{equation}
We obtain
\begin{subequations}
\label{pert_sources}
\begin{align} 
T^{\ \ \ \ t}_{{\rm part}\ t} &= -\frac{m}{4\pi r_0^2} \sqrt{f_0}\, \delta(r-r_0)  
\sum_{\ell=0}^\infty (2\ell+1) P_\ell(\cos\theta), \\
j^t_{\rm part} &= \frac{q}{4\pi r_0^2} \delta(r-r_0) 
\sum_{\ell=0}^\infty (2\ell+1) P_\ell(\cos\theta)
\end{align}
\end{subequations}
after averaging over $\phi$.

We cast the metric perturbation in the Regge-Wheeler gauge \cite{regge-wheeler:57, martel-poisson:05}, and write the perturbed metric as
\begin{subequations}
\label{pert_metric}
\begin{align}
g_{tt} &= f \biggl[ -1 + 2 \sum_{\ell=0}^\infty U_\ell(r)\, P_\ell(\cos\theta) \biggr], \\
g_{rr} &= \frac{1}{f} \biggl[ 1 + 2 \sum_{\ell=0}^\infty U_\ell(r)\, P_\ell(\cos\theta) \biggr],\\
g_{AB} &= r^2 \Omega_{AB} \biggl\{ 1 + 2 \sum_{\ell=0}^\infty \bigl[ U_\ell(r) + K_\ell(r) \bigr]\, P_\ell(\cos\theta) \biggr\},
\end{align}
\end{subequations}
where $f := 1-2M/r+Q^2/r^2$. The upper case latin indices refer to the angular sector of the metric, with $\theta^A := (\theta,\phi)$ and $\Omega_{AB} := \mbox{diag}[1,\sin^2\theta]$. The relation $f^{-1} \delta g_{tt} = f \delta g_{rr}$ between the components of the metric perturbation is a consequence of the angular components of the perturbed Einstein field equations; we incorporate this result from the start to simplify the presentation of the perturbation equations. The form of $\delta g_{AB}$, in terms of the combination $U_\ell + K_\ell$, is adopted for convenience. 

For the vector potential we work in a gauge such that the angular components of the perturbation $\delta A_\alpha$ vanish. The perturbed Maxwell equations further imply that $\delta A_r = 0$, and we are left with
\begin{equation}
A_t = -\frac{Q}{r} - \sum_{\ell=0}^\infty \Phi_\ell(r) P_\ell(\cos\theta)
\label{pert_vecpot}
\end{equation}
as the sole nonvanishing component of the perturbed vector potential.

The perturbation is constrained by the conservation identity
\begin{equation}
\nabla_\beta \bigl( T^{\alpha\beta}_{\rm field} + T^{\alpha\beta}_{\rm part} \bigr) = 0,
\end{equation}
which is formulated in the perturbed spacetime; the field's energy-momentum tensor is given by Eq.~(\ref{T_em}), with $F_{\alpha\beta}$ now standing for the perturbed field tensor associated with the vector potential of Eq.~(\ref{pert_vecpot}). Combined with Maxwell's equations, the identity produces 
\begin{equation}
m(M-Q^2/r_0) = q Q \sqrt{f_0},
\label{balance}
\end{equation} 
where $f_0 := 1 - 2M/r_0 + Q^2/r_0^2$. This is the statement of force balance that was first encountered in Sec.~\ref{sec:exterior} --- refer back to Eq.~(\ref{force}). Equation (\ref{balance}) means that the force required of an external agent to keep the particle in place at $r=r_0$ vanishes; electrostatic repulsion is precisely balanced by gravitational attraction. In the Newtonian limit in which $M/r_0 \to 0$, Eq.~(\ref{balance}) reduces to the expected $mM = qQ$. 

The independent perturbation variables are $\Phi_\ell(r)$, $U_\ell(r)$, and $K_\ell(r)$. They are governed by the perturbed field equations
\begin{equation}
M^\alpha := \nabla_\beta F^{\alpha\beta} - 4\pi j^\alpha_{\rm part} = 0, \qquad
E^{\alpha\beta} := G^{\alpha\beta} - 8\pi \bigl( T^{\alpha\beta}_{\rm field}
+ T^{\alpha\beta}_{\rm part} \bigr) = 0.
\end{equation}
The combination $E^r_{\ r} + E^\theta_{\ \theta} = 0$ produces a decoupled equation for $K_\ell$,
\begin{equation}
r^2 f \frac{d^2 K_\ell}{dr^2} + 4(r-M) \frac{dK_\ell}{dr} - (\ell-1)(\ell+2) K_\ell = 0.
\label{K_eq}
\end{equation}
To decouple the remaining equations we introduce an auxiliary variable $X_\ell(r)$ and write
\begin{subequations}
\label{UPhi_vs_XK}
\begin{align}
U_\ell &= \sqrt{f} X_\ell + \frac{r^2(M-Q^2/r) f}{2(M^2-Q^2)}\, \frac{dK_\ell}{dr}, \\
\Phi_\ell &= (M/Q - Q/r) \sqrt{f} X_\ell + \frac{Q r^2 f^2}{2(M^2-Q^2)}\, \frac{dK_\ell}{dr}.
\end{align}
\end{subequations}
We then observe that all field equations are satisfied when $X_\ell$ is a solution to
\begin{equation}
r^2 f \frac{d^2 X_\ell}{dr^2} + 2(r-M) \frac{dX_\ell}{dr} - \biggl[ \ell(\ell+1)
+ \frac{M^2-Q^2}{r^2 f} \biggr] X_\ell = -(2\ell+1) m\, \delta(r-r_0).
\label{X_eq1}
\end{equation}
The entire perturbation problem has therefore been reduced to finding solutions to Eqs.~(\ref{K_eq}) and (\ref{X_eq1}). To arrive at Eq.~(\ref{X_eq1}) we averaged the Einstein and Maxwell equations with respect to $\phi$, as we did previously for the current density and energy-momentum tensor. 

\subsection{Solution to the perturbation equations}

To proceed it is helpful to introduce the alternative independent variable
\begin{equation}
\xi := (r-M)/L, \qquad
L := \sqrt{M^2-Q^2}.
\label{xi_def} 
\end{equation}
In terms of these we have that $f = (\xi^2-1)/(\xi+\mu)^2$, with $\mu := M/L$. 

We also introduce a second auxiliary variable $Y_\ell$, related to $K_\ell$ by
\begin{equation}
K_\ell = \frac{2L}{r\sqrt{f}}\, Y_\ell = \frac{2}{\sqrt{\xi^2-1}}\, Y_\ell.
\label{K_vs_Y} 
\end{equation}
We then have that Eq.~(\ref{UPhi_vs_XK}) becomes 
\begin{subequations}
\label{UPhi_vs_XY}
\begin{align}
U_\ell &= \frac{\sqrt{\xi^2-1}}{\xi+\mu} \Biggl\{ X_\ell
+ \frac{1+\mu\xi}{\xi^2-1} \biggl[ (\xi^2-1) \frac{dY_\ell}{d\xi} - \xi Y_\ell \biggr] \Biggr\},\\
\Phi_\ell &= \frac{L}{Q} \frac{\sqrt{\xi^2-1}}{(\xi+\mu)^2} \Biggl\{ (1+\mu\xi) X_\ell
+ \frac{Q^2}{L^2} \biggl[ (\xi^2-1) \frac{dY_\ell}{d\xi} - \xi Y_\ell \biggr] \Biggr\}.
\end{align}
\end{subequations} 
The equations (\ref{K_eq}) and (\ref{X_eq1}) become
\begin{equation}
(\xi^2-1) \frac{d^2 X_\ell}{d\xi^2} + 2\xi \frac{d X_\ell}{d\xi}
- \biggl[\ell(\ell+1) + \frac{1}{\xi^2-1} \biggr] X_\ell = -(2\ell+1) \frac{m}{L} \delta(\xi-\xi_0)
\label{X_eq2}
\end{equation}
and
\begin{equation}
(\xi^2-1) \frac{d^2 Y_\ell}{d\xi^2} + 2\xi \frac{d Y_\ell}{d\xi}
- \biggl[\ell(\ell+1) + \frac{1}{\xi^2-1} \biggr] Y_\ell = 0, 
\label{Y_eq}
\end{equation}
where $\xi_0 := (r_0-M)/L$.

The solutions to the homogeneous version of Eq.~(\ref{X_eq2}) are the associated Legendre functions $P^{\m=1}_\ell(\xi)$ and $Q^{\m=1}_\ell(\xi)$. The presence of a delta function on the right-hand side of the equation implies that its solution must satisfy the junction conditions
\begin{equation}
\bigl[ X_\ell \bigr] = 0, \qquad
\bigl[ X'_\ell \bigr] = -(2\ell+1) \frac{m}{L} (\xi_0^2-1)^{-1}
\end{equation}
at $\xi = \xi_0$; here a prime indicates differentiation with respect to $\xi$, and $[\psi] := \lim_{\epsilon\to 0^+} [\psi(\xi=\xi_0+\epsilon) - \psi(\xi=\xi_0-\epsilon)]$ is the jump of $\psi$ across $\xi = \xi_0$. The solution for $\xi > \xi_0$ is required to be well behaved when $\xi \to \infty$, and this selects the function $Q_\ell^{\m=1}(\xi)$. If the exterior spacetime were describing a charged black hole (the situation considered in Ref.~\cite{poisson:21b}), then the solution for $\xi < \xi_0$ would be required to be well behaved at the horizon ($\xi = 1$), and this would select $P_\ell^{\m=1}(\xi)$ as the correct solution. Here the situation is different: there is no horizon, $\xi = 1$ is situated inside the body, and regularity there is no longer required. The solution must therefore be a superposition of associated Legendre functions.

Taking all these aspects into account, the global solution to Eq.~(\ref{X_eq2}) is
\begin{equation}
X_\ell = -\frac{2\ell+1}{\ell(\ell+1)} \frac{m}{L} \left\{
\begin{array}{ll}
  Q_\ell^{\m=1}(\xi_0) P_\ell^{\m=1}(\xi)
  + \lambda_\ell P_\ell^{\m=1}(\xi_0) Q_\ell^{\m=1}(\xi) & \quad \xi < \xi_0 \\
  (1 + \lambda_\ell) P_\ell^{\m=1}(\xi_0) Q_\ell^{\m=1}(\xi) & \quad \xi > \xi_0
\end{array} \right. ,
\label{X_global}
\end{equation}
where $\lambda_\ell = \lambda_\ell(\xi_0)$ is an undetermined constant. Inspection of Eq.~(\ref{X_global}) reveals that $X_\ell$ is continuous at $\xi=\xi_0$; that its first derivative displays the correct discontinuity is guaranteed by the identity
\begin{equation}
\W\bigl( P_\ell^{\m=1}, Q_\ell^{\m=1} \bigr) = \frac{\ell(\ell+1)}{\xi^2-1},
\end{equation}
where $\W(a,b) := ab' - a'b$ is the Wronskian of the functions $a$ and $b$; this is Eq.~(14.2.10) of Ref.~\cite{NIST:10}.

Equation (\ref{X_global}) does not apply when $\ell = 0$, and this case must be handled separately. This was carried out in Sec.~V B of Ref.~\cite{poisson:21b}, and no change is required to account for the replacement of the black hole by a material body. The adjustment of the monopole solution described in Sec.~V E of Ref.~\cite{poisson:21b} also applies unchanged to the case of a material body. Because we are interested in the tidal deformation of the body, which is described by the $\ell \geq 2$ piece of the perturbation, there is no need to review these results here.

We shall be mostly interested in $X_\ell(\xi \leq \xi_0)$, that is, the solution to Eq.~(\ref{X_eq2}) in the domain $R \leq r \leq r_0$, between the body and the particle. To express this in meaningful terms, we introduce the notation
\begin{equation}
{\cal E}^{(\ell)} := \frac{(2\ell+1)!!}{(\ell+1)(\ell-2)!} \frac{m}{L^{\ell+1}} Q_\ell^{\m=1}(\xi_0),
\label{E_defalt}
\end{equation}
and express the constant $\lambda_\ell$ as
\begin{equation}
\lambda_\ell = -2 p_\ell (R/L)^{2\ell+1} \frac{(2\ell-1)!!(2\ell+1)!!}{(\ell-1)!(\ell+1)!}
\frac{Q_\ell^{\m=1}(\xi_0)}{P_\ell^{\m=1}(\xi_0)},
\end{equation}
in terms of a new constant $p_\ell$. The expression for $X_\ell(\xi \leq \xi_0)$ becomes
\begin{equation}
X_\ell(\xi \leq \xi_0) = -\frac{L^\ell {\cal E}^{(\ell)}}{(\ell-1)\ell} \biggl[
\frac{(\ell-1)!}{(2\ell-1)!!}\, P_\ell^{\m=1}(\xi)
- 2 p_\ell (R/L)^{2\ell+1} \frac{(2\ell+1)!!}{(\ell+1)!}\, Q_\ell^{\m=1}(\xi) \biggr].
\label{X_inside}
\end{equation}
The notation ${\cal E}^{(\ell)}$ for the quantity introduced in Eq.~(\ref{E_defalt}) does not conflict with the notation already reserved for the tidal multipole moments of Eq.~(\ref{W_tidal_mass}). We shall confirm in Sec.~\ref{subsec:tidal_moments} below that the tidal moments officially defined by Eq.~(\ref{W_tidal_mass}) are indeed given by Eq.~(\ref{E_defalt}) in the situation considered here. The numerical factors in the relation between $\lambda_\ell$ and  $p_\ell$ were incorporated for later convenience. 

The solution to Eq.~(\ref{Y_eq}) must be well behaved at $\xi = \infty$, and it must therefore be proportional to $Q_\ell^{m=1}(\xi)$. We express it as
\begin{equation}
Y_\ell(\xi) = -\frac{L^\ell {\cal E}^{(\ell)}}{(\ell-1)\ell} \biggl[
2 \frac{q_\ell}{\mu} (R/L)^{2\ell+1} \frac{(2\ell+1)!!}{(\ell+2)!}\, Q_\ell^{\m=1}(\xi) \biggr],
\label{Y_global} 
\end{equation} 
where $q_\ell$ is a constant. Unlike Eq.~(\ref{X_inside}), which gives $X_\ell(\xi)$ in the domain $\xi \leq \xi_0$ only, Eq.~(\ref{Y_global}) is the global solution to Eq.~(\ref{Y_eq}).

\subsection{Potentials}

We insert Eqs.~(\ref{X_inside}) and (\ref{Y_global}) within Eqs.~(\ref{K_vs_Y}) and (\ref{UPhi_vs_XY}), and simplify the results by invoking the recursion relation
\begin{equation}
(\xi^2-1) \frac{dQ_\ell^\m}{d\xi} - \m \xi\, Q_\ell^\m = \sqrt{\xi^2-1}\, Q_\ell^{\m+1},
\end{equation}
which is satisfied by all associated Legendre functions. We obtain our final expressions for the potentials,
\begin{subequations}
\label{potential_final}
\begin{align}
\Phi_\ell &= -\frac{L^\ell {\cal E}^{(\ell)}}{(\ell-1)\ell} \frac{M}{Q}
\frac{(1+\mu\xi) \sqrt{\xi^2-1}}{\mu (\xi+\mu)^2} \biggl[
\frac{(\ell-1)!}{(2\ell-1)!!}\, P_\ell^{\m=1}(\xi)
- 2 p_\ell (R/L)^{2\ell+1} \frac{(2\ell+1)!!}{(\ell+1)!}\, Q_\ell^{\m=1}(\xi)
\nonumber \\ & \quad \mbox{} 
+ 2 q_\ell (R/L)^{2\ell+1} (Q/M)^2 \frac{(2\ell+1)!!}{(\ell+2)!}
  \frac{\mu \sqrt{\xi^2-1}}{1+\mu\xi}\, Q_\ell^{\m=2}(\xi) \biggr], \\
U_\ell &= -\frac{L^\ell {\cal E}^{(\ell)}}{(\ell-1)\ell} \frac{\sqrt{\xi^2-1}}{\xi+\mu} \biggl[
\frac{(\ell-1)!}{(2\ell-1)!!}\, P_\ell^{\m=1}(\xi)
- 2 p_\ell (R/L)^{2\ell+1} \frac{(2\ell+1)!!}{(\ell+1)!}\, Q_\ell^{\m=1}(\xi)
\nonumber \\ & \quad \mbox{} 
  + 2 q_\ell (R/L)^{2\ell+1} \frac{(2\ell+1)!!}{(\ell+2)!}
  \frac{1+\mu\xi}{\mu \sqrt{\xi^2-1}}\, Q_\ell^{\m=2}(\xi) \biggr], \\
K_\ell &= -\frac{L^\ell {\cal E}^{(\ell)}}{(\ell-1)\ell} \frac{2}{\mu\sqrt{\xi^2-1}} \biggl[
2 q_\ell (R/L)^{2\ell+1} \frac{(2\ell+1)!!}{(\ell+2)!}\, Q_\ell^{\m=1}(\xi) \biggr].
\end{align}
\end{subequations}
For $\Phi_\ell$ and $U_\ell$ the expressions are valid for $\xi \leq \xi_0$; the expression for $K_\ell$ applies globally.

For later reference we compute $r d\Phi_\ell/dr = (\xi+\mu) d\Phi_\ell/d\xi$, the radial derivative of the electrostatic potential. We obtain
\begin{align}
r \frac{d\Phi_\ell}{dr} &= 
-\frac{L^\ell {\cal E}^{(\ell)}}{(\ell-1)\ell} \frac{M}{Q} \frac{1}{\mu(\xi+\mu)} \Biggl\{ 
\frac{(\ell-1)!}{(2\ell-1)!!} \biggl[ \frac{2-\mu^2+3\mu\xi+3\mu^2\xi^2+\mu\xi^3}{(\xi+\mu)\sqrt{\xi^2-1}}\, P_\ell^{\m=1}(\xi)
+ (1+\mu\xi)\, P_\ell^{\m=2}(\xi) \biggr]
\nonumber \\ & \quad \mbox{} 
- 2 p_\ell (R/L)^{2\ell+1} \frac{(2\ell+1)!!}{(\ell+1)!} \biggl[ \frac{2-\mu^2+3\mu\xi+3\mu^2\xi^2+\mu\xi^3}{(\xi+\mu)\sqrt{\xi^2-1}}\, Q_\ell^{\m=1}(\xi)
+ (1+\mu\xi)\, Q_\ell^{\m=2}(\xi) \biggr]
\nonumber \\ & \quad \mbox{} 
+ 2 q_\ell (R/L)^{2\ell+1} \mu (Q/M)^2 \frac{(2\ell+1)!!}{(\ell+2)!} \biggl[
(\ell-1)(\ell+2) \sqrt{\xi^2-1}\, Q_\ell^{\m=1}(\xi)
- 2\frac{\xi^2-1}{\xi+\mu}\, Q_\ell^{\m=2}(\xi) \biggr] \Biggr\}
\label{r_dPhi_dr}
\end{align}
after liberal application of recurrence relations for associated Legendre functions. 

\section{Perturbation: interior}
\label{sec:tidal_interior}

Next we turn our attention to the body's interior, which is perturbed by the particle of mass $m \ll M$ and charge $q \ll Q$ at position $r = r_0 > R$. We recall that the body consists of a perfect fluid with a uniform ratio of charge to mass densities and a polytropic equation of state, as was described in Sec.~\ref{sec:interior}.  

\subsection{Perturbation equations}

We make the same choices of gauge as in the exterior problem, and express the perturbed metric as
\begin{subequations}
\label{pert_metric_inside}
\begin{align}
g_{tt} &= e^{2\psi} \biggl[ -1 + 2 \sum_{\ell=0}^\infty U_\ell(r)\, P_\ell(\cos\theta) \biggr], \\
g_{rr} &= \frac{1}{f} \biggl[ 1 + 2 \sum_{\ell=0}^\infty U_\ell(r)\, P_\ell(\cos\theta)\biggr], \\
g_{AB} &= r^2 \Omega_{AB} \biggl\{ 1 + 2 \sum_{\ell=0}^\infty \bigl[ U_\ell(r) + K_\ell(r) \bigr]\, P_\ell(\cos\theta) \biggr\}.
\end{align}
\end{subequations}
The only nonvanishing component of the perturbed vector potential is
\begin{equation}
A_t = -\Upsilon - e^{\psi} f^{-1/2} \sum_{\ell=0}^\infty V_\ell(r)\, P_\ell(\cos\theta),  
\label{pert_vecpot_inside} 
\end{equation}
where $e^{2\psi}$, $f := 1-2m/r$, and $\Upsilon$ are the unperturbed quantities of Sec.~\ref{sec:interior}. Equality of $e^{-2\psi} \delta g_{tt}$ and $f \delta g_{rr}$ is a consequence of the perturbed Einstein equations, and the vanishing of $\delta A_r$ is a consequence of Maxwell's equations. The factor of $e^\psi f^{-1/2}$ in the vector potential is inserted for convenience. 

The fluid's energy momentum tensor of Eq.~(\ref{T_fluid}) is modified by the perturbation. First, the time component of the velocity vector becomes
\begin{equation}
u^t = e^{-\psi} \biggl[ 1 + \sum_{\ell=0}^\infty U_\ell(r)\, P_\ell(\cos\theta) \biggr].
\end{equation}
Second, the energy density and pressure acquire the perturbations 
\begin{equation}
\delta \mu = \sum_{\ell=0}^\infty \mu_\ell(r)\, P_\ell(\cos\theta), \qquad
\delta p = \sum_{\ell=0}^\infty p_\ell(r)\, P_\ell(\cos\theta),
\end{equation}
respectively. We may write the fluid's equation of state in the form $\mu = \mu(p)$, with the pressure $p$ playing the role of independent variable. It follows from this that $\mu_\ell = (d\mu/dp)\, p_\ell$. The current density of Eq.~(\ref{currents2}) is also modified by the perturbation. With an equation of state of the form $\rho_e = \rho_e(p)$, we have that
\begin{equation}
\delta \rho_e = \sum_{\ell=0}^\infty \rho_\ell(r)\, P_\ell(\cos\theta)
\end{equation}
with $\rho_\ell = (d\rho_e/dp)\, p_\ell$. We recall that $\rho_e = \beta \rho_m$ for our charged fluid, where $\beta$ is a constant. 

The only independent matter variable is $p_\ell$, and this can be related to $U_\ell$ and $V_\ell$ by invoking the angular components of the conservation identity, the perturbed version of Eq.~(\ref{en-mom-cons}). We find that
\begin{equation}
p_\ell = (\mu+p)\, U_\ell - f^{-1/2} \rho_e\, V_\ell.
\label{p_vs_UV}
\end{equation}
The radial component of Eq.~(\ref{en-mom-cons}) implies that $\rho_e\, d\mu = (\mu+p)\, d\rho_e$. The assignment $\rho_e = \beta\rho_m$ and the first law of thermodynamics stated in Eq.~(\ref{first_law}) ensure that this constraint is automatically satisfied. 

Equation~(\ref{p_vs_UV}) implies that the independent perturbation variables are $V_\ell$, $U_\ell$, and $K_\ell$. They are governed by the perturbed Maxwell equations $M^\alpha = 0$ --- refer back to Eq.~(\ref{maxwell}) --- and the perturbed Einstein field equations $E^{\alpha\beta} = 0$ --- see Eq.~(\ref{Ein}). The independent equations are $M^t = 0$, $E^r_{\ r} = 0$, and $E^r_{\ \theta} = 0$; the remaining equations are redundant. The explicit listing is 
\begin{subequations} 
\label{eqns_listing} 
\begin{align}
0 &= r^2 \frac{d^2 V_\ell}{dr^2} + \biggl[2r + \frac{4\pi r^3(\mu+p)}{f} \biggr] \frac{d V_\ell}{dr}
- 2 r^2 E \biggl( \frac{d U_\ell}{dr} + \frac{d K_\ell}{dr} \biggr)
\nonumber \\ & \quad \mbox{}
+ \Biggl\{ \frac{32\pi^2 r^4}{f^2} (\mu+p)^2
  + \frac{1}{f^2} \biggl[ 12\pi r^2 - 16 \pi^2 r^4 (3 + d\mu/dp) p + 2\pi r^4(3 + d\mu/dp) E^2 
  - 4\pi r (9 + d\mu/dp) m \biggr] (\mu+p)
\nonumber \\ & \quad \mbox{}
+ \frac{4\pi r^3 (1+d\mu/dp) \rho_e E}{f^{3/2}} - \frac{4\pi r^2 (d\rho_e/dp) \rho_e}{f}
  - \frac{\ell(\ell+1)}{f} \Biggr\} V_\ell
+ \frac{4\pi r^2}{f^{1/2}} \Bigl[ (d\rho_e/dp) (\mu+p) + \rho_e \Bigr] U_\ell,
\label{Max_int} \\ 
0 &= -r^2 E\, \frac{dV_\ell}{dr} + \Bigl( m + 4\pi r^3 p - \tfrac{1}{2} r^3 E^2 \Bigr) \frac{d U_\ell}{dr}
+ \Bigl( r - m + 4\pi r^3 p - \tfrac{1}{2} r^3 E^2 \Bigr) \frac{d K_\ell}{dr}
\nonumber \\ & \quad \mbox{}
+ \biggl[ \frac{4\pi r^2 \rho_e}{f^{1/2}} - \frac{4\pi r^3 E(\mu+p)}{f} \biggr] V_\ell
- \bigl( 4\pi r^2 \mu + 12\pi r^2 p - r^2 E^2 \bigl) U_\ell
- \frac{1}{2} (\ell-1)(\ell+2) K_\ell,
\label{Ein1_int} \\
0 &= (r-2m) \frac{d K_\ell}{dr} + 2 r E\, V_\ell - \bigl( 8\pi r^2 p - r^2 E^2 + 2m/r \bigr) U_\ell. 
\label{Ein2_int}
\end{align}
\end{subequations}
We recall that $E(r)$ is the electric-field variable defined by Eqs.~(\ref{E_def}) and (\ref{E_vs_pot}).

\subsection{Solution to the perturbation equations}

A local analysis of Eqs.~(\ref{eqns_listing}) near $r = 0$ reveals that $V_\ell$ and $U_\ell$ behave as $r^\ell$, while $K_\ell$ behaves as $r^{\ell+2}$. This observation, together with a desire to turn Eqs.~(\ref{eqns_listing}) into a system of first-order differential equations, motivates the introduction of new radial functions $e_j(r)$, $j = \{1, 2, 3, 4 \}$, defined by
\begin{equation}
V_\ell = (r/r_0)^\ell\, e_1, \qquad
r \frac{dV_\ell}{dr} = (r/r_0)^\ell\, e_2, \qquad
U_\ell = (r/r_0)^\ell\, e_3, \qquad
K_\ell = (r/r_0)^{\ell+2}\, e_4,
\label{en_def}
\end{equation}
where $r_0$ is a length scale, soon to be identified with the polytropic scale of Eq.~(\ref{r0_def}). With these assignments the perturbation equations can be cast in the form 
\begin{equation}
r \frac{de_j}{dr} = S_j,
\end{equation}
where $S_j$ are source terms constructed from the radial functions and the variables associated with the unperturbed stellar configuration.

At this stage we return to the polytropic model of Sec.~\ref{subsec:polytrope} and re-introduce the density variable $\vartheta$ of Eq.~(\ref{theta_def}), as well as the $\chi$, $\sigma$, and $\zeta$ variables of Eq.~(\ref{dimless}). In terms of these we have that
\begin{equation}
\frac{d\mu}{dp} = n \biggl[ 1 + \frac{1}{(n+1) b \vartheta} \biggr], \qquad
\frac{d\rho_e}{dp} = \frac{n\beta}{(n+1) b \vartheta},
\end{equation}
where $n$ is the polytropic index and $b := p_c/\rho_c$ is the ratio of central pressure to central mass density. The explicit form of the perturbation equations is too lengthy to be displayed here.

We refine the local analysis near $r=0$ by expressing the perturbation variables as expansions in powers of $\zeta$,
\begin{equation}
e_j = e_j^0 + e_j^1\, \zeta + e_j^2\, \zeta^2 + e_j^3\, \zeta^3 + \cdots, 
\end{equation}
where $e_j^k$ are constants, and making the substitution within the perturbation equations. We simultaneously expand $\vartheta$, $\chi$, and $\sigma$ in powers of $\zeta$, and insert the expansions within the structure equations (\ref{structure}). All this reveals that $e_1^0 := e_1(\zeta=0)$ and $e_3^0 := e_3(\zeta=0)$ can be assigned arbitrarily, and that all other coefficients are determined in terms of these two constants.

The solution to the perturbation equations can therefore be determined up to two arbitrary constants. To form a basis of solutions we define
\begin{equation}
\bm{e}^{\rm I} := \bm{e}\bigl( e_1^0 = 1, e_3^0 = 0 \bigr),\qquad
\bm{e}^{\rm II} := \bm{e}\bigl( e_1^0 = 0, e_3^0 = 1 \bigr),
\label{solution_basis} 
\end{equation}
where the vector $\bm{e}$ stands for the set $e_j$ of radial functions. The general solution to the perturbation equations is then
\begin{equation}
\bm{e} = \alpha_{\rm I}\, \bm{e}^{\rm I} + \alpha_{\rm II}\, \bm{e}^{\rm II},
\label{general_solution}
\end{equation}
where $\alpha_{\rm I}$ and $\alpha_{\rm II}$ are arbitrary amplitudes. The physically correct solution will come with specific values for these amplitudes. They are determined by connecting the interior and exterior solutions at the surface $r=R$.

\subsection{Junction conditions}

The exterior solution to the perturbation equations, constructed in Sec.~\ref{sec:tidal_exterior} and displayed in Eqs.~(\ref{potential_final}) and (\ref{r_dPhi_dr}), depends on two unknown constants, $p_\ell$ and $q_\ell$. The common multiplicative factor ${\cal E}^{(\ell)}$, defined by Eq.~(\ref{E_defalt}), is a known function of the particle's mass $m$ and position $r_0$. Because this factor merely provides an overall normalization to the solution, it can be assigned arbitrarily; for our purposes here it is convenient to set
\begin{equation}
\frac{L^\ell {\cal E}^{(\ell)}}{(\ell-1) \ell} = 1.
\end{equation}
The interior solution to the perturbation equations, as displayed in Eq.~(\ref{general_solution}), also depends on two unknown constants, $\alpha_{\rm I}$ and $\alpha_{\rm II}$.

We have at our disposal four junction conditions to determine the four unknown constants,
\begin{subequations}
\label{matching} 
\begin{align} 
(R/r_0)^\ell\, e_1(r=R) &= \Phi_\ell(r=R), \\
(R/r_0)^\ell\, e_2(r=R) &= r \frac{d\Phi_\ell}{dr} \biggr|_{r=R}, \\
(R/r_0)^\ell\, e_3(r=R) &= U_\ell(r=R), \\
(R/r_0)^{\ell+2}\, e_4(r=R) &= K_\ell(r=R).
\end{align} 
\end{subequations}
These equations follow from the definitions of the gravitational and electromagnetic perturbation variables --- refer back to Eqs.~(\ref{pert_metric}), (\ref{pert_vecpot}), (\ref{pert_metric_inside}), and (\ref{pert_vecpot_inside}) --- as well as the definition of the $e_j$ functions --- see Eq.~(\ref{en_def}). The first two equations rely on the facts that $e^\psi f^{-1/2} = 1$ and $d (e^\psi f^{-1/2})/dr = 0$ at $r=R$. The potentials on the right-hand side of Eqs.~(\ref{matching}) refer to the exterior solutions of Sec.~\ref{sec:tidal_exterior}.

The system of Eq.~(\ref{matching}) can written in matrix form, ${\sf M} {\sf x} = {\sf b}$, with ${\sf x} = [p_\ell, q_\ell, \alpha_{\rm I}, \alpha_{\rm II}]$ denoting the vector of unknowns, and with ${\sf M}$ and ${\sf b}$ constructed from known quantities. The matrix is then inverted with standard numerical methods. 

\begin{figure}
\includegraphics[width=0.7\linewidth]{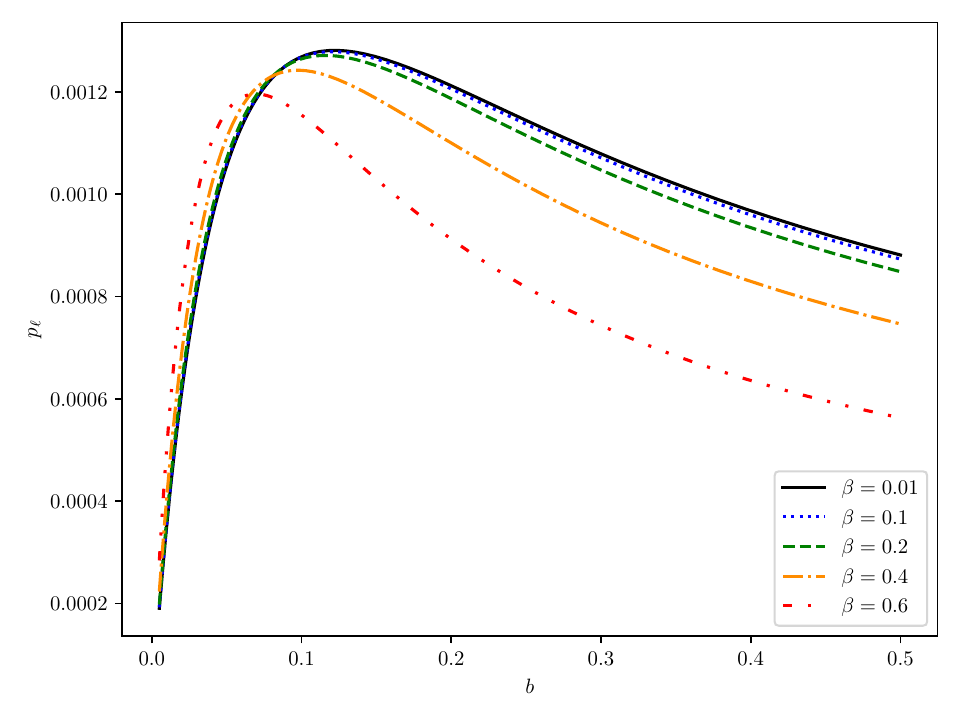}
\caption{Tidal constant $p_\ell$ as a function of $b := p_c/\rho_c$ for $n=1$, $\ell = 2$, and selected values of $\beta := \rho_e/\rho_m$. } 
\label{fig:fig7} 
\end{figure} 

\begin{figure}
\includegraphics[width=0.7\linewidth]{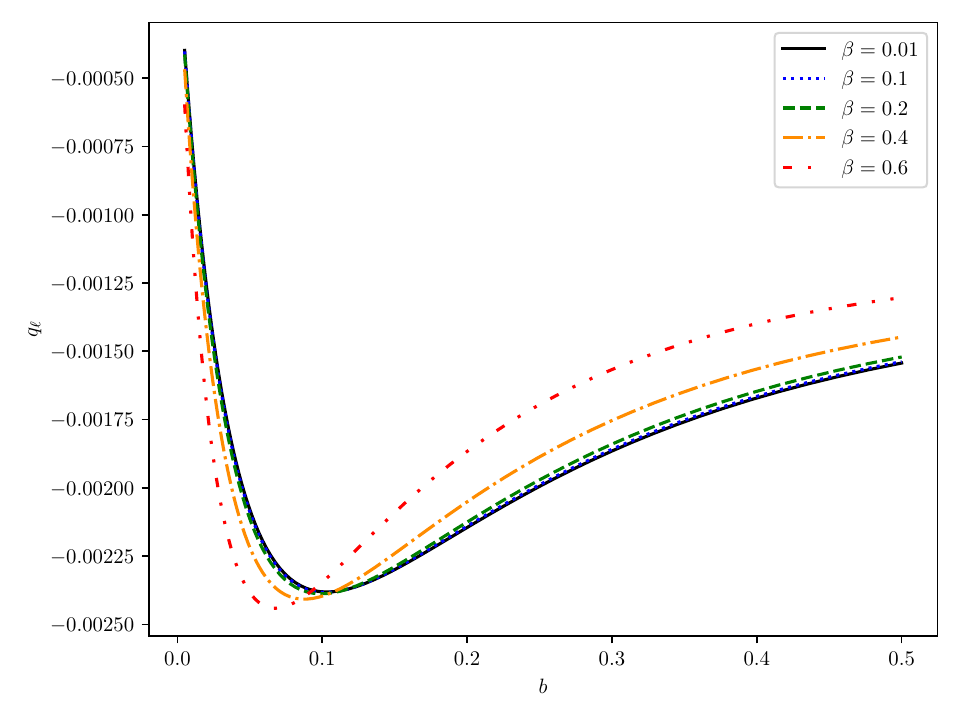}
\caption{Tidal constant $q_\ell$ as a function of $b := p_c/\rho_c$ for $n=1$, $\ell = 2$, and selected values of $\beta := \rho_e/\rho_m$.} 
\label{fig:fig8} 
\end{figure} 

A sample of our numerical results is displayed in Figs.~\ref{fig:fig7} and \ref{fig:fig8}; the tidal constants $p_\ell$ and $q_\ell$ are computed for $n = 1$ and $\ell = 2$, for selected values of $\beta$. The plots reveal that $p_\ell > 0$ while $q_\ell < 0$.

\section{Harmonic coordinates: exterior}
\label{sec:harmonic_exterior}

As was emphasized in Secs.~\ref{sec:pN} and \ref{sec:strategy}, the tidal multipole moments ${\cal E}^{(\ell)}$ and the mass multipole moments ${\cal Q}^{(\ell)}$ are properly defined in a post-Newtonian setting that relies on harmonic coordinates. The perturbed metric of Sec.~\ref{sec:tidal_exterior}, however, is not presented in harmonic coordinates, but in the standard $(t,r,\theta,\phi)$ coordinates (for which $r$ is an areal radius) and in Regge-Wheeler gauge. In this section we take the necessary step of transforming the perturbed metric to a system of harmonic coordinates.  

\subsection{Harmonic coordinates of the Reissner-Nordstr\"om spacetime} 

A set of four independent scalar fields $X^\Omega$, $\Omega = \{0, 1, 2, 3 \}$, that satisfy the wave equation
\begin{equation} 
g^{\alpha\beta} \nabla_\alpha \nabla_\beta\, X^\Omega = 0
\label{wave-eqn} 
\end{equation}
in the Reissner-Nordstr\"om (RN) spacetime forms a system of harmonic coordinates for this spacetime. Here $\Omega$ serves to label the scalars; it is not to be interpreted as a vector index.

We re-introduce the notation
\begin{equation}
L := \sqrt{M^2-Q^2}, \qquad \mu := M/L, \qquad \xi := (r-M)/L,
\label{L_def}
\end{equation}
so that $r = L(\xi + \mu)$, and look for solutions of the form 
\begin{equation}
X^0 := t, \qquad
X^1 := L\, z(\xi)\sin\theta\cos\phi, \qquad
X^2 := L\, z(\xi) \sin\theta\sin\phi, \qquad
X^3 := L\, z(\xi) \cos\theta,  
\label{harm_coordinates}
\end{equation}
so that $\bar{r} = L z(\xi)$ is the harmonic radial coordinate. It may be verified that $X^0 = t$ is an immediate solution to Eq.~(\ref{wave-eqn}), and this shall be our harmonic time coordinate. For $X^1$, $X^2$, and $X^3$ to be solutions, we must have that $z(\xi)$ satisfies
\begin{equation}
(\xi^2-1) z'' + 2\xi z' - 2z = 0,
\label{harm_eq}
\end{equation}
in which a prime indicates differentiation with respect to $\xi$. The linearly independent solutions are Legendre functions,
\begin{equation}
P_1(\xi) = \xi, \qquad
Q_1(\xi) = \frac{1}{2} \xi \ln \frac{\xi+1}{\xi-1} - 1,
\end{equation}
and the appropriate solution to the differential equation is the linear superposition
\begin{equation}
z(\xi) = \xi + C Q_1(\xi),
\label{z_sol}
\end{equation}
where the constant in front of $\xi$ was chosen so that $\bar{r} \sim r$ at large distances from the body, and $C$ is an unknown dimensionless constant. For a black hole we would set $C = 0$ to avoid a singular behavior at the event horizon (situated at $\xi = 1$); for a material body $C$ does not vanish, and as we will see in Sec.~\ref{sec:harmonic_interior}, it depends on the body's internal structure.

Equation (\ref{z_sol}) cannot be inverted to yield a simple explicit expression for $\xi(z)$. We can find an approximate inversion for large values of $z$ if we expand Eq.~(\ref{z_sol}) as
\begin{equation}
z = \xi + C \biggl( \frac{1}{3\xi^2} + \frac{1}{5\xi^4} + \frac{1}{7\xi^6} + \frac{1}{9\xi^8} + \cdots \biggr)
\end{equation}
and invert the power series. We arrive at
\begin{equation}
\xi = z - C \biggl( \frac{1}{3 z^2} + \frac{1}{5 z^4} + \frac{2C}{9 z^5} + \frac{1}{7 z^6}
+ \frac{2C}{5 z^7} + \frac{3 + 7C^2}{27 z^8} + \cdots \biggr).
\label{xi_vs_z}
\end{equation}
This expansion will be required in subsequent calculations. 

\subsection{Harmonic gauge}

The coordinates of Eq.~(\ref{harm_coordinates}) are harmonic in the background RN spacetime, and we now demand that they stay harmonic in the perturbed spacetime. Mathematically, this means that in addition to Eq.~(\ref{wave-eqn}), in which the metric $g_{\alpha\beta}$ and the covariant derivative $\nabla_\alpha$ refer to the background spacetime, the scalar fields $X^\Omega$ are also required to satisfy the wave equation 
\begin{equation}
\hat{g}^{\alpha\beta} \hat{\nabla}_\alpha \hat{\nabla}_\beta X^\Omega = 0
\end{equation}
in the perturbed spacetime, with $\hat{g}_{\alpha\beta} = g_{\alpha\beta} + p_{\alpha\beta}$ denoting the perturbed metric, and $\hat{\nabla}_\alpha$ the covariant derivative compatible with this metric. A routine calculation reveals that both requirements are met when the metric perturbation $p_{\alpha\beta}$ satisfies
\begin{equation}
\nabla_\alpha \bigl( \bar{p}^{\alpha\beta} \nabla_\beta X^\Omega \bigr) = 0,
\label{harm_gauge_cond}
\end{equation}
where
\begin{equation}
\bar{p}_{\alpha\beta} := p_{\alpha\beta} - \frac{1}{2} g_{\alpha\beta}\, p
\end{equation}
is the trace-reversed perturbation. In Eq.~(\ref{harm_gauge_cond}) and other equations below, all indices are raised and lowered with the background metric, and $p := g^{\alpha\beta} p_{\alpha\beta}$. The four equations contained in Eq.~(\ref{harm_gauge_cond}) constitute a set of gauge conditions on the metric perturbation. A perturbation $p_{\alpha\beta}$ that satisfies these conditions shall be said to be in the {\it harmonic gauge}. With a perturbation presented in harmonic gauge, the fields $X^\Omega$ are harmonic coordinates in both the background and perturbed spacetimes.  

The perturbation constructed in Sec.~\ref{sec:tidal_exterior} was cast in the Regge-Wheeler gauge, and it does not satisfy the conditions of Eq.~(\ref{harm_gauge_cond}). However, a gauge transformation
\begin{equation}
p^{\rm H}_{\alpha\beta} = p^{\rm RW}_{\alpha\beta} - \nabla_\alpha \Xi_\beta
- \nabla_\beta \Xi_\alpha
\label{RW-H} 
\end{equation}
generated by a suitable vector field $\Xi^\alpha$ will produce a physically equivalent perturbation that does satisfy Eq.~(\ref{harm_gauge_cond}). As the label indicates, the new perturbation $p^{\rm H}_{\alpha\beta}$ shall be in the harmonic gauge. 

The search for harmonic coordinates in the perturbed spacetime therefore becomes a search for $\Xi^\alpha$. The set of equations that determine this vector is obtained by inserting Eq.~(\ref{RW-H}) within Eq.~(\ref{harm_gauge_cond}); this eventually gives rise to
\begin{equation}
\bigl( g^{\beta\gamma} \nabla_\beta \nabla_\gamma \Xi^\alpha
+ R^\alpha_{\ \beta} \Xi^\beta \bigr) \nabla_\alpha X^\Omega
+ 2\nabla^\alpha \Xi^\beta\, \nabla_\alpha \nabla_\beta X^\Omega
= \nabla_\alpha \bigl( \bar{p}^{\alpha\beta}_{\rm RW} \nabla_\beta X^\Omega \bigr), 
\label{gauge_vec_cond}
\end{equation}
where $R^\alpha_{\ \beta}$ is the Ricci tensor of the unperturbed spacetime. With $\Xi^\alpha$ a solution to these equations, the transformation of Eq.~(\ref{RW-H}) brings a metric perturbation in Regge-Wheeler gauge to one in harmonic gauge.  

\subsection{Gauge equations}

We shall now work on turning Eq.~(\ref{gauge_vec_cond}) into a concrete set of differential equations for the gauge vector $\Xi^\alpha$, which we decompose as 
\begin{equation}
\Xi_t = 0, \qquad
\Xi_r = L\, R_\ell(r)\, P_\ell(\cos\theta), \qquad
\Xi_A = L\, r S_\ell(r)\, \partial_A P_\ell(\cos\theta); 
\label{gauge_vec} 
\end{equation}
the length $L := (M^2-Q^2)^{1/2}$ was introduced to make $R_\ell$ and $S_\ell$ dimensionless. The vector's time component is taken to vanish at the outset, because a more complete computation carried out with $\Xi_t \neq 0$ reveals that this component completely decouples from the remaining ones, and that it can always be set to zero. The metric perturbation in Regge-Wheeler gauge was constructed in Sec.~\ref{sec:tidal_exterior} and was expressed as
\begin{subequations}
\begin{align}
p^{\rm RW}_{tt} &= 2 f U_\ell(r) P_\ell(\cos\theta), \\ 
p^{\rm RW}_{rr} &= 2 f^{-1} U_\ell(r) P_\ell(\cos\theta), \\ 
p^{\rm RW}_{AB} &= 2 r^2 \Omega_{AB} \bigl[ U_\ell(r) + K_\ell(r) \bigr] P_\ell(\cos\theta). 
\end{align}
\end{subequations}
We recall that $f := 1-2M/r+Q^2/r^2$, $\Omega_{AB} := \mbox{diag}[1,\sin^2\theta]$, and that $U_\ell$, $K_\ell$ are given by Eq.~(\ref{potential_final}).

The equations for $R_\ell$ and $S_\ell$ that result from Eq.~(\ref{gauge_vec_cond}) are long and will not be displayed here. A remarkable simplification occurs when we perform the change of variables
\begin{subequations}
\label{RS_vs_GH}
\begin{align} 
R_\ell &= \frac{1}{f z'} \bigl[ (\ell+1) G_\ell + \ell H_\ell \bigr], \\
S_\ell &= \frac{r}{Lz} \bigl[ -G_\ell + H_\ell \bigr],
\end{align}
\end{subequations}
where $z(\xi)$ is defined by Eq.~(\ref{z_sol}), a prime continues to indicate differentiation with respect to $\xi$, and $f = (\xi^2-1)/(\xi+\mu)^2$. The transformation produces the decoupled set of equations
\begin{subequations}
\label{gauge_eqns_GH}
\begin{align}
0 &= (\xi^2-1) G_\ell'' + 2\xi G_\ell' - (\ell+1)(\ell+2) G_\ell + T_\ell, \\
0 &= (\xi^2-1) H_\ell'' + 2\xi H_\ell' - (\ell-1) \ell H_\ell + T_\ell,
\end{align}
\end{subequations}
where $T_\ell := T_\ell^0 + C T_\ell^1$ and
\begin{subequations}
\begin{align}
T_\ell^0 &:= \frac{2}{2\ell+1} \bigl[ (\xi^2-1) K_\ell' + 2\xi K_\ell \bigr], \\
T_\ell^1 &:= \frac{2}{2\ell+1} \bigl[ (\xi^2-1) Q_1' K_\ell' + 2 Q_1 K_\ell \bigr].
\end{align}
\end{subequations}
With $K_\ell$ given (globally) by Eq.~(\ref{potential_final}), more explicit expressions of the source terms are
\begin{subequations}
\begin{align}
T_\ell^0 &= 2\ell(\ell+1) {\cal A}_\ell\, Q_\ell, \\
T_\ell^1 &= 2 {\cal A}_\ell \biggl[ \frac{2}{\xi^2-1} Q'_\ell + \ell(\ell+1) Q_1'\, Q_\ell \biggr],
\end{align}
\end{subequations}
where
\begin{equation}
{\cal A}_\ell := -\frac{4}{(\ell-1)\ell} \frac{(2\ell-1)!!}{(\ell+2)!} \frac{q_\ell}{\mu} 
(R/L)^{2\ell+1} L^\ell {\cal E}^{(\ell)}. 
\label{Aell_def} 
\end{equation}
To arrive at these expressions we exploited the familiar recursion relations among Legendre functions. 

\subsection{Integration of the gauge equations}

The general solution to each one of Eqs.~(\ref{gauge_eqns_GH}) consists of a particular solution to the equation with $T_\ell = T_\ell^0$, added to a particular solution to the equation with $T_\ell = C T_\ell^1$, added to a linear combination of solutions to the homogeneous equation. For $G_\ell$ we have that these are given by $\{ P_{\ell+1}, Q_{\ell+1} \}$, while they are $\{ P_{\ell-1}, Q_{\ell-1} \}$ for $H_\ell$; in either case we eliminate $P_{\ell\pm 1}$ from  the solution, to ensure that it is well behaved at infinity. Particular solutions to the equations with $T_\ell = S_\ell^0$ are obtained by direct inspection; we have that 
\begin{equation}
G_\ell^0 = \ell {\cal A}_\ell\, Q_\ell, \qquad
H_\ell^0 = -(\ell+1) {\cal A}_\ell\, Q_\ell.
\end{equation}
The equations with $T_\ell = T_\ell^1$ are not so easy to integrate, but formal solutions are given by
\begin{subequations}
\label{particular_sols} 
\begin{align}
G_\ell^1 &= c_\ell Q_{\ell+1} - P_{\ell+1} \int T_\ell^1\, Q_{\ell+1}\, d\xi
+ Q_{\ell+1} \int T_\ell^1\, P_{\ell+1}\, d\xi, \\
H_\ell^1 &= c'_\ell Q_{\ell-1} - P_{\ell-1} \int T_\ell^1\, Q_{\ell-1}\, d\xi
+ Q_{\ell-1} \int T_\ell^1\, P_{\ell-1}\, d\xi;
\end{align}
\end{subequations}
the freedom to include a $Q_{\ell\pm 1}$ contribution to the particular solution is exploited to ensure that the expansion of $G_\ell^1$ and $H_\ell^1$ in powers of $\xi^{-1}$ begins with the largest admissible power. 

Happily we find that the integrals of Eq.~(\ref{particular_sols}) can be evaluated when we insert specific values of $\ell$. An examination of several cases, ranging from $\ell = 2$ to $\ell = 7$, allows us to discern a pattern in the resulting expressions, and we conjecture that the particular solutions for any $\ell$ are given by
\begin{equation}
G_\ell^1 = {\cal A}_\ell\, M_\ell, \qquad H_\ell^1 = {\cal A}_\ell\, N_\ell
\end{equation}
with
\begin{equation}
M_\ell = \ell Q_0 Q_\ell - \frac{(2\ell+1)(2\ell+3)}{\ell+1} Q_{\ell+1} + Z_\ell, \qquad
N_\ell = -(\ell+1) Q_0 Q_\ell + Z_\ell,
\label{MN_def}
\end{equation}
where
\begin{equation}
Z_\ell = -\sum_{p=0}^{\frac{1}{2}(\ell-2)} (4p + 3) Q_{2p+1} + \frac{1}{\xi^2-1}
\label{P_def1} 
\end{equation}
for even $\ell$, while
\begin{equation}
Z_\ell = -\sum_{p=0}^{\frac{1}{2} (\ell-1)} (4p + 1) Q_{2p} + \frac{\xi}{\xi^2-1}
\label{P_def2} 
\end{equation}
for odd $\ell$. 
We were not able to devise a formal proof that $G^1_\ell$ and $H^1_\ell$ are indeed solutions to the differential equations, but the expressions of Eqs.~(\ref{MN_def}) were successfully tested for a large sample of values for $\ell$, from small to large. Based on this test, we are confident that the expressions are reliable for any value of $\ell$. We have that
\begin{equation}
M_\ell = O(\xi^{-(\ell+4)}), \qquad
N_\ell = O(\xi^{-(\ell+4)})
\end{equation}
when $\xi$ is large. 

We express the final solutions to Eqs.~(\ref{gauge_eqns_GH}) as 
\begin{subequations}
\label{GH_sol}
\begin{align}
G_\ell &= -\frac{L^\ell {\cal E}^{(\ell)}}{(\ell-1)\ell} (R/L)^{2\ell+1} \biggl\{
8 s_\ell \frac{(2\ell+1)!!}{(\ell+2)!} Q_{\ell+1}(\xi)
+ \frac{4}{\mu} q_\ell \frac{(2\ell-1)!!}{(\ell+2)!} \bigl[ \ell Q_\ell(\xi) + C M_\ell(\xi) \bigr] \biggr\}, \\  
H_\ell &= -\frac{L^\ell {\cal E}^{(\ell)}}{(\ell-1)\ell} (R/L)^{2\ell+1} \biggl[
8 r_\ell \frac{(2\ell-3)!!}{\ell!} Q_{\ell-1}(\xi)
+ \frac{4}{\mu} q_\ell \frac{(2\ell-1)!!}{(\ell+2)!} \bigl[ -(\ell+1) Q_\ell(\xi) + C N_\ell(\xi) \bigr] \biggr\}, 
\end{align}
\end{subequations}
where $s_\ell$ and $r_\ell$ are constants of integration.

\subsection{Perturbation in harmonic gauge}

From Eq.~(\ref{RW-H}) we find that the perturbation in harmonic gauge is given by
\begin{subequations}
\label{pert_H}
\begin{align}
p^{\rm H}_{tt} &= 2\frac{\xi^2-1}{(\xi+\mu)^2} \biggl[ U_\ell
+ \frac{\mu\xi+1}{(\xi+\mu)^3}\, R_\ell \biggr] P_\ell(\cos\theta), \\
p^{\rm H}_{rr} &= 2 \frac{(\xi+\mu)^2}{\xi^2-1} \biggl[ U_\ell
- \frac{\xi^2-1}{(\xi+\mu)^2} \frac{d R_\ell}{d\xi}
- \frac{\mu\xi+1}{(\xi+\mu)^3}\, R_\ell \biggr] P_\ell(\cos\theta), \\
p^{\rm H}_{rA} &= -L \biggl[ (\xi+\mu) \frac{dS_\ell}{d\xi} - S_\ell + R_\ell \biggr]
\partial_A P_\ell(\cos\theta), \\
p^{\rm H}_{AB} &= 2 L^2 \Biggl\{
(\xi+\mu)^2 \biggl[ U_\ell + K_\ell - \frac{\xi^2-1}{(\xi+\mu)^3}\, R_\ell
+ \frac{\ell(\ell+1)}{2(\xi+\mu)}\, S_\ell \biggr] \Omega_{AB} P_\ell(\cos\theta)
- (\xi+\mu) S_\ell\, D_{\langle AB \rangle} P_\ell(\cos\theta) \Biggr\},
\end{align}
\end{subequations}
where
\begin{equation}
D_{\langle AB \rangle} P_\ell(\cos\theta)
:= \biggl( D_A D_B - \frac{1}{2} \Omega_{AB} D^2 \biggr)  P_\ell(\cos\theta)
= \biggl[ D_A D_B + \frac{1}{2} \ell(\ell+1) \Omega_{AB} \biggr] P_\ell(\cos\theta) 
\end{equation}
is the tracefree projection of second derivatives of the Legendre polynomials --- tracefree with respect to $\Omega_{AB}$,  the metric on a unit two-sphere. Here $D_A$ is the covariant-derivative operator compatible with $\Omega_{AB}$, and $D^2 := \Omega^{AB} D_A D_B$ is the Laplacian operator on the sphere. Explicitly, we have that
\begin{equation}
D_{\langle \theta\theta \rangle} P_\ell = -\biggl[ \frac{\cos\theta}{\sin\theta} \frac{d}{d\theta}
+ \frac{1}{2} \ell(\ell+1) \biggr] P_\ell, \qquad
D_{\langle \phi\phi \rangle} P_\ell = \sin^2\theta \biggl[ \frac{\cos\theta}{\sin\theta} \frac{d}{d\theta}
+ \frac{1}{2} \ell(\ell+1) \biggr] P_\ell, 
\end{equation}
where we made use of Legendre's equation to eliminate the second derivatives. 

From Eq.~(\ref{pert_H}) we find that the trace of the metric perturbation is given by 
\begin{equation}
p^{\rm H} = 2 \biggl[ 2 U_\ell + 2 K_\ell
- \frac{\xi^2-1}{(\xi+\mu)^2} \frac{dR_\ell}{d\xi}
- \frac{2\xi}{(\xi+\mu)^2}\, R_\ell
+ \frac{\ell(\ell+1)}{\xi+\mu}\, S_\ell \biggr] P_\ell(\cos\theta).
\label{pertH_trace}
\end{equation}
This result will be needed in forthcoming calculations. 

\section{Harmonic coordinates: interior} 
\label{sec:harmonic_interior}

In this section we construct the harmonic coordinates $X^\Omega$ for the perturbed stellar interior. Connecting the interior and exterior coordinates at the stellar surface determines the constants $C$, $s_\ell$, and $r_\ell$ introduced in Sec.~\ref{sec:harmonic_exterior}. 

\subsection{Unperturbed spacetime}

We first define harmonic coordinates $X^\Omega$ for the unperturbed interior. We set
\begin{equation}
X^0 = t, \qquad
X^1 = s(r) \sin\theta\cos\phi, \qquad 
X^2 = s(r) \sin\theta\sin\phi, \qquad
X^3 = s(r) \cos\theta,
\label{harm_int}
\end{equation}
and find that the radial function must satisfy
\begin{equation}
r^2 f \frac{d^2 s}{dr^2} + \Bigl( r^2 f \frac{d\psi}{dr} - r \frac{dm}{dr}
+ 2r - 3m \Bigr) \frac{ds}{dr} - 2 s = 0
\label{radial_int} 
\end{equation}
if each $X^\Omega$ is to be a solution to Eq.~(\ref{wave-eqn}); we recall that $f := 1-2m/r$, and that the functions $m(r)$ and $\psi(r)$ are related to the metric of Eq.~(\ref{metric_int}).

\begin{figure}
\includegraphics[width=0.49\linewidth]{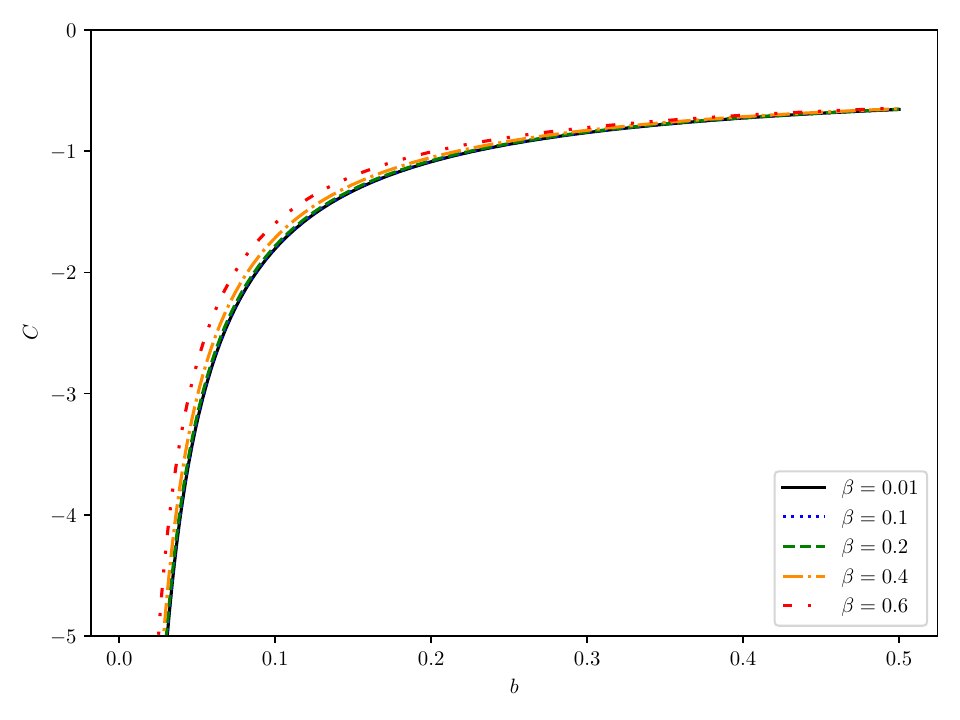}
\includegraphics[width=0.49\linewidth]{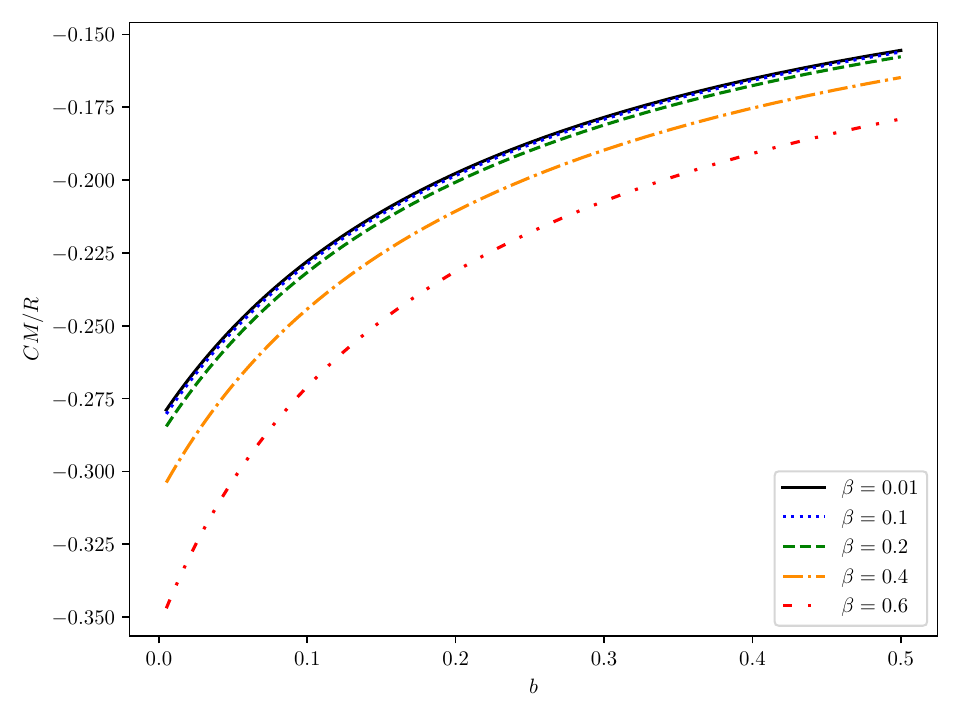}
\caption{Left: Harmonic constant $C$ as a function of $b := p_c/\rho_c$. Right: Rescaled constant $C M/R$ as a function of $b$. In both cases the calculation is performed for a $n = 1$ polytropic model, for selected values of $\beta$.} 
\label{fig:fig9} 
\end{figure}

We integrate Eq.~(\ref{radial_int}) for our polytropic stellar models by making use of the techniques described in Sec.~\ref{subsec:polytrope}. The interior coordinates must join smoothly with the exterior coordinates of Eq.~(\ref{harm_coordinates}), and this requires
\begin{equation}
s = L z, \qquad
\frac{ds}{dr} = \frac{dz}{d\xi}
\label{harm_junction} 
\end{equation}
at $r = R$; we recall that $\xi := (r-M)/L$, $L := (M^2-Q^2)^{1/2}$, and that $z(\xi)$ is given by Eq.~(\ref{z_sol}). Equations (\ref{harm_eq}) and (\ref{radial_int}) then imply that $d^2 s/dr^2 = L^{-1}\, d^2 z/d\xi^2$ at $r=R$, so that the harmonic radial coordinate is actually twice differentiable at the stellar surface. The junction conditions of Eq.~(\ref{harm_junction}) determine the constant $C$ that appears in $z(\xi)$, given a choice of stellar model. We display a sample of our numerical results in Fig.~\ref{fig:fig9}. These reveal that $C$ is negative, and that it diverges when $b := p_c/\rho_c$ goes to zero; but the numerics indicate that $C M/R$ tends to a finite value in the limit. Analytical insights into this behavior are provided in Appendix \ref{sec:Cshell}. 

\subsection{Perturbed spacetime}

The coordinates $X^\Omega$ will remain harmonic in the perturbed spacetime if the perturbation is presented in a gauge that satisfies Eq.~(\ref{harm_gauge_cond}). We carry out a transformation from the original Regge-Wheeler gauge employed in Sec.~\ref{sec:tidal_interior} to this new harmonic gauge by means of a vector field $\Xi_\alpha$ that satisfies Eq.~(\ref{gauge_vec_cond}).

We express the vector field as in Eq.~(\ref{gauge_vec}), 
\begin{equation}
\Xi_t = 0, \qquad
\Xi_r = r_0\, \hat{R}_\ell(r)\, P_\ell(\cos\theta), \qquad
\Xi_A = r_0\, r \hat{S}_\ell(r)\, \partial_A P_\ell(\cos\theta), 
\end{equation}
except that we now scale the radial functions with the length $r_0$ introduced in Eq.~(\ref{r0_def}). The gauge vector is continuous and differentiable at $r=R$.

Inspired by Eq.~(\ref{RS_vs_GH}) we introduce new variables $\hat{G}_\ell$ and $\hat{H}_\ell$ defined by
\begin{subequations}
\label{RS_vs_GH_int}
\begin{align}
\hat{R}_\ell &= \frac{1}{f (ds/dr)} \bigl[ (\ell+1) \hat{G}_\ell + \ell \hat{H}_\ell \bigr], \\ 
\hat{S}_\ell &= \frac{r}{s} \bigl[ -\hat{G}_\ell + \hat{H}_\ell \bigr].
\end{align}
\end{subequations}
The factors in front of the square brackets are equal to those of Eq.~(\ref{RS_vs_GH}) at the stellar surface, and because they are also differentiable there, we have that
\begin{equation}
r_0 \hat{G}_\ell = L G_\ell, \qquad
r_0 \hat{H}_\ell = L H_\ell
\label{GH-junction1} 
\end{equation}
and
\begin{equation}
r_0 \frac{d \hat{G}_\ell}{dr} = \frac{dG_\ell}{d\xi}, \qquad
r_0 \frac{d \hat{H}_\ell}{dr} = \frac{dH_\ell}{d\xi}
\label{GH-junction2} 
\end{equation}
at $r=R$. 

By virtue of Eq.~(\ref{gauge_vec_cond}), we find that the differential equations satisfied by the new radial functions are
\begin{equation}
r^2 f\, \frac{d^2 \hat{G}_\ell}{dr^2}
+ \bigl[ 2r - 2m - q^2/r - 4\pi r^3 (\mu-p) \bigr]\, \frac{d \hat{G}_\ell}{dr}
- (\ell+1)(\ell+2) \hat{G}_\ell + \hat{T}_\ell = 0
\label{Ghat_eqn}
\end{equation}
and
\begin{equation}
r^2 f\, \frac{d^2 \hat{H}_\ell}{dr^2}
+ \bigl[ 2r - 2m - q^2/r - 4\pi r^3 (\mu-p) \bigr]\, \frac{d \hat{H}_\ell}{dr}
- (\ell-1)\ell \hat{H}_\ell + \hat{T}_\ell = 0, 
\label{Hhat_eqn}
\end{equation}
with
\begin{equation}
\hat{T}_\ell := \frac{1}{(2\ell+1) r_0} \biggl[ -4 q \frac{ds}{dr}\, V_\ell
+ (4 m - 2q^2/r + 16\pi r^3 p) \frac{ds}{dr}\, U_\ell + 4 s K_\ell \biggr].
\end{equation}
As we observed in Sec.~\ref{sec:harmonic_exterior}, the equations are decoupled and implicate the same source term. We recall that $q(r)$ is the charge within radius $r$ introduced in Eq.~(\ref{q_in_r}), and that $U_\ell$, $V_\ell$, and $K_\ell$ are the perturbation fields first encountered in Sec.~\ref{sec:tidal_interior}, in the original Regge-Wheeler gauge.

\subsection{Integration of the gauge equations}

An analysis of Eqs.~(\ref{Ghat_eqn}) and (\ref{Hhat_eqn}) near $r=0$ reveals that regular solutions to these equations behave as $\hat{G}_\ell \sim r^{\ell+1}$ and $\hat{H}_\ell \sim r^{\ell-1}$. We peel off this leading behavior and turn the differential equations into a first-order system by introducing yet another set of variables $c_j(r)$, $j = \{ 1, 2, 3, 4 \}$, defined by
\begin{equation}
\hat{G}_\ell = (r/r_0)^{\ell+1}\, c_1, \qquad
r \frac{d \hat{G}_\ell}{dr} = (r/r_0)^{\ell+1}\, c_2, \qquad
\hat{H}_\ell = (r/r_0)^{\ell-1}\, c_3, \qquad
r \frac{d \hat{H}_\ell}{dr} = (r/r_0)^{\ell-1}\, c_4.
\end{equation}
Each function $c_j$ approaches a constant as $r \to 0$. The differential equations impose constraints among these constants, and we find that only $c_{10} := c_1(r=0)$ and $c_{30} := c_3(r=0)$ are freely specifiable.

The system of differential equations must be integrated numerically, and we employ the same techniques as in Sec.~\ref{sec:tidal_interior}. It is again helpful to introduce a basis of solutions to identify the correct solution to the equations. We illustrate the construction for the decoupled subsystem that concerns $c_1$ and $c_2$, which we collect under the vector $\bm{c}$; the construction is entirely similar in the case of $c_3$ and $c_4$.

The basis of solutions is denoted $\bm{c}^{\rm I}$ and $\bm{c}^{\rm II}$, and it is defined by
\begin{subequations}
\begin{align}
\bm{c}^{\rm I} &:= \frac{1}{2} \bigl[ \bm{c}(c_{10} = 1) + \bm{c}(c_{10} = -1) \bigr], \\ 
\bm{c}^{\rm II} &:= \frac{1}{2} \bigl[ \bm{c}(c_{10} = 1) - \bm{c}(c_{10} = -1) \bigr],
\end{align}
\end{subequations}
in which we make different assignments for the arbitrary constant $c_{10} := c_1(r=0)$. It is easy to show that $\bm{c}^{\rm I}$ is a particular solution to the first-order system of differential equations equivalent to Eq.~(\ref{Ghat_eqn}), and that $\bm{c}^{\rm II}$ is a solution to the homogeneous system in which the source term $\hat{T}_\ell$ is set to zero. For any specified value of $c_{10}$, the solution to the equations is
\begin{equation}
\bm{c} = \bm{c}^{\rm I} + c_{10}\, \bm{c}^{\rm II}. 
\end{equation}
The correct value of $c_{10}$, however, remains to be determined.

The interior solutions for $\hat{G}_\ell$ and $\hat{H}_\ell$ are connected to the exterior solutions for $G_\ell$ and $H_\ell$ using the junction conditions of Eqs.~(\ref{GH-junction1}) and (\ref{GH-junction2}). The interior solutions depend on the two unknown constants $c_{10}$ and $c_{30}$, while the external solutions of Eq.~(\ref{GH_sol}) depend on $p_\ell$ and $q_\ell$. The junction conditions give us four equations for the four unknowns, and we obtain a unique solution. 

\begin{figure}
\includegraphics[width=0.7\linewidth]{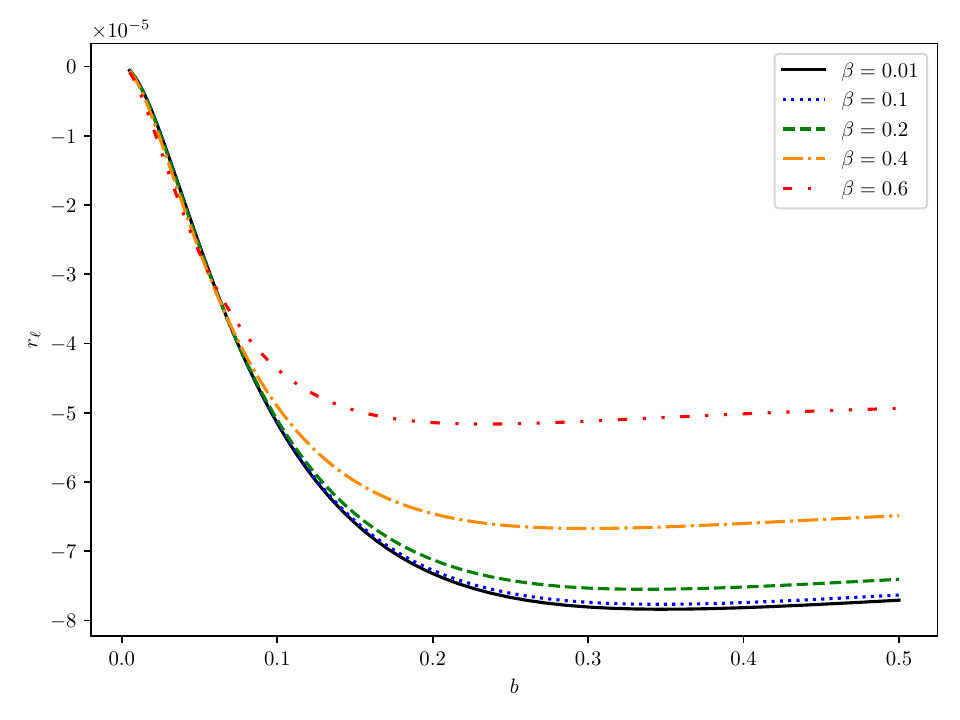}
\caption{Harmonic-gauge constant $r_\ell$ as a function of $b := p_c/\rho_c$ for $n = 1$, $\ell=2$, and selected values of $\beta$.} 
\label{fig:fig10} 
\end{figure} 

\begin{figure}
\includegraphics[width=0.7\linewidth]{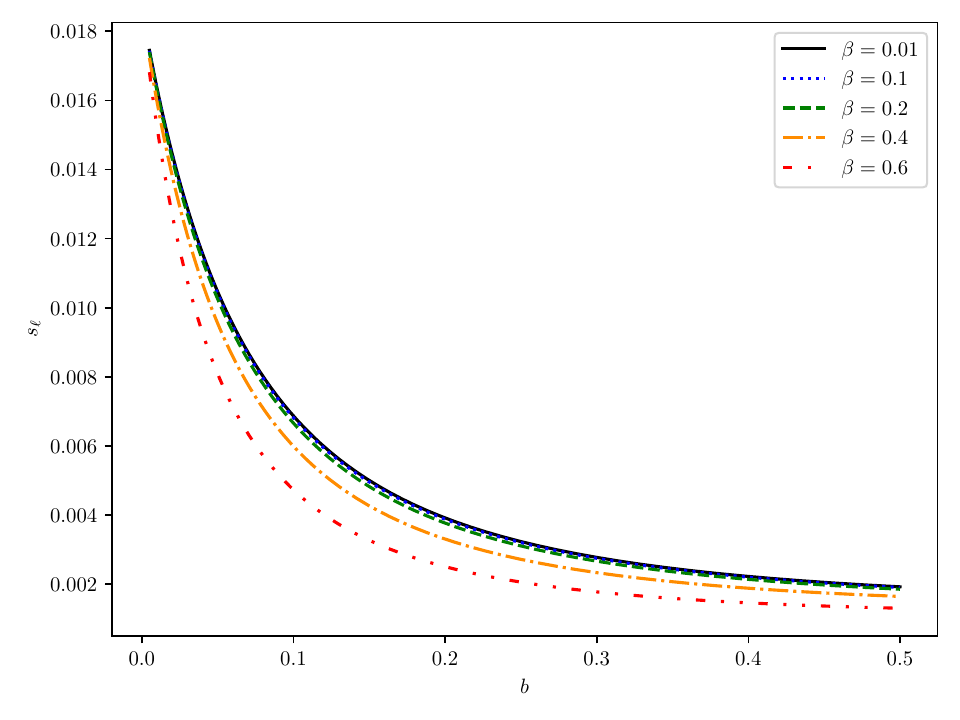}
\caption{Harmonic-gauge constant $s_\ell$ as a function of $b := p_c/\rho_c$ for $n = 1$, $\ell=2$, and selected values of $\beta$.} 
\label{fig:fig11} 
\end{figure} 

We display a sample of our numerical results for the harmonic-gauge constants $r_\ell$ and $s_\ell$ in Figs.~\ref{fig:fig10} and \ref{fig:fig11}, respectively. We do not present results for $c_{10}$ and $c_{30}$, because these interior constants are not particularly interesting. 

\section{Love numbers}
\label{sec:love} 

In this section we collect our results and finally calculate the Love numbers of a tidally deformed body of mass $M$, charge $Q$, and radius $R$. The deformation is produced by a particle of mass $m \ll M$ and charge $q \ll Q$ situated at $r = r_0 > R$. We have a situation of balanced gravitational and electrostatic forces acting on the particle, as described by Eq.~(\ref{balance}).   

\subsection{Tidal and mass multipole moments: Definitions} 

As was explained in Sec.~\ref{sec:pN}, the quantity that allows us to properly identify the tidal multipole moments ${\cal E}^{(\ell)}$ and mass multipole moments ${\cal Q}^{(\ell)}$ is
\begin{equation}
W := \gothg^{tt} := \sqrt{-\hat{g}}\, \hat{g}^{tt},
\end{equation}
where $\hat{g}_{\alpha\beta}$ is the metric of the perturbed spacetime expressed in harmonic coordinates, and $\hat{g}$ is its determinant. Specifically, according to Eq.~(\ref{W_tidal_mass}),
\begin{subequations}
\label{W_tidal_mass_repeat}
\begin{align}
W_\elltide &= -\frac{4}{(\ell-1)\ell}\, {\cal E}^{(\ell)}\, \bar{r}^\ell P_\ell(\cos\theta),\\
W_\ellmass &= \frac{4(2\ell-1)!!}{\ell!}\, {\cal Q}^{(\ell)}\, \bar{r}^{-(\ell+1)} P_\ell(\cos\theta),  
\end{align}
\end{subequations}
and the multipole moments are identified by the coefficients of those terms in $W$ that behave as $\bar{r}^\ell P_\ell(\cos\theta)$ and $\bar{r}^{-(\ell+1)} P_\ell(\cos\theta)$ when $W$ is expanded in inverse powers of the harmonic radial coordinate $\bar{r}$.

There are substantial differences between the (general) discussion of Sec.~\ref{sec:pN} and the (specific) implementation of these ideas carried out here. In Sec.~\ref{sec:pN} we imagined a $W$ computed order by order in a post-Newtonian sequence. Here $W$ is computed in full general relativity, and the post-Newtonian approximation is obtained after the fact, by carrying out the expansion in inverse powers of $\bar{r}$. (It is understood that this expansion is carried out in the region between the body and the perturbing particle.) In Sec.~\ref{sec:pN} we had in mind a general situation in which the tidal deformation is nonlinear and dynamical, such that ${\cal Q}^{(\ell)}$ could be expressed as a simultaneous nonlinearity and time-derivative expansion. Here we have a dramatic simplification of this description. First, our spacetime is static and all moments are independent of time. Second, the tidal deformation is computed to first order in the perturbation, so that ${\cal Q}^{(\ell)}$ is proportional to ${\cal E}^{(\ell)}$.

In the current context, therefore, the relation of Eq.~(\ref{Q_vs_E}) becomes exact,
\begin{equation}
{\cal Q}^{(\ell)} = -\frac{2 (\ell-2)!}{(2\ell-1)!!}\, k_\ell\, R^{2\ell+1}\, {\cal E}^{(\ell)}, 
\end{equation}
and it is this relation that provides an official definition for the Love numbers $k_\ell$. With this we have that 
\begin{equation}
W_\ellmass = -\frac{4}{(\ell-1)\ell}\, 2 k_\ell R^{2\ell+1} {\cal E}^{(\ell)}\,
\bar{r}^{-(\ell+1)} P_\ell(\cos\theta),
\label{Wellmass} 
\end{equation}
which supplies an alternative definition. In our situation the harmonic radial coordinate is given by $\bar{r} = Lz$, with $z(\xi)$ defined by Eq.~(\ref{z_sol}). 

\subsection{Computation of $W$} 

To compute $W$ we need
\begin{equation} 
\hat{g}^{tt} = g^{tt} - p^{tt}_{\rm H} = -f^{-1} \bigl( 1 + f^{-1} p^{\rm H}_{tt} \bigr),
\end{equation}
where $f = (\xi^2-1)/(\xi+\mu)^2$ and $p^{\rm H}_{tt}$ is given by Eq.~(\ref{pert_H}). We also require the determinant of the perturbed metric expressed in the harmonic coordinates of Eq.~(\ref{harm_coordinates}). To get this we note first that in the original coordinates $x^\alpha = (t,r,\theta,\phi)$, and in the harmonic gauge, the metric determinant is $-r^4 \sin^2\theta (1 + p^{\rm H})$, where $p^{\rm H}$ is the trace of the metric perturbation, as given by Eq.~(\ref{pertH_trace}). A transformation to the harmonic coordinates $X^\Omega$ changes this by a factor of $J^{-2}$, where $J := \mbox{det}[\partial X^\Omega/\partial x^\alpha] = L^2 z^2 z' \sin\theta$ is the Jacobian determinant. This gives
\begin{equation}
\sqrt{-\hat{g}} = \frac{(\mu + \xi)^2}{z^2 z'} \Bigl( 1 + \frac{1}{2} p^{\rm H} \Bigr),
\end{equation}
Gathering these results, we arrive at
\begin{equation}
W = \frac{(\mu + \xi)^4}{(\xi^2-1) z^2 z'} 
\Bigl[ 1 + \Gamma_\ell(\xi)\, P_\ell(\cos\theta) \Bigr]
\label{W_final}
\end{equation} 
with 
\begin{equation}
\Gamma_\ell := 4 U_\ell + 2 K_\ell - \frac{\xi^2-1}{(\mu+\xi)^2} \frac{dR_\ell}{d\xi}
- \frac{2(\xi^2-1)}{(\mu+\xi)^3} R_\ell + \frac{\ell(\ell+1)}{\mu+\xi} S_\ell.
\label{Gamma_def}
\end{equation}
In this we may insert Eqs.~(\ref{potential_final}) for $U_\ell$ and $K_\ell$, and Eq.~(\ref{RS_vs_GH}) to express $R_\ell$ and $S_\ell$ in terms of $G_\ell$ and $H_\ell$, given by Eq.~(\ref{GH_sol}). 

\subsection{Tidal multipole moments}
\label{subsec:tidal_moments}

According to Eq.~(\ref{W_tidal_mass_repeat}), the tidal multipole moment ${\cal E}^{(\ell)}$ is identified by the term of order $z^\ell$ in $\Gamma_\ell$, when this quantity is expanded in powers of $z^{-1}$. We recall that $z := \bar{r}/L$ is a dimensionless version of the harmonic radial coordinate, and that in this context of large distances from the body, $\xi$ can be expressed in terms of $z$ by means of Eq.~(\ref{xi_vs_z}). Looking closely at Eq.~(\ref{Gamma_def}), we conclude that such a term necessarily comes from the piece of $U_\ell$ that is proportional to $P_\ell^{\m=1}(\xi)$, which behaves as 
\begin{equation}
P_\ell^{\m=1}(\xi) = \frac{(2\ell-1)!!}{(\ell-1)!}\, \xi^\ell\, \bigl[1 + O(\xi^{-2}) \bigr]
\end{equation}
when $\xi$ is large. Writing $\xi$ in terms of $z$, we find that the relevant contribution to $W$ is  
\begin{equation} 
W_\elltide = -\frac{4}{(\ell-1)\ell} L^\ell {\cal E}^{(\ell)}\, z^\ell P_\ell(\cos\theta),
\end{equation}
in agreement with Eq.~(\ref{W_tidal_mass_repeat}). This means that the quantity ${\cal E}^{(\ell)}$ introduced previously in Eq.~(\ref{E_defalt}) is indeed to be identified with the tidal multipole moment, defined properly by Eq.~(\ref{W_tidal_mass_repeat}). This is given by
\begin{equation}
{\cal E}^{(\ell)} = \frac{(2\ell+1)!!}{(\ell+1)(\ell-2)!} \frac{m}{L^{\ell+1}} Q_\ell^{m=1}(\xi_0), 
\end{equation}
and it is obtained in full general relativity, at all post-Newtonian orders. A post-Newtonian approximation of these moments can be produced by expanding them in inverse powers of $\xi_0$. To leading order we have that 
\begin{equation}
{\cal E}^{(\ell)} = -(\ell-1) \ell\, \frac{m}{\bar{r}_0^{\ell+1}} \bigl[ 1 + O(L/\bar{r}_0, M/\bar{r}_0) \bigr], 
\end{equation}
in agreement with the Newtonian expression of Eq.~(\ref{EL_Newt2}). 

\subsection{Mass multipole moments and Love numbers} 

Next we examine the mass-moment contribution to $W$, given by Eq.~(\ref{Wellmass}). We are looking for terms at order $z^{-(\ell+1)}$ in $W$, and to find all contributions we rely on the asymptotic relations
\begin{equation}
Q_\ell^{\m=1} \sim -\frac{(\ell+1)!}{(2\ell+1)!!}\, \xi^{-(\ell+1)}, \qquad
Q_\ell^{\m=2} \sim \frac{(\ell+2)!}{(2\ell+1)!!}\, \xi^{-(\ell+1)},
\end{equation}
which we insert within Eq.~(\ref{potential_final}) for $U_\ell$ and $K_\ell$, as well as 
\begin{equation}
Q_{\ell-1}(\xi) \sim \frac{(\ell-1)!}{(2\ell-1)!!}\, \xi^{-\ell}, \qquad
Q_\ell(\xi) \sim \frac{\ell!}{(2\ell+1)!!}\, \xi^{-(\ell+1)}, \qquad
Q_{\ell+1}(\xi) \sim \frac{(\ell+1)!}{(2\ell+3)!!}\, \xi^{-(\ell+2)},
\end{equation}
which we substitute within Eq.~(\ref{GH_sol}) for $G_\ell$ and $H_\ell$. A complication arises from the fact that growing terms in $U_\ell$, issued from $P_\ell^{\m=1}(\xi)$, become decaying terms once $\xi$ is re-expressed in terms of $z$ by means of Eq.~(\ref{xi_vs_z}). These contributions must be examined carefully and included in the final result.   

Incorporating all these ingredients, we find that the contribution to $W_\ellmass$ from $U_\ell$ alone is  
\begin{align}
W_\ellmass[U_\ell] &= -\frac{4}{(\ell-1)\ell} L^\ell {\cal E}^{(\ell)} \biggl[
2(p_\ell+q_\ell) (R/L)^{2\ell+1}\, z^{-(\ell+1)}
\nonumber \\ & \quad \mbox{} 
+ \mbox{$z^{-(\ell+1)}$ term in}\ \frac{(\ell-1)!}{(2\ell-1)!!}
\frac{(\mu+\xi)^3}{\sqrt{\xi^2-1}\, z^2 z'} P_\ell^{\m=1}(\xi) \biggr] P_\ell(\cos\theta).
\end{align}
We also find that $K_\ell$ makes no contribution, and that the contributions from $R_\ell$ and $S_\ell$ are
\begin{equation}
W_\ellmass[R_\ell, S_\ell] = -\frac{4}{(\ell-1)\ell} L^\ell {\cal E}^{(\ell)} \Bigl[
2 r_\ell (R/L)^{2\ell+1}\, z^{-(\ell+1)} \Bigr] P_\ell(\cos\theta); 
\end{equation}
these come entirely from $H_\ell$. 

\begin{figure}
\includegraphics[width=0.49\linewidth]{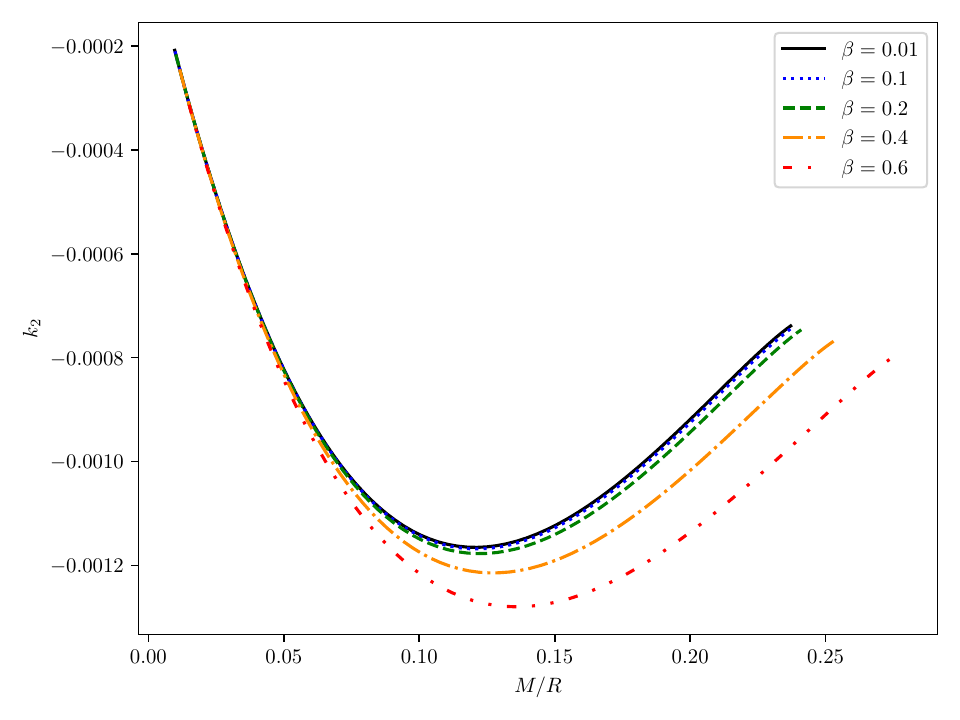}
\includegraphics[width=0.49\linewidth]{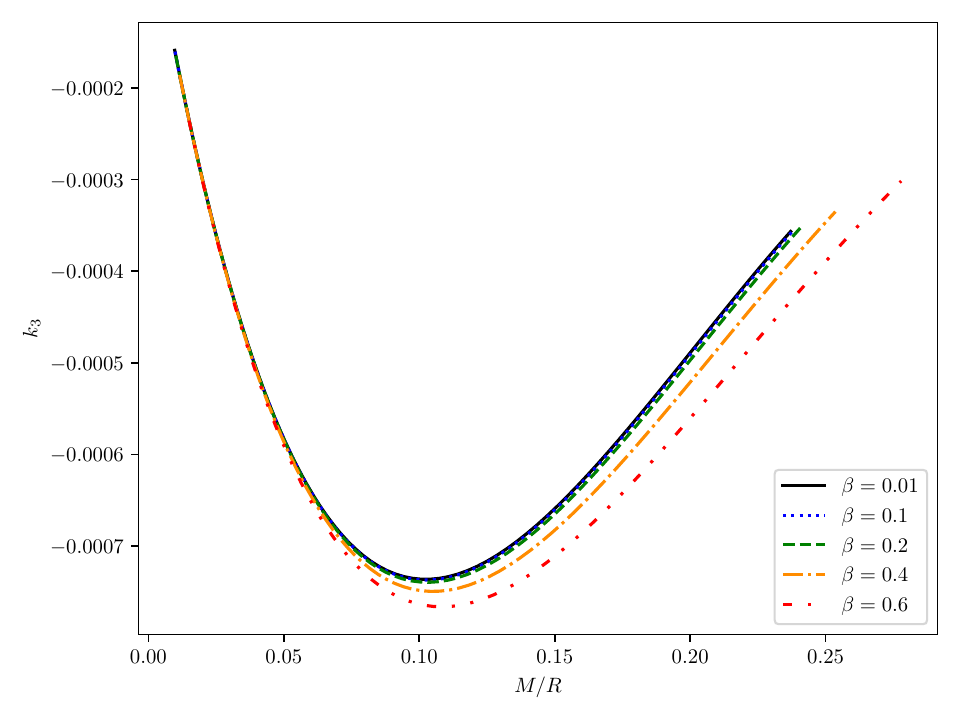}
\caption{Love numbers $k_\ell$ for $\ell=2$ (left) and $\ell=3$ (right) as functions of $M/R$. These are computed for a $n=1$ polytrope and for selected values of $\beta$.}
\label{fig:fig12} 
\end{figure}

\begin{figure}
\includegraphics[width=0.49\linewidth]{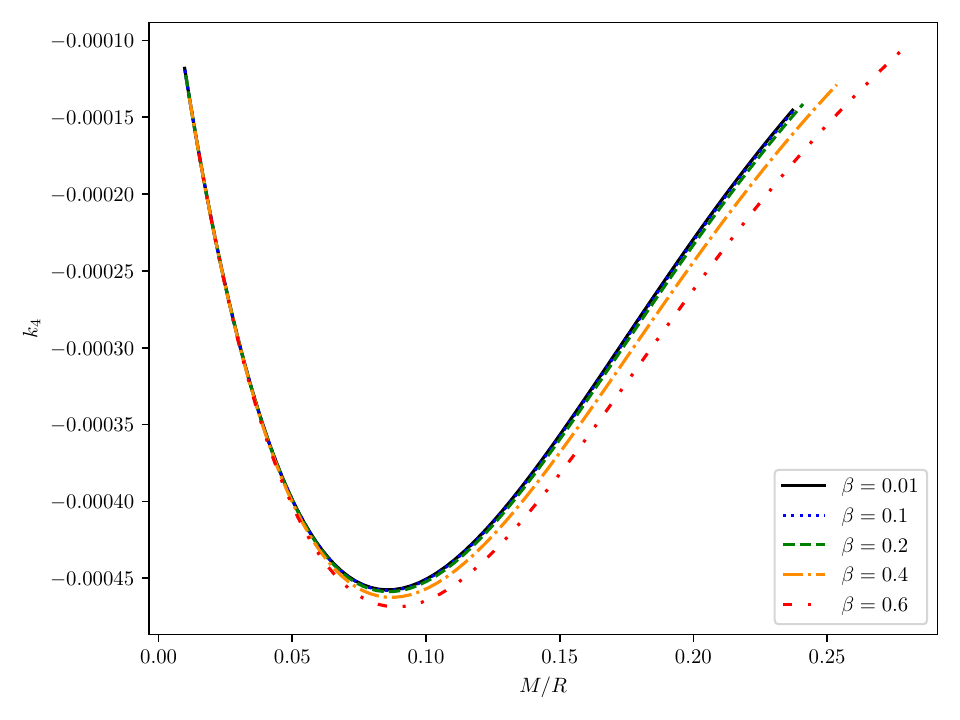}
\includegraphics[width=0.49\linewidth]{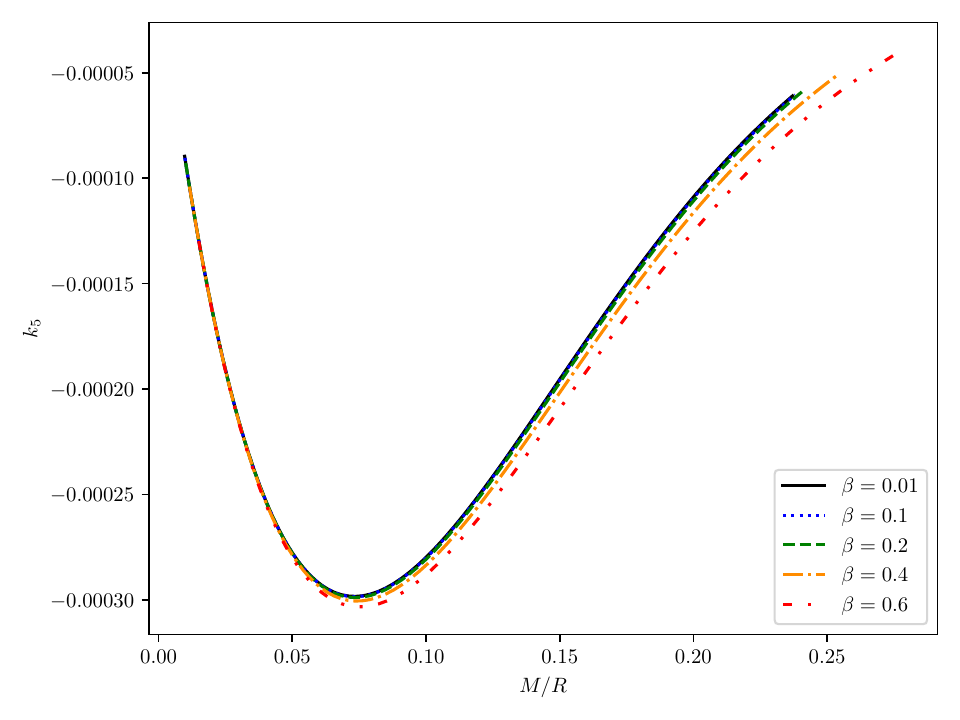}
\caption{Love numbers $k_\ell$ for $\ell=4$ (left) and $\ell=5$ (right) as functions of $M/R$. These are computed for a $n=1$ polytrope and for selected values of $\beta$.}
\label{fig:fig13} 
\end{figure}

With all this we arrive at
\begin{equation}
k_\ell = p_\ell + q_\ell + r_\ell + t_\ell (L/R)^{2\ell+1} 
\label{k_vs_pqrt}
\end{equation}
for the Love numbers, where
\begin{equation}
t_\ell := \mbox{coefficient of $z^{-(\ell+1)}$ in}\
\frac{1}{2} \frac{(\ell-1)!}{(2\ell-1)!!}
\frac{(\mu+\xi)^3}{\sqrt{\xi^2-1}\, z^2 z'} P_\ell^{\m=1}(\xi).
\label{tell_def}
\end{equation} 
We recall that $p_\ell$ and $q_\ell$ are the tidal constants that appear directly in the perturbation of the metric and vector potential (in Regge-Wheeler gauge); they were introduced in Sec.~\ref{sec:tidal_exterior} and calculated in Sec.~\ref{sec:tidal_interior}. We recall also that $r_\ell$ is attributed to the transformation to the harmonic gauge; it first appeared in Eq.~(\ref{GH_sol}) and was computed in Sec.~\ref{sec:harmonic_interior}. The final constant $t_\ell$ is an identifiable function of $\mu := M/L$ and $C$ defined by $z(\xi) = \xi + C Q_1(\xi)$; an explicit listing is 
\begin{subequations}
\label{tell_low}
\begin{align}
t_2 &= 0, \\
t_3 &= \biggl( \frac{9}{175} + \frac{1}{5} \mu^2 \biggr) C, \\
t_4 &= \biggl( \frac{22}{441} + \frac{6}{35} \mu^2 \biggr) C
+ \biggl( \frac{9}{70} + \frac{1}{18} \mu^2 \biggr) \mu C^2, \\
t_5 &= \biggl( \frac{172}{4851} + \frac{4}{35} \mu^2 \biggr) C
+ \biggl( \frac{122}{525} + \frac{4}{45} \mu^2 \biggr) \mu C^2
+ \biggl( \frac{17}{810} + \frac{1}{18} \mu^2 \biggr) C^3
\end{align}
\end{subequations}
for the first few values of $\ell$.

\subsection{Numerical results}
\label{subsec:love_results}

\begin{figure}
\includegraphics[width=0.49\linewidth]{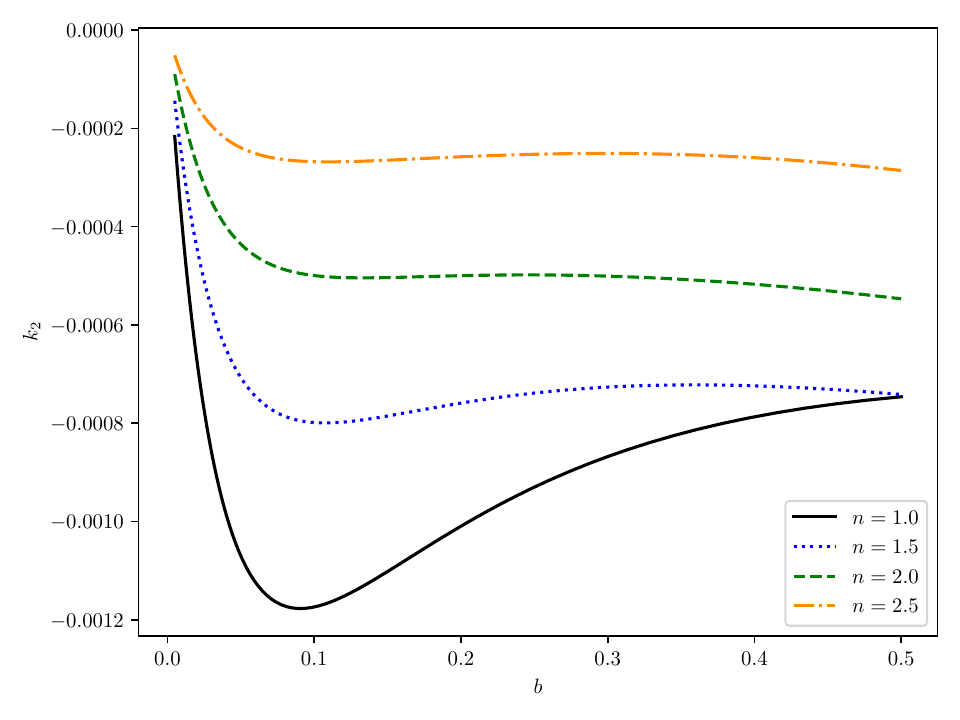}
\includegraphics[width=0.49\linewidth]{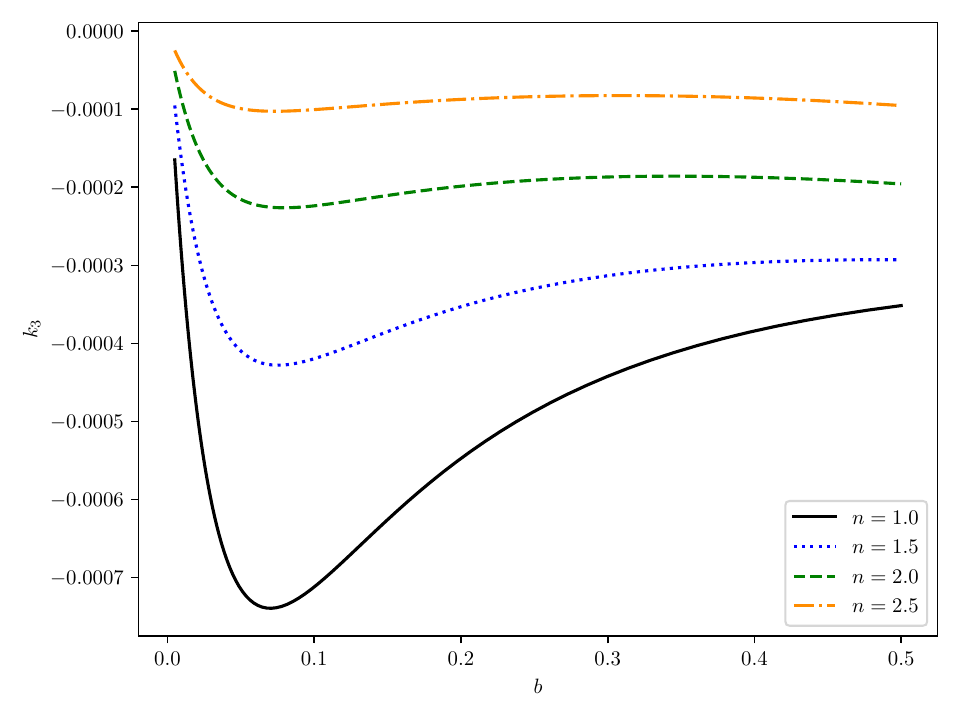}
\caption{Love numbers $k_\ell$ for $\ell=2$ (left) and $\ell = 3$ (right), computed for different polytropic models with $\beta = 0.2$. The Love numbers are plotted as functions of $b := p_c/\rho_c$.}
\label{fig:fig14} 
\end{figure}

\begin{figure}
\includegraphics[width=0.49\linewidth]{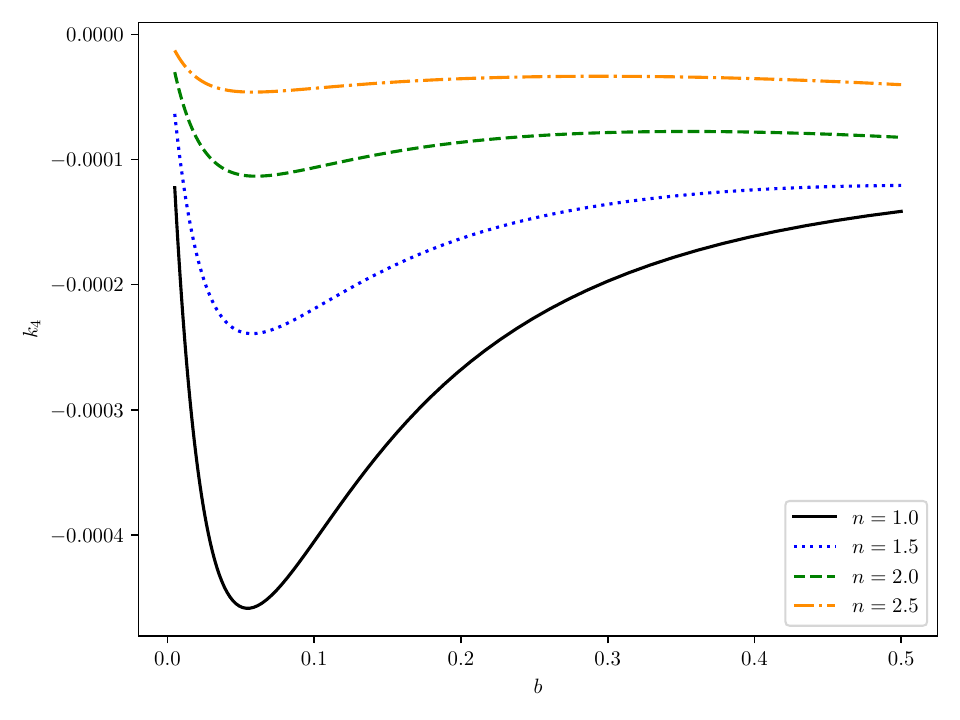}
\includegraphics[width=0.49\linewidth]{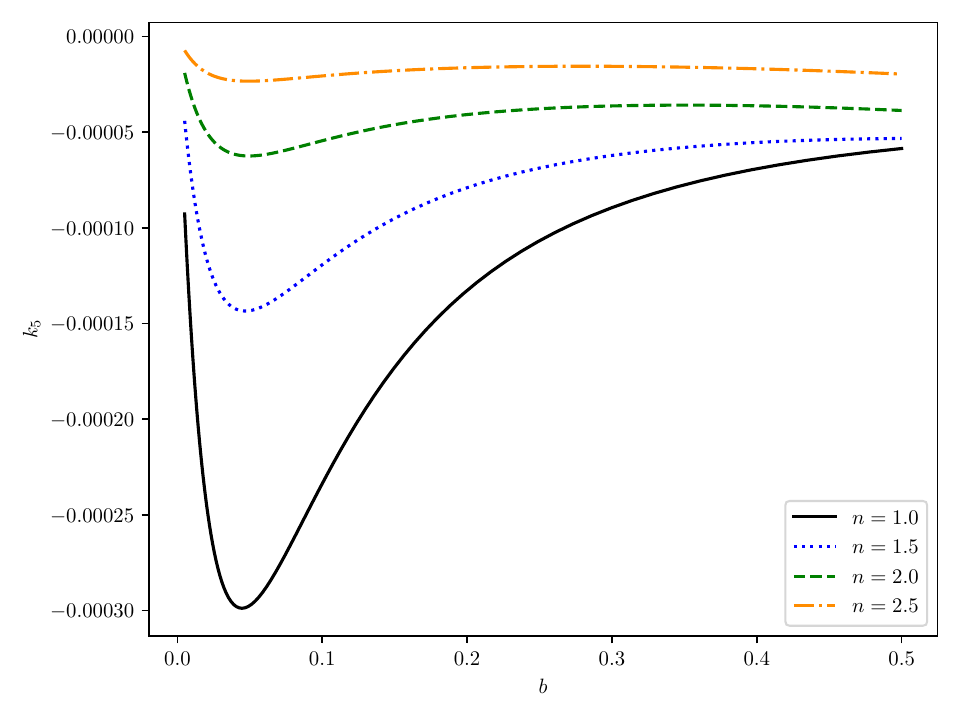}
\caption{Love numbers $k_\ell$ for $\ell=4$ (left) and $\ell = 5$ (right), computed for different polytropic models with $\beta = 0.2$.}
\label{fig:fig15} 
\end{figure}

The constants $p_\ell$, $q_\ell$, and $r_\ell$ were obtained previously by numerical integration of the interior equations and joining the solutions with the exterior solutions at the stellar surface; the constants $t_\ell$ are listed in Eq.~(\ref{tell_low}). We combine them as in Eq.~(\ref{k_vs_pqrt}) and finally obtain the Love numbers of a charged fluid body.

A sample of our results are displayed in Figs. ~\ref{fig:fig12} and \ref{fig:fig13}, in which $k_\ell$ is plotted as a function of $M/R$; the computations are carried out for $\ell = \{ 2, 3, 4, 5 \}$, for a $n=1$ polytropic model, and for selected values of $\beta := \rho_e/\rho_m$. There are three salient points to emphasize. First, the plots reveal that {\it the Love numbers of a charged star in a situation of balanced gravitational and electrostatic forces are negative}. Second, the plots suggest that the limit $\beta \to 0$ of the Love numbers is well defined, in spite of the fact that in a situation of balanced forces, the limit can only be reached by sending $q/m$ to infinity. Third, our numerical results indicate that $k_\ell = -(\mbox{constant}) (M/R) + O(M^2/R^2)$ when $M/R$ is small, with the constant depending on $\ell$ and the fluid's equation of state. The significance of these results was discussed in Sec.~\ref{subsec:intro_work}. 

In Figs.~\ref{fig:fig14} and \ref{fig:fig15} we show $k_\ell$ for $\ell = \{ 2, 3, 4, 5 \}$ and $\beta = 0.2$ for selected values of the polytropic index $n$. The Love numbers are plotted as functions of $b := p_c/\rho_c$, and the figures offer a different view of the results first shown in Figs.~\ref{fig:fig1} and \ref{fig:fig2}, in which the Love numbers were presented as functions of $M/R$. We see that $k_\ell$ is a single-valued function of $b$, which is monotonic in the central density $\rho_c := \rho_m(r=0)$, while it can be a multi-valued function of $M/R$. This change of behavior is explained by Fig.~\ref{fig:fig6}, which reveals that for larger values of the polytropic index $n$, $M/R$ is not monotonic in $b$; the compactness first increases along the sequence of equilibrium configurations, until it reaches a maximum and then starts decreasing. 

The plots of Figs.~\ref{fig:fig14} and \ref{fig:fig15} reveal that for a given value of $b$, the Love numbers tend to decrease (in absolute value) as the polytropic index $n$ increases. This behavior is typical, and has to do with the fact that when $n$ is small, the equation of state is stiff, and the density profile tends to be fairly uniform; such a body comes with large Love numbers (again, in absolute value). On the other hand, when $n$ is large, the equation of state is soft, and the body tends to be centrally dense; such a body comes with small Love numbers.   

\begin{acknowledgments} 
We thank Berend Schneider for helpful conversations. This work was supported by the Natural Sciences and Engineering Research Council of Canada.  
\end{acknowledgments} 

\appendix

\section{Newtonian model}
\label{sec:Newtmodel} 

The computations presented in the bulk of the paper are horrendously complicated. It is helpful to capture their essence in an augmented Newtonian model in which electrostatic field energy is allowed to act as a source of gravity. We present these much simpler calculations in this appendix, and show that the model returns negative Love numbers that are proportional to $M/R$ when the compactness is small. These features are shared with the relativistic results displayed in Figs.~\ref{fig:fig12} and \ref{fig:fig13}.

It turns out to be crucial for our purposes here to allow the electrostatic field energy to act as a source of gravity. In a strictly Newtonian model, in which only mass creates gravity, the calculation would return vanishing Love numbers --- refer back to Sec.~\ref{subsec:intro_work}. This happens because the condition of balanced gravitational and electrostatic forces on the perturbing particle necessarily applies also to each fluid element within the star, and there is no tidal deformation. In the augmented model, the distributions of field energy are different inside and outside the star, and we no longer have balanced forces within the star; there is now a nonvanishing tidal deformation. Furthermore, the multipole moments obtained in this setting account for the deformed distributions of both mass and field energy. These features are automatically incorporated in the relativistic calculation presented in the bulk of the paper; they should also be present in the ``Newtonian'' model to achieve comparable results. 

\subsection{Model}

We consider a spherical body of mass $M$, charge $Q$, and radius $R$ perturbed by a particle of mass $m$ and charge $q$ at a distance $r_0$ from the body. For simplicity we take the body to possess a constant density of mass $\rho_m$, and a constant density of charge $\rho_e$; we have that
\begin{equation}
M = \frac{4\pi}{3} \rho_m R^3, \qquad
Q = \frac{4\pi}{3} \rho_e R^3, 
\end{equation}
and $\beta := \rho_e/\rho_m = Q/M$ is the charge-to-mass ratio.

The body and particle create an electric field $\bm{E}$ that is related to a potential $V$ via $E_a = -\nabla_a V$. They also create a gravitational field $\bm{g}$ related to a potential $U$ by $g_a = \nabla_a U$. The electrostatic energy density ${\bm E}^2/(8\pi)$ is added to the mass density $\rho_m$ as a source of gravity. The field equations are
\begin{subequations}
\label{Nfield_eqns} 
\begin{align} 
\nabla^2 V &= -4\pi \rho_e - 4\pi q \delta(\bm{x} - \bm{r}_0), \\
\nabla^2 U &= -4\pi \rho_m - \tfrac{1}{2} \bm{\nabla} V \cdot \bm{\nabla} V - 4\pi m \delta(\bm{x} - \bm{r}_0),
\end{align}
\end{subequations}
in which $\nabla^2 := g^{ab} \nabla_a \nabla_b$ is the Laplacian operator in Euclidean, three-dimensional space, $\bm{x}$ is the position of an arbitrary field point, and $\bm{r}_0$ denotes the particle's position. Working in spherical polar coordinates, we associate $\bm{x}$ with the coordinates $(r,\theta,\phi)$, while $\bm{r}_0$ is associated with $(r_0, 0, \phi_0)$, as in Sec.~\ref{sec:exterior}; the particle is situated on the polar axis, and $\phi_0$ is an arbitrarily assigned angle that plays no role in the calculation.

The equation of hydrostatic equilibrium is taken to be
\begin{equation}
\bm{\nabla} p = \rho_m \bm{\nabla} U - \rho_e \bm{\nabla} V,
\label{N_hydro}
\end{equation}
where $p$ is the pressure; the densities of gravitational and electrostatic forces appear on the right-hand side. We were tempted to replace $\rho_m$ with $\rho_m + \bm{E}^2/(8\pi)$ in this equation to be consistent with Eq.~(\ref{Nfield_eqns}), but decided against this unnecessary complication. The only essential modification to the standard Newtonian theory is that the electric field energy acts as a source of gravity.

\subsection{Unperturbed structure}

We first examine the structure of the unperturbed body, in the absence of the point particle. In spherical symmetry the field equations become
\begin{subequations}
\begin{align}
0 &= V'' + \frac{2}{r} V' + \frac{3Q}{R^3}, \\
0 &= U'' + \frac{2}{r} U' + \frac{3M}{R^3} + \frac{1}{2} (V')^2
\end{align}
\end{subequations}
inside the body, and
\begin{subequations}
\begin{align}
0 &= V'' + \frac{2}{r} V', \\
0 &= U'' + \frac{2}{r} U' + \frac{1}{2} (V')^2
\end{align}
\end{subequations}
outside; a prime indicates differentiation with respect to $r$. The potentials must be regular at $r=0$ and $r=\infty$, and they must be continuous and differentiable at the stellar surface. The solution that satisfies all these requirements is
\begin{subequations}
\label{Nunpert_potentials}
\begin{align} 
V^{\rm out} &= \frac{Q}{r}, \\
U^{\rm out} &= \biggl( M + \frac{3}{5} \frac{Q^2}{R} \biggr) \frac{1}{r} - \frac{Q^2}{4 r^2}, \\
V^{\rm in} &= \frac{Q}{2R} ( 3 - r^2/R^2 ), \\
U^{\rm in} &= \frac{M}{2R} ( 3 - r^2/R^2 ) + \frac{Q^2}{40R^2} (15 - r^4/R^4).
\end{align}
\end{subequations}
It may be noted that while $M$ denotes the body's {\it material mass}, the {\it gravitational mass} $M_{\rm grav}$ differs from this by virtue of the distribution of electrostatic energy; we have that $M_{\rm grav} = M + 3Q^2/(5 R)$.

For a body of constant densities, Eq.~(\ref{N_hydro}) implies that $p = \rho_m U - \rho_e V + \mbox{const}$, where the constant is determined by the requirement that the pressure must vanish at $r=R$. This yields
\begin{equation}
p = \biggl[ \frac{3(M^2-Q^2)}{8\pi R^4} + \frac{3M Q^2}{160\pi R^5} (1 + r^2/R^2) \biggr] (1 - r^2/R^2).
\label{Nunpert_pressure} 
\end{equation}

\subsection{Perturbation equations} 

We now insert the particle, and let it create a perturbation of the solution constructed previously. We write the perturbed potentials as $V + \delta V$ and $U + \delta U$, with $\delta V$ and $\delta U$ denoting the perturbations, and $p + \delta p$ is the perturbed pressure. The densities $\rho_m$ and $\rho_e$ remain at their constant value, and are therefore not perturbed. After linearization with respect to the perturbations, we find that the field equations become
\begin{equation}
\nabla^2 \delta V = 0, \qquad
\nabla^2 \delta U = -\bm{\nabla} V \cdot \bm{\nabla} \delta V
\label{Npert_in}
\end{equation}
inside the body, and
\begin{equation}
\nabla^2 \delta V =-4\pi q \delta(\bm{x} - \bm{r}_0), \qquad
\nabla^2 \delta U = -\bm{\nabla} V \cdot \bm{\nabla} \delta V - 4\pi m \delta(\bm{x} - \bm{r}_0)
\label{Npert_out} 
\end{equation}
outside the body. The perturbation in the pressure is given by
\begin{equation}
\delta p = \rho_m\, \delta U - \rho_e\, \delta V.
\label{Npertp} 
\end{equation}

The perturbed potentials are continuous and differentiable at the deformed surface, which is described by
\begin{equation}
r = R + \delta R,
\end{equation}
where $\delta R$ is a function of the polar angle $\theta$. These requirements produce the junction conditions
\begin{equation}
\bigl[ \delta V \bigr]_R = 0, \qquad
\bigl[ \delta U \bigr]_R = 0
\label{Njunction1} 
\end{equation}
and
\begin{equation}
\bigl[ \partial_r \delta V \bigr]_R + \bigl[ V'' \bigr]_R\, \delta R = 0, \qquad 
\bigl[ \partial_r \delta U \bigr]_R + \bigl[ U'' \bigr]_R\, \delta R = 0,  
\label{Njunction2} 
\end{equation}
where $[\psi]_R := \psi(r=R^+) - \psi(r=R^-)$ denotes the jump of a quantity $\psi$ across $r=R$. The perturbed pressure vanishes at the deformed surface, and this requirement yields
\begin{equation}
\delta p(r=R) + p'(r=R)\, \delta R = 0.
\label{Njunction3}
\end{equation}
This equation determines $\delta R$, which is then inserted within Eqs.~(\ref{Njunction1}) and (\ref{Njunction2}).

We decompose all perturbations in Legendre polynomials,
\begin{subequations}
\begin{align}
\delta V &= \sum_{\ell=0}^\infty V_\ell(r)\, P_\ell(\cos\theta), \\
\delta U &= \sum_{\ell=0}^\infty U_\ell(r)\, P_\ell(\cos\theta), \\
\delta p &= \sum_{\ell=0}^\infty p_\ell(r)\, P_\ell(\cos\theta), \\
\delta R &= \sum_{\ell=0}^\infty R_\ell\, P_\ell(\cos\theta), 
\end{align}
\end{subequations}
insert these within Eqs.~(\ref{Npert_in}) and (\ref{Npert_out}), write $\delta(\bm{x}-\bm{r}_0) = r^{-2} \delta(r-r_0) \delta(\cos\theta-1) \delta(\phi-\phi_0)$, make use of Eq.~(\ref{delta_identity}), and average the equations with respect to $\phi$. We arrive at the system of equations
\begin{subequations}
\label{Npert_ell_in} 
\begin{align}
0 &= r^2 V_\ell'' + 2r V_\ell' - \ell(\ell+1) V_\ell, \\
0 &= r^2 U_\ell'' + 2r U_\ell' - \ell(\ell+1) U_\ell + r^2 V' V_\ell'
\end{align}
\end{subequations}
inside the body, and
\begin{subequations}
\label{Npert_ell_out} 
\begin{align}
0 &= r^2 V_\ell'' + 2r V_\ell' - \ell(\ell+1) V_\ell + (2\ell+1) q\, \delta(r-r_0), \\
0 &= r^2 U_\ell'' + 2r U_\ell' - \ell(\ell+1) U_\ell + r^2 V' V_\ell' + (2\ell+1) m\, \delta(r-r_0)
\end{align}
\end{subequations}
outside. Equation (\ref{Npertp}) becomes $p_\ell = \rho_m U_\ell - \rho_e V_\ell$, and Eqs.~(\ref{Njunction1}), (\ref{Njunction2}), and (\ref{Njunction3}) can be projected in a similar manner.

In addition to the junction conditions at $r=R$, the delta functions in Eq.~(\ref{Npert_ell_out}) give rise to another set of junction conditions at $r=r_0$. We have that
\begin{equation}
\bigl[ V_\ell \bigr]_{r_0} = 0, \qquad
\bigl[ V_\ell' \bigr]_{r_0} = -(2\ell+1)\, \frac{q}{r_0^2}
\label{Njump_r0_1} 
\end{equation}
and
\begin{equation}
\bigl[ U_\ell \bigr]_{r_0} = 0, \qquad
\bigl[ U_\ell' \bigr]_{r_0} = -(2\ell+1)\, \frac{m}{r_0^2}.
\label{Njump_r0_2} 
\end{equation}

\subsection{Solution} 

It is a straightforward matter to find solutions to Eqs.~(\ref{Npert_ell_in}) and (\ref{Npert_ell_out}) subjected to the junction conditions of Eqs.~(\ref{Njunction1}), (\ref{Njump_r0_1}), and (\ref{Njump_r0_2}); notice that we momentarily exclude Eq.~(\ref{Njunction2}) from the complete list of conditions. When we also impose regularity at $r=0$ and $r=\infty$, we arrive at
\begin{subequations}
\label{NV_solution}
\begin{align}
V^{\rm in}_\ell(r) &=\biggl( c_1 \frac{Q}{R^{\ell+1}} + \frac{q}{r_0^{\ell+1}} \biggr) r^\ell, \\
V^{\rm out}_\ell(r < r_0) &= c_1 Q R^\ell\, \frac{1}{r^{\ell+1}} + \frac{q}{r_0^{\ell+1}}\, r^\ell, \\
V^{\rm out}_\ell(r > r_0) &= ( c_1 Q R^\ell + q r_0^\ell )\, \frac{1}{r^{\ell+1}}
\end{align}
\end{subequations}
and
\begin{subequations}
\label{NU_solution}
\begin{align}
U^{\rm in}_\ell(r) &= \frac{\ell}{2(2\ell+3)} \frac{Q}{R^3} \biggl( c_1 \frac{Q}{R^{\ell+1}}
+ \frac{q}{r_0^{\ell+1}} \biggr)\, r^{\ell+2}
\nonumber \\ & \quad \mbox{} 
+ \biggl[ -\frac{3(\ell+1)}{2(2\ell+3)} c_1 \frac{Q^2}{R^{\ell+2}}
+ c_2 \frac{M}{R^{\ell+1}} + \frac{m - \frac{3}{2} \frac{\ell+1}{2\ell+3} \frac{qQ}{R}}{r_0^{\ell+1}}
+ \frac{qQ}{2 r_0^{\ell+2}} \biggr]\, r^\ell, \\
U^{\rm out}_\ell(r < r_0) &= c_2 M R^\ell\, \frac{1}{r^{\ell+1}}
+ \frac{m + \frac{qQ}{2r_0}}{r_0^{\ell+1}}\, r^\ell
- \frac{1}{2} c_1 Q^2 R^\ell\, \frac{1}{r^{\ell+2}} - \frac{qQ}{2 r_0^{\ell+1}}\, r^{\ell-1}, \\
U^{\rm out}_\ell(r > r_0) &= \Biggl[ c_2 M R^\ell + \biggl( m + \frac{qQ}{2 r_0} \biggr) r_0^\ell \Biggr]\, \frac{1}{r^{\ell+1}}
- \frac{1}{2} Q (c_1 Q R^\ell + q r_0^\ell)\, \frac{1}{r^{\ell+2}},
\end{align}
\end{subequations}
where $c_1$ and $c_2$ are dimensionless constants. These are determined by finally enforcing Eq.~(\ref{Njunction2}), in which we insert $[V''] = 3Q/R^3$, $[U''] = 3M/R^3$, and
\begin{equation}
p'(r=R) = -\frac{3(M^2-Q^2)}{4\pi R^5} - \frac{3M Q^2}{40\pi R^6}.
\end{equation}
We get
\begin{align}
c_1 &= \biggl[ 20(\ell-1)(2\ell+1)(2\ell+3) (M^2-Q^2) R + 2(\ell-1)(4\ell^2+14\ell+21) MQ^2 \biggr]^{-1}
\nonumber \\ & \quad \mbox{} \times 
\biggl\{ 30(2\ell+1)(2\ell+3)(mM-qQ) \frac{R^{\ell+2}}{r_0^{\ell+1}}
+ 15(2\ell+1)(2\ell+3) qMQ \frac{R^{\ell+2}}{r_0^{\ell+2}}
- 60\ell (\ell+2) qMQ \frac{R^{\ell+1}}{r_0^{\ell+1}} \biggr\}
\end{align}
and
\begin{align}
c_2 &= \biggl[ 20(\ell-1)(2\ell+1)(2\ell+3) (M^2-Q^2) R + 2(\ell-1)(4\ell^2+14\ell+21) MQ^2 \biggr]^{-1}
\nonumber \\ & \quad \mbox{} \times 
\biggl\{ 30(2\ell+1)(2\ell+3)(mM-qQ) \frac{R^{\ell+2}}{r_0^{\ell+1}}
+ 15(2\ell+1)(2\ell+3) qMQ \frac{R^{\ell+2}}{r_0^{\ell+2}}
\nonumber \\ & \quad \mbox{} 
- 30 \frac{Q}{M} \Bigl[ (\ell+1)(2\ell+1) qM^2 + (\ell+2)(2\ell+1) q Q^2
- (2\ell^2+4\ell+3) mMQ \Bigr] \frac{R^{\ell+1}}{r_0^{\ell+1}}
\nonumber \\ & \quad \mbox{} 
+ 15 (2\ell^2+4\ell+3) q Q^3 \frac{R^{\ell+1}}{r_0^{\ell+2}}
- 3(2\ell+1)(5\ell+7) q Q^3 \frac{R^\ell}{r_0^{\ell+1}} \biggr\}.
\end{align} 
We now have a complete solution to the system of perturbation equations. 

\subsection{Force balance}

Our solution is currently valid for any choice of parameters $M$, $Q$, $m$, $q$, and $r_0$. The situation considered in the main text, however, has the particle subjected to balanced gravitational and electrostatic forces. We choose to impose this condition here as well.

The forces are balanced when $m U' = q V'$ at $r=r_0$. Using the unperturbed potentials of Eq.~(\ref{Nunpert_potentials}), we find that this equation produces
\begin{equation}
qQ = m \biggl( M + \frac{3}{5} \frac{Q^2}{R} \biggr) - \frac{m Q^2}{2 r_0}.
\label{Nbalance}
\end{equation}
In conventional Newtonian gravity we would of course get the expected $qQ = mM$, but this is modified in the current context, in which electrostatic energy is a source of gravity.

\subsection{Love numbers}

The discussion of Sec.~\ref{sec:newton} reveals that the tidal multipole moment is obtained from the piece of $U_\ell(r < r_0)$ that is proportional to $r^\ell$. Comparison between Eqs.~(\ref{tidal_potential2}) and (\ref{NU_solution}) yields
\begin{equation}
{\cal E}^{(\ell)} = -(\ell-1)\ell\, \frac{m + \frac{qQ}{2r_0}}{r_0^{\ell+1}},
\end{equation}
a modification of Eq.~(\ref{EL_Newt2}). The discussion reveals also that the mass multipole moment is provided by the piece of $U_\ell(r < r_0)$ that is proportional to $r^{-(\ell+1)}$. Comparison between Eqs.~(\ref{resp_potential2}) and (\ref{NU_solution}) yields
\begin{equation}
{\cal Q}^{(\ell)} = \frac{\ell!}{(2\ell-1)!!}\, c_2 M R^\ell.
\end{equation}
The Love numbers are then defined by Eq.~(\ref{Q_vs_E2}).

We simplify the final expression for $k_\ell$ by assuming that $R/r_0 \ll 1$ and taking the formal limit $R/r_0 \to 0$. We also assume that $\beta := Q/M \ll 1$, and express the Love number as a Taylor series in powers of this quantity. Finally, we write the result in terms of $M/R$, the body's compactness. This returns
\begin{align}
k_\ell &\simeq -\frac{3(\ell+1)}{4(\ell-1)(2\ell+3)}\, \frac{M}{R}
\nonumber \\ & \quad \mbox{} 
- \frac{3}{20(\ell-1)(2\ell+1)(2\ell+3)^2} \biggl[ (2\ell+3)(22\ell^2 + 44\ell + 9)
+ (20\ell^3 + 61\ell^2 + 51\ell + 9) \frac{M}{R} \biggr] \frac{M}{R} \beta^2 + O(\beta^4). 
\label{Nlove}
\end{align}
This expression informs us that $k_\ell$ is negative and proportional to $M/R$ when the compactness is small. It also reveals that the $\beta \to 0$ limit of the Love number is well defined, in spite of the fact that
the limit can only be reached by sending $q/m$ to infinity. 

\section{Harmonic radius for a thin-shell spacetime}
\label{sec:Cshell} 

It was observed in Fig.~\ref{fig:fig9} that $(M/R) C$ approaches a finite value when the stellar compactness $M/R$ tends to zero; here $C$ is the harmonic constant defined by Eq.~(\ref{z_sol}). In this appendix we demonstrate this behavior through an analytical calculation that implicates an idealized body instead of a polytropic star. We take the body to consist of an infinitely thin shell of mass $M$ and radius $R$. For simplicity we set the charge $Q$ to zero, as it has a minimal impact on the behavior of $C$ in the limit $M/R \to 0$.  

The metric outside the shell (for $r > R$) is given by the Schwarzschild solution,
\begin{equation}
ds^2_{\rm out} = -f\, dt^2 + f^{-1}\, dr^2 + r^2 \bigl( d\theta^2 + \sin^2\theta\, d\phi^2 \bigr),
\end{equation}
where $f := 1-2M/r$. The harmonic radial coordinate $\bar{r}$ is given there by
\begin{equation}
\bar{r}_{\rm out} = r - M - C \bigl[ \tfrac{1}{2} (r-M) \ln f + M \bigr];
\end{equation}
this is Eq.~(\ref{z_sol}) in which we substituted $\xi = r/M - 1$. We recall that the harmonic radius is defined so that $X^1 := \bar{r}\, \sin\theta\cos\phi$, $X^2 := \bar{r}\, \sin\theta\sin\phi$, and $X^3 := \bar{r}\, \cos\theta$ satisfy Eq.~(\ref{wave-eqn}) in the Schwarzschild spacetime. It is normalized so that $\bar{r} \sim r$ when $r \to \infty$.

Inside the shell (for $r < R$) we have the Minkowski metric expressed in spherical coordinates,
\begin{equation}
ds^2_{\rm in} = -F\, dt^2 + dr^2 + r^2 \bigl( d\theta^2 + \sin^2\theta\, d\phi^2 \bigr),
\end{equation}
where $F := 1-2M/R$. The time coordinate was rescaled so that it is continuous at $r=R$. In this region of spacetime we have that the harmonic radial coordinate is given by
\begin{equation}
\bar{r}_{\rm in} = B r,
\end{equation}
where $B$ is a constant of integration. It is easy to verify that with this assignment, $X^1$, $X^2$, and $X^3$ are solutions to Eq.~(\ref{wave-eqn}) in Minkowski spacetime.

The radial harmonic coordinate must be continuous and differentiable at $r=R$:
\begin{equation}
\bigl[ \bar{r} \bigr] = 0, \qquad
\bigl[ n^\alpha \nabla_\alpha \bar{r} \bigr] = 0,
\label{junction_harmonic} 
\end{equation}
where $[ \psi ] := \psi_{\rm out}(r=R) - \psi_{\rm in}(r=R)$ is the jump of a quantity $\psi$ across the shell, and $n^\alpha$ is the unit normal to the shell. A derivation of these junction conditions is provided below. For our situation they become
\begin{equation}
\bar{r}_{\rm out} = \bar{r}_{\rm in}, \qquad
\sqrt{f}\, \frac{d \bar{r}_{\rm out}}{dr} = \frac{d \bar{r}_{\rm in}}{dr},
\end{equation}
where both sides of each equation are evaluated at $r=R$.

The junction conditions provide us with two equations for the two unknowns $B$ and $C$. The solutions are
\begin{equation}
B = \biggl\{ \frac{R^3}{2M^3} \Bigl[ 1 - (1-M/R) \sqrt{F} \Bigr] \ln F
+ \frac{R^2}{M^2} \Bigl( 1 - M/R - \sqrt{F} \Bigr) \biggr\}^{-1}
\end{equation}
and
\begin{equation}
C = \frac{R^3}{M^3} \Bigl[ F - (1-M/R) \sqrt{F} \Bigr]\, B.
\end{equation}
When $M/R \ll 1$ the constants admit the series expansions
\begin{equation}
B = 1 - M/R + O(M^2/R^2), \qquad
C = -\frac{R}{2M} \bigl[ 1 - M/R + O(M^2/R^2) \bigr].
\end{equation}
We see that $(M/R) C$ does indeed approach a finite value when $M/R \to 0$. In this idealized case, the limiting value is equal to $-1/2$. In the case of a polytropic star, the numerical value depends on the parameters that enter the description of the equation of state.

To establish Eq.~(\ref{junction_harmonic}) we exploit the distributional techniques utilized in Sec.~3.7 of Ref.~\cite{poisson:b04}, which provides a derivation of the Israel junction conditions \cite{israel:66} across a singular hypersurface. We invoke the existence of a coordinate system $x$ in which all components of the metric $g$ are continuous across the hypersurface. (We omit all tensorial indices to simplify the notation in this discussion.) The final result, the conditions of Eq.~(\ref{junction_harmonic}), are independent of these coordinates; they are scalar equations that can be formulated in any coordinate system, continuous or discontinuous.

In the coordinates $x$ the metric can be expressed as the distribution $g = g_+\, \Theta(\ell) + g_-\, \Theta(-\ell)$, in which $\ell$ is the proper distance to the hypersurface as measured along geodesics that cross it orthogonally, and $\Theta$ is the Heaviside step function. The coordinates are such that $[g] = 0$; the metric is continuous across the hypersurface. This property ensures that the connection can be expressed as $\Gamma = \Gamma_+\, \Theta(\ell) + \Gamma_-\, \Theta(-\ell)$, without a term proportional to $\delta(\ell)$, with $\delta := d\Theta/d\ell$ denoting the Dirac distribution.

In a similar manner we express the harmonic coordinates as the distribution $X = X_+\, \Theta(\ell) + X_-\, \Theta(-\ell)$. We impose continuity at the hypersurface, so that $[X] = 0$. When we then insert this within Eq.~(\ref{wave-eqn}), we arrive at
\begin{equation}
0 = \Box X = (\Box X_+)\, \Theta(\ell) + (\Box X_-)\, \Theta(-\ell)
+ \bigl[ n \cdot \nabla X \bigr]\, \delta(\ell),
\label{wave-eqn_distribution} 
\end{equation}
where $\Box$ is the wave operator that occurs in Eq.~(\ref{wave-eqn}), and $n \cdot \nabla = d/d\ell$ is the derivative in the direction of the unit normal. To arrive at this we made use of the geometric definition of $\ell$ to write $n_\alpha = \nabla_\alpha \ell$.

Equation (\ref{wave-eqn_distribution}) is satisfied as a distribution when $\Box X_\pm = 0$ on each side of the hypersurface, and when $[X] = 0 = [n \cdot \nabla X]$. These junction conditions implicate spacetime scalars that can be expressed in any system of coordinates. In our application, we work with the coordinates $(t,r,\theta,\phi)$, in which the metric is discontinuous across the thin shell. We find that the junction conditions become Eq.~(\ref{junction_harmonic}).

\bibliography{love_charged}
\end{document}